\documentclass[twocolumn]{aastex6}
\usepackage{hyperref}
\usepackage{amsmath}
\newcommand\sinopsis{{\sc sinopsis}}
\def\ls{{_<\atop^{\sim}}}
\def\gs{{_>\atop^{\sim}}}

\slugcomment{}

\shorttitle{GASP: the JO36 test case}
\shortauthors{Fritz et al.}

\begin{document}

\title{GASP III. JO36: a case of multiple environmental effects at play?}

\author{Jacopo Fritz\altaffilmark{1}$^*$, Alessia Moretti\altaffilmark{2}, Marco Gullieuszik\altaffilmark{2}, Bianca Poggianti\altaffilmark{2}, Gustavo Bruzual\altaffilmark{1}, Benedetta Vulcani\altaffilmark{3,2}, Fabrizio Nicastro\altaffilmark{4}, Yara Jaff\'e\altaffilmark{5}, Bernardo Cervantes Sodi\altaffilmark{1}, Daniela Bettoni\altaffilmark{2}, Andrea Biviano\altaffilmark{5}, Giovanni Fasano\altaffilmark{2}, St\'ephane Charlot\altaffilmark{6}, Callum Bellhouse\altaffilmark{7,8}, and George Hau\altaffilmark{8}}

\altaffiltext{1}{Instituto de Radioastronom\'\i a y Astrof\'\i sica, UNAM, Campus Morelia, A.P. 3-72, C.P. 58089, Mexico}
\altaffiltext{2}{INAF-Osservatorio Astronomico di Padova, Vicolo dell'Osservatorio 5, Padova, Italy}
\altaffiltext{3}{School of Physics, The University of Melbourne Swanston St \& Tin Alley Parkville VIC 3010}
\altaffiltext{4}{INAF-Osservatorio Astronomico di Roma, Via di Frascati 33, I-00040 Monte Porzio Catone, RM, Italy}
\altaffiltext{5}{INAF-Osservatorio Astronomico di Trieste, via G. B. Tiepolo 11, I-34131, Trieste, Italy}
\altaffiltext{6}{Sorbonne Universit\'es, UPMC-CNRS, UMR7095, Institut d'Astrophysique de Paris, F-75014 Paris, France}
\altaffiltext{7}{University of Birmingham School of Physics and Astronomy, Edgbaston, Birmingham, England}
\altaffiltext{8}{European Southern Observatory, Alonso de Cordova 3107, Vitacura, Casilla 19001, Santiago de Chile, Chile}
\altaffiltext{*}{E-mail: j.fritz@crya.unam.mx}

\begin{abstract}
The so--called jellyfish galaxies are objects exhibiting disturbed morphology, mostly in the form of tails of gas stripped from the main body of the galaxy. Several works have strongly suggested ram pressure stripping to be the mechanism driving this phenomenon. Here, we focus on one of these objects, drawn from a sample of optically selected jellyfish galaxies, and use it to validate \sinopsis, the spectral fitting code that will be used for the analysis of the GASP (GAs Stripping Phenomena in galaxies with MUSE) survey, and study the spatial distribution and physical properties of gas and stellar populations in this galaxy. We compare the model spectra to those obtained with {\sc gandalf}, a code with similar features widely used to interpret the kinematic of stars and gas in galaxies from IFU data. We find that \sinopsis \ can reproduce the pixel--by--pixel spectra of this galaxy at least as good as {\sc gandalf} does, providing reliable estimates of the underlying stellar absorption to properly correct the nebular gas emission. Using these results, we find strong evidences of a double effect of ram pressure exerted by the intracluster medium onto the gas of the galaxy. A moderate burst of star formation, dating between 20 and 500 Myr ago and involving the outer parts of the galaxy more strongly than the inner regions, was likely induced by a first interaction of the galaxy with the intracluster medium. Stripping by ram pressure, plus probable gas depletion due to star formation, contributed to create a truncated ionized gas disk. The presence of an extended stellar tail on only one side of the disk, points instead to another kind of process, likely a gravitational interaction by a fly--by or a close encounter with another galaxy in the cluster.
\end{abstract}


\section{Introduction} \label{sec:intro}
The evolution of galaxies is driven by physical mechanisms of either internal or external nature. Among the first ones are the processes related to stellar evolution (e.g. the star formation activity \citealt{kennicutt98,kennicutt12,madau14}, supernovae explosions, \citealt{burrows00,france10,marasco15,fielding17}), nuclear activity (accretion on a supermassive black hole and the related release of mechanical energy, \citealt{silk98,fabian03,croton06,mcnamara07}), and to the whole structural configuration of the different components (e.g. angular momentum reconfiguration by stellar bars, \citealt{hohl71,weinberg85,debattista00,athanassoula02,martinez06}). As for the external ones, interactions with galaxies, with the gravitational potential of large, massive structures (such as galaxy groups or clusters), and with the dense, hot gas of the intracluster medium, are among those playing the major role.

Several of such environment--dependent processes have been identified and proposed to explain the different evolutive paths that galaxies in clusters follow with respect to isolated ones, both regarding their stellar content (or, equivalently, their star formation history) and to their morphology. These include harassment (repeated high velocity encounters with galaxies in the cluster, \citealt{moore96}), starvation/strangulation (the removal, during the cluster collapse, of the galactic gas halo which fuels the star formation, \citealt{larson80,balogh00}), ram pressure stripping (the removal of the interstellar gas by means of high velocity interactions with the intracluster medium, e.g. \citealt{gunn72,faltenbacher06}, \citealt{takeda84}), thermal evaporation \citep{cowie77}, major and/or minor mergers \citep[e.g.][]{toomre77,tinsley79,mihos94,springel00}, or tidal effects of the cluster as a whole \citep[e.g.][]{byrd90,valluri93}.

As the star formation history of a galaxy crucially depends on the amount of gas available, processes removing, adding, or even perturbing the gas, are ultimately determining the evolution and the fate of a galaxy, at least as far as the stellar content is concerned.

Evidences of abruptly interrupted star formation due to gas removal \citep[e.g.][]{steinhauser16} as well as of enhancement of star formation \citep{boselli06} are found in the cluster galaxy population. The latter phenomenon in particular, is believed to be caused by the early effect of the ram pressure of the hot intracluster medium that compresses the gas of the galaxy providing the dynamical instabilities needed to kick--start a star formation event \citep[see, e.g.][]{crowl08,steinhauser12,ebeling14,bischko15,merluzzi16}. 

Spectacular examples of distorted morphologies due to gas losses, are the so--called jellyfish galaxies. Firstly dubbed as such by \cite{smith10_1} to describe the appearance of filaments and knots departing from the main body of the galaxy, these objects are mostly found in clusters both locally \citep[see, e.g.][]{fumagalli14,merluzzi16,abramson16} and at high redshift \citep[e.g.][]{cortese07,ebeling14,mcpartland16}. The availability of new generation Integral Field Units (IFU), such as the Multi Unit Spectroscopic Explorer (MUSE) on 8m class telescopes, has opened a new window to study the physical processes at play in these galaxies. 

GASP\footnote{\url{http://web.oapd.inaf.it/gasp/index.html}} (GAs Stripping Phenomena in galaxies with MUSE) is an ESO large program (P.I. B. Poggianti) that uses the second generation IFU MUSE mounted on the Nasmyth focus of the UT4 at the VLT, to observe a sample of 124 low redshift ($z=0.04-0.07$) galaxies with evidence of disturbed morphology in optical images of clusters from the WINGS/OmegaWINGS project \citep{fasano06,gullieuszik15}. GASP was granted 120 hours of time spread over four semesters from Period 96 (October 2015), and the second half of the observational campaign is currently being performed.

The ultimate goal of this project is to take a step forward in the understanding of the processes that remove gas in galaxies, halting the ongoing star formation processes. To which extent is the environment playing a role in gas stripping? Where is this more efficient? Why is it occurring and by which mechanism(s)? These are the most urgent questions that this project tries to address. We refer the reader to \cite{poggianti17} for a more detailed presentation of the survey, its characteristics, and its goals. 

In this work we focus on JO36, a galaxy drawn from the GASP sample. In the first part of the paper, we present an updated and improved version of \sinopsis, the spectrophotometric fitting code we adopt for the spectral analysis of the whole survey, and use MUSE data of this object as a test case to validate the code. To this aim, we perform a comparison between \sinopsis \ and {\sc gandalf} \citep{sarzi06}, a similar code that has been widely used to interpret the kinematic of stars and gas in galaxies from IFU data. 

In the second part of the paper, we use the outputs of \sinopsis \ to characterise the properties and distribution of the stellar populations in the galaxy, and give an interpretation of its observed characteristics in relation to its position and dynamical status within its host cluster. Exploiting archival data, we calculate the dust mass and use this to derive an estimate of the total gas mass, while X--ray observations are used to constrain the possible presence of an active galactic nucleus (AGN). 

Like in all the papers of the GASP series, we will assume a standard $\Lambda$CDM cosmology, with H$_0=70$, $\Omega_M=0.3$, and $\Omega_\Lambda=0.7$. Similarly, stellar masses and star formation rates are calculated assuming a \cite{chabrier03} Initial Mass Function (IMF). An observed redshift of 0.04077 like that of the galaxy under investigation, in this cosmology, corresponds to a luminosity distance of 180.0 Mpc and to an angular scale of 0.81$''$/kpc, which results in a physical spaxel size of about 160 pc/spaxel for MUSE.

\section{The spectral fitting code}\label{sec:sinopsis}
In this section we summarise the main features of \sinopsis, and describe new implementations and improvements with respect to its older versions.

\subsection{Modeling details}\label{ssec:model}
\sinopsis\footnote{\sinopsis \ is publicly available under the MIT open source licence, and can be downloaded from \url{http://www.irya.unam.mx/gente/j.fritz/JFhp/SINOPSIS.html}.} (SImulatiNg OPtical Spectra wIth Stellar population models), is a spectrophotometric fitting code that reproduces the main features of galaxy spectra in the ultra--violet to near--infrared spectral range. Here we summarize the most important aspects of \sinopsis. We refer the reader to previous papers describing in detail the code's approach, main characteristics, and reliability of its performance \citep{fritz07,fritz11}. The reader who is not interested in the technical details can safely skip this Section and go directly to Sect.~\ref{sec:data}.

\sinopsis \ has its roots on the spectral fitting code used by \cite{poggianti01} to reproduce the stacked optical spectra of a sample of Luminous Infrared Galaxies of different spectral types. Since then, it has been successfully applied to derive the physical properties (stellar mass, dust attenuation, star formation history, mean stellar ages, etc.) of galaxies in various samples \citep{dressler09,fritz11,vulcani15,guglielmo15,paccagnella16,cheung16}. The code has been validated both by fitting  simulated spectra of galaxies \citep{fritz07} and by comparison with the results from other datasets and models \citep{fritz11}. Nowadays \sinopsis \ has been used to fit several thousands of optical spectra.

A number of other codes can be found in the literature that serve similar purposes and are commonly used to derive the properties of the stellar populations and extinction in galaxies from their optical spectra, including e.g.  {\sc starlight} \citep{cidfernandes05}, {\sc steckmap} \citep{ocvirk06}, {\sc vespa} \citep{tojeiro07}, {\sc gossip} \citep{franzetti08}, ULySS \citep{koleva09}, {\sc popsynth} \citep{macarthur09}, {\sc firefly} \citep{wilkinson15}, {\sc fit3d} \citep{sanchez16} (but this list is most likely incomplete). \sinopsis \ shares similar features with some of these codes, while including substantial improvements. 

In order to reproduce an observed spectrum, the code calculates the average value of the observed flux in a pre--defined set of spectral bands (see Table~\ref{tab:bands} for the set used in the MUSE data analysis), accurately chosen for the lack of prominent spectral features such as emission and absorption lines, and the equivalent width values of significant lines (i.e. the hydrogen lines of the Balmer series, the calcium H and K lines, plus the [{\sc Oii}] 3727 \AA \ line, if present within the observed wavelength range), both in emission and in absorption. It then compares them to the same features in a theoretical model which is created as follows. 

From a set of $\sim 200$ mono--metallicities SSP spectra  with ages spanning the range between $10^4$ and $14\times 10^9$ years, \sinopsis \ creates a new set, with a reduced number of model spectra, by binning the models of the original grid with respect to the SSP's age. In this way, the number of theoretical spectra shrinks to only 12, for any given metallicity value. The choice of the age bins is made based on the presence and intensity of spectral features as a function of age \citep[see][for more details]{fritz07}.
\begin{table}
\centering
\begin{tabular}{rcc}
\hline
\#  & $\lambda_{inf}$ & $\lambda_{sup}$\\
\hline
  1  &  4600   &     4750  \\
  2  &  4845   &     4853  \\
  3  &  4858   &     4864  \\
  4  &  4870   &     4878  \\
  5  &  5040   &     5140  \\
  6  &  5210   &     5310  \\
  7  &  5400   &     5500  \\
  8  &  5650   &     5800  \\
  9  &  5955   &     6055  \\
10  &  6150   &     6250  \\
11  &  6400   &     6490  \\
12  &  6620   &     6690  \\
13  &  6820   &     6920  \\
14  &  7110   &     7210  \\
\hline
\end{tabular}
\caption{List of photometric windows, defined by the respective lower and upper wavelength, where the continuum flux is calculated to compare observed and model spectrum.}
\label{tab:bands}
\end{table}

Each of these spectra is multiplied by an appropriate guess value of the stellar mass, and then dust attenuation is applied before the spectra are summed together to yield the final model. The combination of the parameters which minimises the differences between the constraints in the observed and model spectra, is randomly explored by means of a simulated annealing algorithm. The range of values within which the search is performed is given in Table~\ref{tab:priors}. The large range of extinction values that we allow, is mainly meant to give a high degree of flexibility to the code, making it able to deal with galaxies having even ``extreme'' properties, without the need of further tuning.

\subsection{The treatment of dust extinction}\label{sect:dust}
One of the distinctive features of \sinopsis \ is that it is possible to allow for differential extinction as a function of the stellar age. In this way, the code simulates a selective extinction effect \citep{calzetti94}, where the light emitted by the youngest stellar populations is most likely to be affected by the presence of dust which is typically abundant in star forming molecular complexes. Once a stellar population ages, it progressively gets rid of this interstellar medium envelope, either by means of supernova explosions, which will blow it away, or because of the proper motions of the star clusters, or by a combination of the two effects.

Dust is found in the interstellar medium and is well mixed with the stars. A proper treatment of its extinction effect on the starlight would require the use of radiative transfer models, which can fully take into account the 3-D geometry of dust and stars, and their relative distribution \cite[see, e.g., the review by][]{steinacker13}. This is prohibitive for two reasons: one is the computational effort required to calculate such kind of models, and the second is the lack of a detailed enough knowledge of the spatial distribution of these two components in any given galaxy.
\begin{table}
\centering
\begin{tabular}{rcccc}
\hline
Age  & E(B-V)$_{min}$ & E(B-V)$_{max}$& SFR$_{min}$ & SFR$_{max}$\\
\hline
       $2\times 10^6$    &     0.0  &   1.50   &    0.0    &    3  \\
       $4\times 10^6$    &     0.0  &   1.50   &    0.0    &    3  \\
    $6.9\times 10^6$    &     0.0  &   1.50   &    0.0    &    3  \\
       $2\times 10^7$    &     0.0  &   1.00   &    0.0    &    3  \\
    $5.7\times 10^7$    &     0.0  &   0.80   &    0.0    &    2  \\
      $2\times 10^8$     &     0.0  &   0.40   &    0.0    &    2  \\
    $5.7\times 10^8$    &     0.0  &   0.40   &    0.0    &    2  \\
                   $10^9$     &     0.0  &   0.40   &    0.0    &    2  \\
      $3\times 10^9$     &     0.0  &   0.20   &    0.0    &    1  \\
    $5.7\times 10^9$    &     0.0  &   0.20   &    0.0    &    1  \\
               $10^{10}$    &     0.0  &   0.20   &    0.0    &    1  \\
$1.4\times 10^{10}$    &     0.0  &   0.08   &    0.0    &    1  \\
\hline
\end{tabular}
\caption{Maximum and minimum values allowed for the extinction, parametrised by the color excess E(B-V) and the SFR as a function of the SSP's age (the latter expressed in years). Note that the upper values for the SFR are normalised to the oldest SSP.}
\label{tab:priors}
\end{table}

Just like many other spectral fitting codes, \sinopsis \ includes the effect of dust extinction by modelling it as a uniform dust layer in front of the source. While this is indeed a simplification, \cite{liu13} have shown that a foreground dust screen reproduces well the effects of dust on starlight at large scales. Furthermore, the mix of stellar ages and extinction can be naturally taken into account by the age--dependent way of treating dust attenuation allowed by \sinopsis.

Different extinction and attenuation laws can be chosen including, among others, the attenuation law from \cite{calzetti94}, the average Milky Way extinction curve \citep{cardelli89}, or the Small and Large Magellanic Clouds curves \citep{fitzpatrick86}. Throughout this work and in all the papers of the GASP series, we adopt the Milky Way extinction curve ($R_V=3.1$).

\subsection{Spectral lines}
Another key feature of \sinopsis \ is the use of SSP models for which we have calculated the effect of nebular gas emission. Other models in the literature combine the effect of stellar and nebular emission, including e.g. the works by \cite{gutkin16,byler17} \citep[but see also the pioneering work of][]{charlot01}, who present models with both components, and \cite{pacifici12,chevellard16}, who describe an application of such kind of models to observed data.

\sinopsis \ has nebular emission lines included since its very first version (\citealt{poggianti01}, but see also \citealt{berta03} and \citealt{fritz07}), which now have been re--calculated for the new SSP models.

Including nebular emission lines in SSP spectra is a great advantage for a number of reasons: emission lines in the observed spectra need not to be masked for the fitting, a reliable value for dust extinction can be calculated (even when H$\beta$ is not observed), and star formation rates can be automatically estimated as well. Last but not least, especially for the purposes of the GASP project, correction of the underlying absorption in Balmer lines is performed in a self--consistent way, by simultaneously taking into account both the absorption and emission components.

The calculation of the lines intensities is obtained by pre--processing the SSPs spectral energy distribution (SED) with ages $\le 5\times 10^7$ years through the photoionisation code {\sc cloudy} \citep[][]{ferland93,ferland98,ferland13}. The adopted parameters are those typical of an {\sc Hii} region \citep[see also][]{charlot01}: hydrogen average density of $10^2$ atoms cm$^{-3}$, a gas cloud with a inner radius of $10^{-2}$ pc, and a metal abundance corresponding to the metallicity of the relative SSP.  

Note that only SSPs with ages less than $2\times 10^7$ years have a strong enough UV continuum to produce detectable emission lines and are, hence, the only ones for which gas emission is included.

The luminosity in the following Hydrogen line series is computed: Balmer (from H$\alpha$ to H$\epsilon$), Paschen (from Pa$\alpha$ to Pa$\delta$), Brackett (from Br$\alpha$ to Br$\delta$), and Lyman (Ly$\alpha$ and Ly$\beta$) series. Luminosity of UV and optical forbidden lines from various other elements (such as {\sc [Oi], \sc [Oii]}, and {\sc [Oiii]}, {\sc [Nii]}, {\sc [Sii]} and {\sc [Siii]}) is calculated as well. The latter are not used as constraints in the fitting procedure, as their intensities are dependent on several physical parameters (such as gas metallicity, geometry, electron temperature, electron density, ionisation source, dust depletion) whose determination is, at the moment, well beyond the scope of \sinopsis. 

Table~\ref{tab:lines} reports the list of spectral lines that are used as constraints in the spectral fits. The choice of the lines to be reproduced by the model is dictated mostly by the availability of a good physical characterisation and understanding of the physical processes driving their intensities. This is why forbidden lines are not included, with the exception of the {\sc[Oii]} doublet at 3726,3729 \AA. Other absorption lines, such as the NaD doublet at 5890, 5896 \AA, and the Mg line at 5177 \AA, have a strong dependence on the $\alpha$ enhancement \citep[e.g.][]{wallerstein62,thomas99} and on the presence of dust \citep[this is the case of the sodium doublet, see e.g.][]{poznanski12}. As these features are not included in the theoretical models, we do not attempt to reproduce them.

On the other hand, the intensity of the {\sc [Oii]} doublet was found to correlate with the intensity of H$\alpha$ \citep[e.g.][for studies at various redshifts]{moustakas06,weiner07,hayashi13}, such that the former is often used to quantify the SFR in distant galaxies, where H$\alpha$ falls out of the observed spectral range. This is why the {\sc [Oii]} line is used as well.
\begin{table}
\centering
\begin{tabular}{rcc}
\hline
\#  & Line & $\lambda_{c}$\\
\hline
  1  &  {\sc [Oii]}   		&     3727     \\
  2  &  CaK  			&     3933.6  \\
  3  &  CaH+H$\epsilon$    &     3969    \\
  4  &  H$\delta$   		&     4101.7  \\
  5  &  H$\gamma$   		&     4340.5  \\
  6  &  $^*$H$\beta$   	&     4861.3  \\
  7  &  $^*$H$\alpha$   	&     6562.8  \\
  \hline
\end{tabular}
\caption{List of spectral emission and absorption lines that are used, when available, to constrain the model's parameters. Lines indicated with a $^*$ are the ones contained within the wavelength range sampled by MUSE, and are hence the only ones we use in this work, and in all of the papers from the GASP series.}
\label{tab:lines}
\end{table}

The observed intensity of hydrogen emission lines, particularly those found in the optical range, are widely exploited to calculate the recent star formation rate, and the amount of dust extinction. Reproducing these observables with a theoretical spectrum gives hence strong constraints on these two quantities. For this reason, as a sanity check, we have calculated the luminosity of the H$\alpha$ line from our {\sc cloudy} modelling, corresponding to a constant star formation rate over $10^7$ years, and checked that this value is consistent with the factor typically used to convert an H$\alpha$ luminosity into a star formation rate value. We found a good agreement  when considering a \cite{chabrier03} IMF \citep[see][for a recalibration of this SFR indicator]{kennicutt12}, as it is the case for the SSPs version (discussed below) currently implemented in \sinopsis.

As for the dust extinction calculations, the ratio between the observed intensity of the H$\alpha$ and H$\beta$ lines is commonly used, exploiting the fact that, in normal star--forming and {\sc Hii} regions, its theoretical expected value is $\sim 2.86$ \citep[see e.g.][]{osterbrock06}. Indeed, the ratio of the luminosities of the two lines we have calculated in spectra of various ages is 2.88, which is very close to the aforementioned theoretical value.

\subsection{Recovered parameters and uncertainties}
As we fully embrace the selective extinction hypothesis, the parameter space that \sinopsis \ explores includes 12 values for the SFR and 12 values for the dust extinction, one for each age bin we consider. As extensively explained by \cite{cidfernandes07}, using an over--dimensioned parameters space is an expression of a principle of maximum ignorance and, when the results are to be taken into account, the properties calculated over the initial 12 age bins must be compressed to a lower time resolution SFH. This, in our case, results in considering as a reliable result the SFR calculated in 4 times intervals (see below), and the extinction in 2, namely ``young'' (i.e. for SSPs displaying emission lines) and ``global'' (i.e. calculated as an average over all the stellar ages).

The use of this non--parametric approach, as compared to e.g. the assumption of an analytic prescription for the SFH (such as a tau--model, a lognormal, or a declining exponential), has the obvious advantage of limiting the number of priors the model needs to assume. Furthermore, it is a more fair representation of the stellar populations evolution, the SFH of galaxies being in general characterized by various episodes of star formation of different intensity at various ages, especially when galaxies in dense environments are considered \citep[see e.g.][]{boselli16}.

Reconstructing the evolution of the stellar populations in a galaxy, by means of a non parametric SFH as we do here, is a methodology that is by all means very similar to that embraced by other codes and works tackling similar issues \citep[see e.g.][]{ocvirk06,merluzzi13,merluzzi16}, and has proven effective to this task.

The choice of the number of age bins and their definition is based on simulations we have performed in \cite{fritz07} for integrated spectra of the WINGS survey. While, on the one hand, the quality of the spectra, especially in terms of signal--to--noise (S/N), is much better for MUSE data, on the other side these spectra are sampling the rest--frame spectral region between $\sim 4700$ and $\sim 9000$ \AA, hence missing some of those features which are normally used to constrain the stellar population properties. This is why we have decided not to push our interpretation to a higher age resolution, despite the excellent quality of the data. However, we are still satisfied by the modelling and the provided results, since it is extremely difficult to disentangle the contribution to the integrated light of stellar populations in the 7-14 Gyr range. This is especially true when non--resolved spectroscopy is used, where in one single spectrum stars of all possible ages are superimposed. Furthermore, well--known effects such as the age--metallicity degeneracy, plus dust extinction, conspire to make the spectra of simple stellar populations very similar in this age range.

Another viable approach would be to use an analytical prescription for the SFH (such as a log-normal, a double exponential, or a so--called $\tau$--model), which is equivalent to imposing an arbitrary prior on the shape of the SFR as a function of time. This is why we choose to follow this ``free'' approach, which we consider more fair with respect to the complexity of the problem.

Derived physical parameters include the total stellar mass, the SFR in the 4 age bins, luminosity and mass --weighted ages, and dust extinction \citep[see][for a complete list]{fritz11}. The latter is calculated as the ratio between the dust--free and the best--fit model, as explained in \cite{fritz11} (see their equation 6).

The estimation of uncertainties in the physical parameters that are given as outputs, follows a Monte Carlo-like approach, described in detail in \cite{fritz07}. The best fit is searched within the parameter space by randomly exploring a large number of models. As the choice of the trial point (which hence results in a given model spectrum) performed at each step depends on the model calculated at the previous step, starting from a different set of initial conditions will always result in a different set of best-fit parameters (with minimal differences between best fits). The properties of the different best-fit models are hence used to calculate uncertainties on the physical parameters.

\subsection{Improvements and adjusts for IFU data dealing}
\begin{figure*}
\begin{tabular}{ll}
\centering
\includegraphics[height=0.46\textwidth]{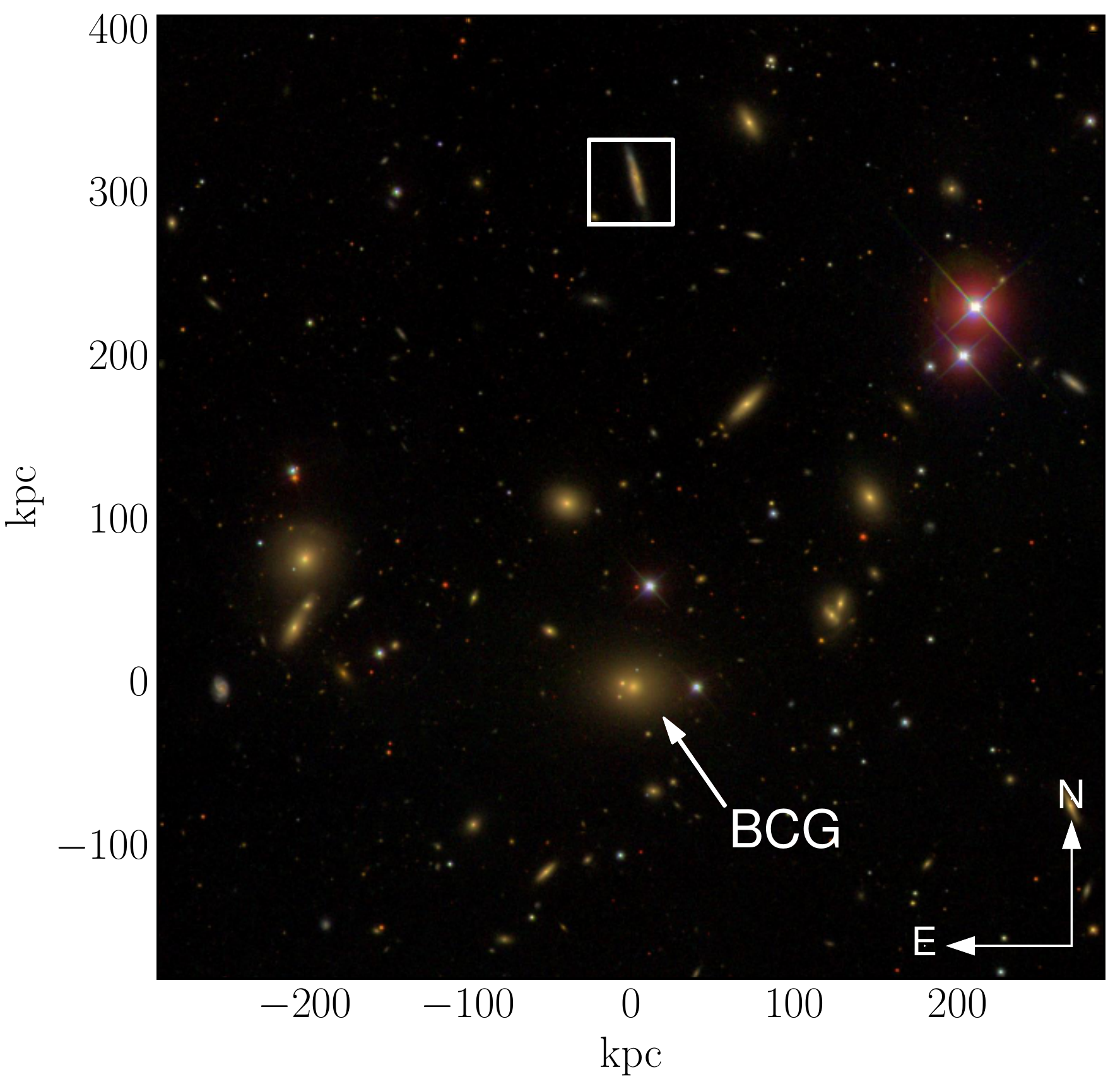} &
\includegraphics[height=0.46\textwidth]{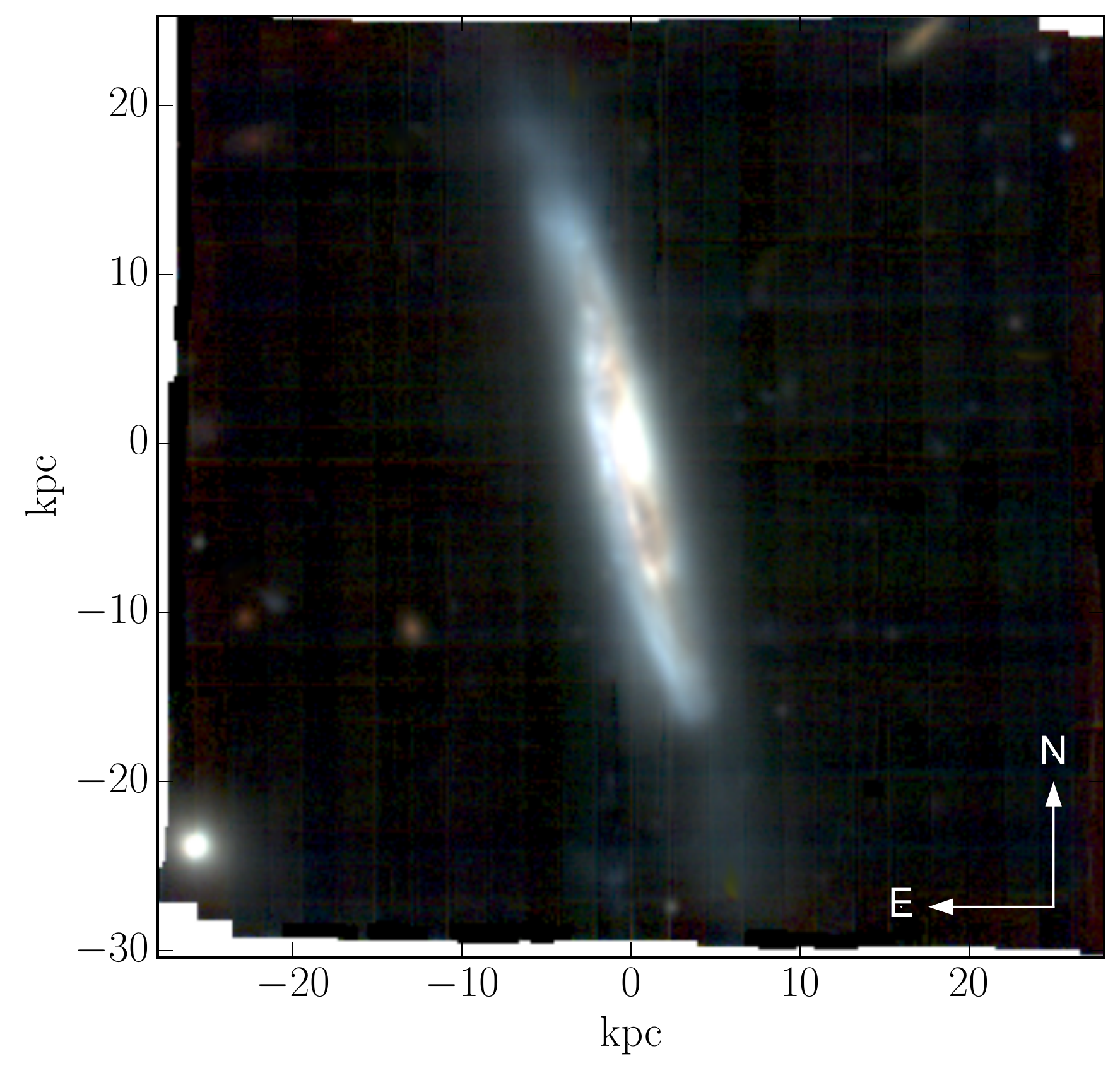} \\
\end{tabular}
\caption{Left panel: SDSS color image of the central region of the cluster Abell 160 with the position of the BCG indicated, and the MUSE Field of View around the galaxy. Right panel: RGB image of JO36, as provided by the MUSE data reduction pipeline. This is based on the $grz$--bands, where the $g$ filter is included at a 50\% \ level, due to the incomplete spectral coverage.}
\label{fig:opt}
\end{figure*}
In order for \sinopsis \ to properly deal with IFU datacubes, a number of changes and improvements were implemented with respect to the versions presented in \cite{fritz07} and \cite{fritz11}. These are very briefly described hereafter:
\begin{itemize}
\item \sinopsis \ can now ingest observed spectra in fits format. Data format can be either 1D (a single spectrum), 2D (a series of spectra, as e.g. provided by multi-slit or fibre fed spectrographs), or 3D (for e.g. IFU such as MUSE);
\item when data in ``cube'' format are used, most of the results are now saved on datacubes in fits format, with each plane containing one of the properties typically derived from this kind of analysis (e.g. pixel--by--pixel stellar mass, extinction, star formation rate, stellar age, etc. See \citealt{fritz11} for a detailed description of the meaning of each parameter);
\item a new set of SSP models by Charlot \& Bruzual (2018, in prep.) is used, which has higher spectral and age resolution, and a larger number of metallicity values (namely 13, from Z=0.0001 to Z=0.04, as compared to the 3 default values used before). This new models include the most recent version of the PADOVA evolutionary tracks from \cite{bressan12} (PARSEC), and have been coupled with stellar atmosphere libraries from several sources depending on the wavelength coverage, luminosity, and effective temperature \citep[see][for the full compilation of the adopted stellar spectra]{gutkin16}. For the wavelength range of interest for this paper, and for GASP in general, the stellar spectra are mostly from the Miles stellar library \citep{sanchez06,falcon11}. The evolutionary tracks include the treatment of the Wolf--Rayet phase, for stars typically more massive than 25 M$_\odot$. The assumed IMF is \cite{chabrier03} with masses in the range 0.1 to 100 M$_\odot$;
\item one of the outputs includes now the purely stellar emission, that is the model spectrum without the nebular emission lines component. These are calculated from the best fit parameters but using instead the SSP set with pure stellar emission;
\item when spectra from different regions of a galaxy are considered, it is possible that the velocities of the gas and of the stars are different. For our purposes, this means that during the spectral fitting, when using redshifts calculated from absorption lines (i.e. that of the stellar component), the center of emission lines could be displaced with respect to the absorption component. This might turn into a miscalculation of the equivalent width of the lines, or sometimes even to a non--detection. To overcome this possible issue, we now allow the simultaneous use of redshifts calculated from the two components. If no emission lines are detected, only the stellar redshift is used, while if emission lines are present, the measurement of the equivalent width is performed using the emission--line redshift for lines in emission;
\item \sinopsis \ has been optimised from the computational efficiency point of view, and can successfully reproduce one optical spectrum in less than 1 second on a 3.5 GHz Intel Core i7 machine (running Mac OS X Version 10.10.5). The code is currently not parallelised and can only use one core at the time. We are planning to implement multithreading to exploit the full resource power of multi--core computers for the analysis of multiple spectra/IFU data, which has proven to be quite computational demanding.
\end{itemize}

\section{Data}\label{sec:data}
\begin{figure}
\centering
\includegraphics[height=0.48\textwidth]{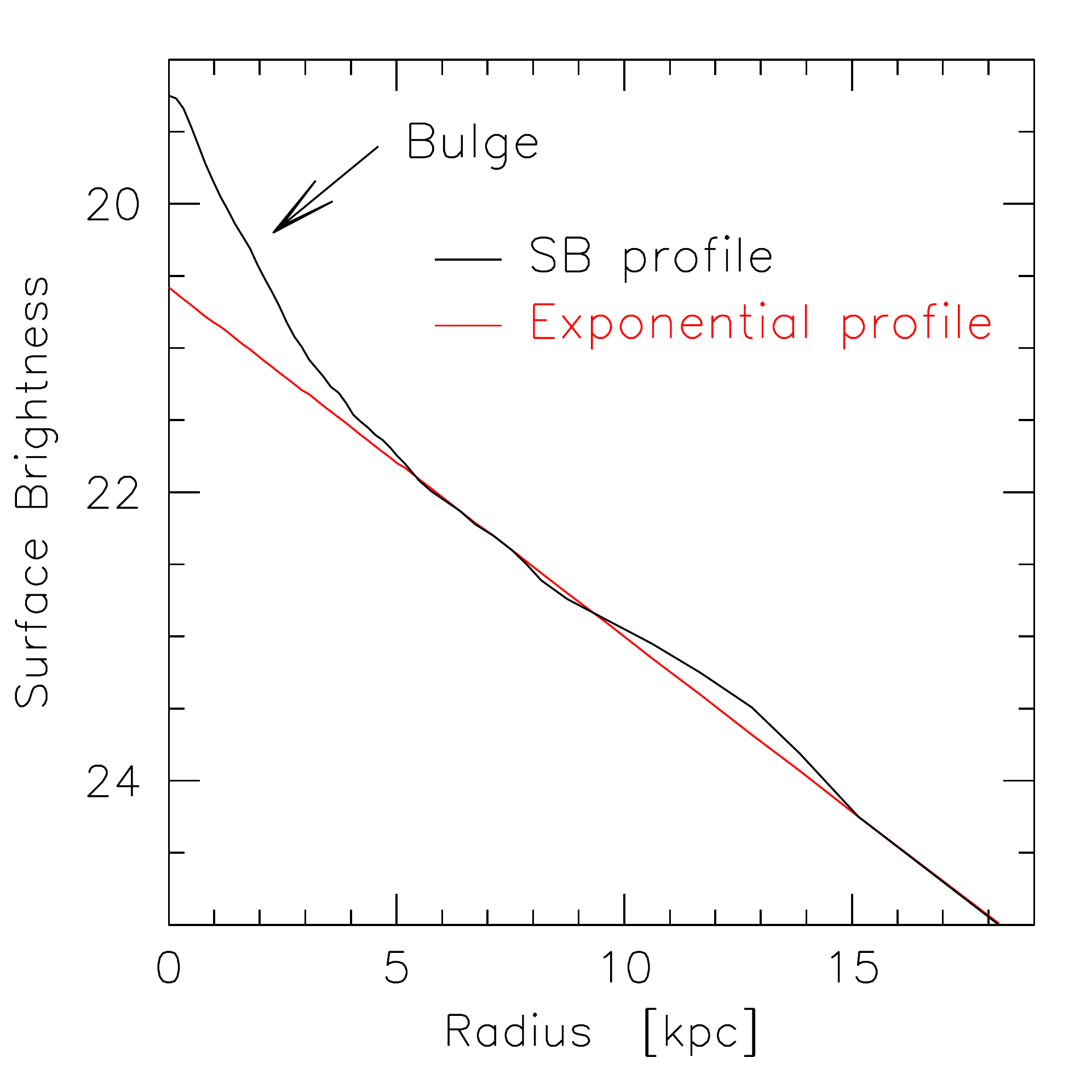}
\caption{Surface brightness profile of the galaxy (black line), highlighting the presence of light excess, with respect to an exponential profile (red line), in the central regions. This is what we identify as the bulge.}
\label{fig:bulge}
\end{figure}
As already outlined in the Introduction, this work provides a detailed analysis of a single galaxy drawn from the GASP sample. JO36 was selected from the sample of jellyfish candidates of \cite{poggianti16} found in the OmegaWINGS database \citep{gullieuszik15}. Also known as 2MFGC00903, or WINGSJ011259.41+153529.5, this galaxy was chosen for testing and validating \sinopsis \ because its SED is dominated by the emission of the stellar populations as opposed to the nebular one. In Fig.~\ref{fig:opt} we present a $g$--$r$--$z$ color composite image of the field where the galaxy is located, and of the galaxy itself as derived from the MUSE cube. 

JO36 (RA=01h12m59.4s; DEC=+15d35m29s) is a disturbed galaxy with an assigned stripping class value JClass=3 (on a 1 to 5 scale, where 5 represents the maximum morphological disturbance in the optical, see \citealt{poggianti16}) belonging to the Abell cluster A160. With a V band magnitude of 15.5, JO36 is located at a projected radial distance of about 310 kpc from the Brightest Cluster Galaxy (BCG), and was recognised in optical images because of the presence of a bright optical tail, both in the V and B band, departing from the galaxy disk towards the south. MUSE data for this object were taken in October 10th, 2015, with an exposure of 2700 s.

The galaxy is classified as a Sc spiral seen almost edge--on: its apparent axial ratio of about 0.15 is in fact consistent with the intrinsic flattening value usually assumed for galaxies with such morphological classification. 

A bulge can be identified both photometrically and kinematically. The surface brightness profile, which we report in Fig.~\ref{fig:bulge}, shows the bulge component as a light excess with respect to the exponential profile representing the disk (red line in the Figure). As clearly visible in the plot, the bulge becomes dominant in the innermost 4 kpc. Similarly, the highest values of the stellar velocity dispersion are found in the central regions at similar galactocentric distances.

The data reduction for the whole GASP project is described in the survey's presentation paper, \cite{poggianti17}, and we refer the reader to this work for all the relevant details.

\section{Comparison with {\sc gandalf}}
We now briefly describe the main differences between \sinopsis \ and {\sc gandalf}, by \cite{sarzi06}, a code which is commonly used to perform similar analysis on IFU data \citep[see, e.g.][and other papers of the SAURON survey]{bacon01,dezeeuw02}. We then compare the performances of the two codes, with a focus on the aspect which is the main driver of the comparison: the derivation of an emission line--free model. This is done by analysing the very same dataset with both codes.

\subsection{Differences between the two codes}
The choice of a comparison with {\sc gandalf} among many other similar codes, is dictated by the need of subtracting, for a major part of the analysis of galaxies in GASP, the stellar component from the nebular lines when performing spatially resolved analysis on the gas properties. {\sc gandalf} is one of the most used tools to perform such subtraction, and was hence chosen as a reference.

As already outlined in Sect.~\ref{ssec:model}, \sinopsis \ has been used to analyze spectra from different instruments and various surveys. Still, an application to integral field data was so far missing. Even though, in principle, it all comes down to correctly reproduce the most significant features of an optical spectrum, we have performed a number of tests to check the reliability of the results when dealing with spatially resolved data. This was done by comparing our outcomes to those obtained with {\sc gandalf}, an IDL tool which shares similar features with \sinopsis, but  that focuses mostly on the analysis and interpretation of the emission and absorption lines characteristics to derive the stellar and gas kinematics, even though the stellar population properties can be inferred as well.

Both codes attempt to reproduce, by means of theoretical spectra, the observed features of an optical spectrum. The underlying models are very similar, as they both use stellar atmosphere from MILES (\citealt{vazdekis10} for {\sc gandalf} and the similar version of \citealt{sanchez06} for \sinopsis), at a spectral resolution of $\sim 2.5$ \AA.  They both provide an emission line free model spectrum.

The main differences between the two codes can be summarised as follows:
\begin{itemize}
\item the models used by \sinopsis \ already include the nebular emission, which has been self--consistently calculated using the SSP spectra as an input source fed into the photoionisation code {\sc cloudy}. {\sc gandalf}, instead, reproduces them as gaussian functions, fitting not only their intensity, but their width and central wavelengths as well;
\item \sinopsis \ assumes an extinction curve (either from models or derived from observations) to account for dust reddening in the models before matching to the observed ones, while {\sc gandalf} corrects for the effect of dust by multiplying the model spectra by $n_{th}$--order Legendre polynomials;
\item \sinopsis \ adopts a ``selective extinction'' approach, where the amount of dust attenuation is  considered to be age--dependent.
\end{itemize}

\subsection{Direct comparison}
We have run the two fitting tools on a subset of pixels of the original MUSE cube, specifically on a rectangle of $109\times 255$ spaxels which encompasses the full disk of the galaxy, and where a redshift value, either stellar or from the gas, was available (see also Sect.~\ref{sec:results}). Furthermore, we have limited the wavelength range to the spectral window between $4750$ and $\sim 7650$ \AA, discarding the red end of the spectrum, much less rich in the kind of features that are crucial for the purposes of this study. 

We have analyzed the performances of the two codes by calculating, a posteriori, the goodness of the fit to the continuum emission, hence not taking into account any spectral line (a further check is done on the equivalent width values of the H$\alpha$ and H$\beta$ lines, in a separate comparison). To do so, we have exploited the same 14 spectral windows which are used to constrain the fit for \sinopsis. These windows have, in general, a width of $\sim 100$ \AA, except in a few cases in which they are narrower due to the need of sampling a specific continuum emission region, while at the same time avoiding nearby emission or absorption features. The windows were chosen in order to homogeneously sample the whole spectral range (see Table~\ref{tab:bands} for the details).

For both codes, we have calculated a goodness index, $\Gamma$, both ``global'' and for each of the aforementioned bands, defined as:
\begin{equation}\label{eqn:gi}
\Gamma=\sum_{j=1}^{N}\Gamma^j=\sum_{j=1}^{N}\left(\frac{F^j_o-F^j_m}{\sigma^j}\right)^2 \; ,
\end{equation}
where $F^j_o$ and $F^j_m$ are the average fluxes calculated over the $j-th$ band of the observed and model spectrum, respectively. $\sigma^j$ is the uncertainty on the observed flux in that band, calculated as the standard deviation of the flux. Hence, $\Gamma^j$ is the goodness index for the {\it j-th} band. Note that, with the definition of the flux errorbars we have chosen, we might be slightly overestimating the uncertainties on the flux in the highest S/N spectra. This is, however, irrelevant for the relative comparison as the errors are the same when the $\Gamma$ index is calculated for \sinopsis \ and {\sc gandalf}. In principle, one would expect that values of $\Gamma^j$ lower than 1 (that is, with the model flux being within $1 \sigma$ from the observed one), are to be considered acceptable fits. In reality, the values are always much smaller in the vast majority of the cases.

We have constructed maps of both the $\Gamma$ and $\Gamma^j$ values for each of the 14 bands, so that we can check for the presence of systematic differences in any of the spectral ranges defined above. 

The value of the goodness index, averaged over all the spaxels, was found to be 0.85 and 1.37 in the case of \sinopsis \ and {\sc gandalf}, respectively, indicating that the two codes provide, globally and on average, very satisfying fits to the observed data, with \sinopsis \ performing slightly better than {\sc gandalf}.

We note that, in order for \sinopsis \ to provide satisfactory fits, in particular towards the bulge of the galaxy, we needed to relax the constraints on the maximum values of dust extinction, in particular for the oldest stellar populations. In \sinopsis \ this parameter is allowed to vary freely with stellar age, as explained in Sect.~\ref{sect:dust}. Normally, the upper limits of the values that dust extinction can reach for each stellar population are an inverse function of their age. This can be viewed on a equal footing with a prior which helps limiting the effect of possible degeneracies. 

The maximum value of the color excess was increased to 0.6 for the two oldest stellar populations, and to 0.8 for the others up to $\sim50 Myr$ (see Table~\ref{tab:priors} for a comparison to the standard values).

Allowing older stars to be more heavily affected by dust extinction, is not a mere matter of increasing the degree of freedom of the parameters, but has an actual physical meaning: the central part of the galaxy is the most crowded area, and the light emitted by old stars is easily contaminated both by other stellar populations with a whole range of stellar ages, but also by the presence of dust, located anywhere along the line of sight. The orientation of the galaxy is, in fact, very far from being face on (see Sect.~\ref{sec:data}), a configuration which would minimise the dust reddening effect \citep[see, e.g.][]{delooze14}. Hence, the light reaching us from the innermost part has a contribution from both the bulge and the disk, which is likely to contain the majority of the dust. Allowing higher values of dust extinction even for the older stars that are dominating the bulge, is hence needed to account for the effect of the dust lane.

\begin{figure}[!t]
\centering
\includegraphics[height=0.42\textwidth]{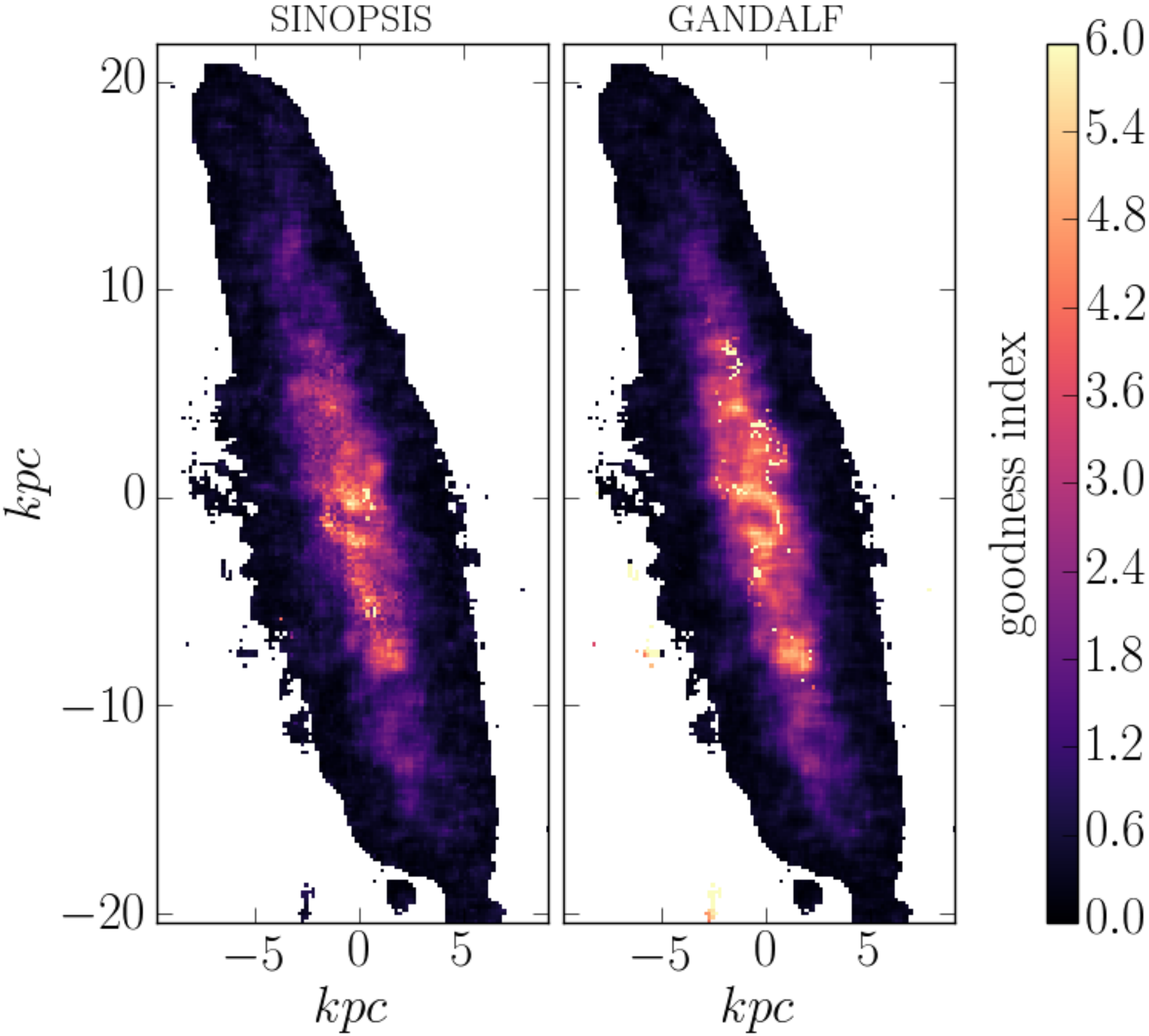}
\caption{Maps of the goodness index (see Eqn.~\ref{eqn:gi}) for \sinopsis \ (left--hand panel) and {\sc gandalf} (right--hand panel), for the central spaxels (with pixel coordinates within the $106 < x < 217$ and $60 < y < 307$ range) of JO36, where the two codes were run for the comparison. The values of $\Gamma$ are displayed in a linear scale ranging form 0 to 6.}
\label{fig:GvsSGamma}
\end{figure}

In Fig.~\ref{fig:GvsSGamma} we show the $\Gamma$ map calculated as explained above for both codes (note that the values for each spaxel are calculated according to Eqn.~\ref{eqn:gi}, hence not normalised to the total number of observables). A visual comparison of the two maps in Fig.~\ref{fig:GvsSGamma}, where only pixels having a stellar redshift or a S/N$>5$ for the H$\alpha$ emission line are shown, confirms the aforementioned result, highlighting that \sinopsis \ performs slightly better with respect to {\sc gandalf}, at least as far as the continuum emission is concerned. 

It can be easily noted that the $\Gamma$ values decrease, on average, as a function of the galactocentric distance for both models. This is due to the fact that the spectra become fainter as we approach the galaxy outskirts, and the S/N hence gets lower. This increases the observed uncertainties on the fluxes, and it consequently makes the fits more degenerate giving, as a result, lower values of the goodness index.

As for the other observed features, namely the spectral lines, an automatic comparison with the data is much less straightforward, due to the complexity of properly measuring emission and absorption lines, especially in the lowest S/N spectra. Hence, we have performed two quality checks: in the first one, we have compared observed and model spectra, while in the second we focussed on possible differences between the models provided by the two codes. 

For the first one, we have visually inspected the fits to the observed spectra provided by the two codes around the H$\alpha$ and H$\beta$ lines, checking for differences. This was done in a subset of $\sim 150$ spaxels along the major and minor axis of the galaxy, which allowed not only to include the full range of S/N values found in the datacube (i.e. from $\sim 80$ to $\sim10$), but also to sample different spectral properties, such as continuum shape and line intensities.

No significant differences were found between the two codes. On a minor fraction of the spectra in this control sample (less than 10\%), \sinopsis \ better recovers the H$\beta$ emission, especially when deeply embedded within the absorption profile. In these cases, {\sc gandalf} was usually overestimating the emission line intensity, but no systematic trend could be found e.g. with respect to the S/N or to the spectral properties (given also the small number of spectra where this discrepancy was spotted).

\begin{figure}
\centering
\includegraphics[height=0.48\textwidth]{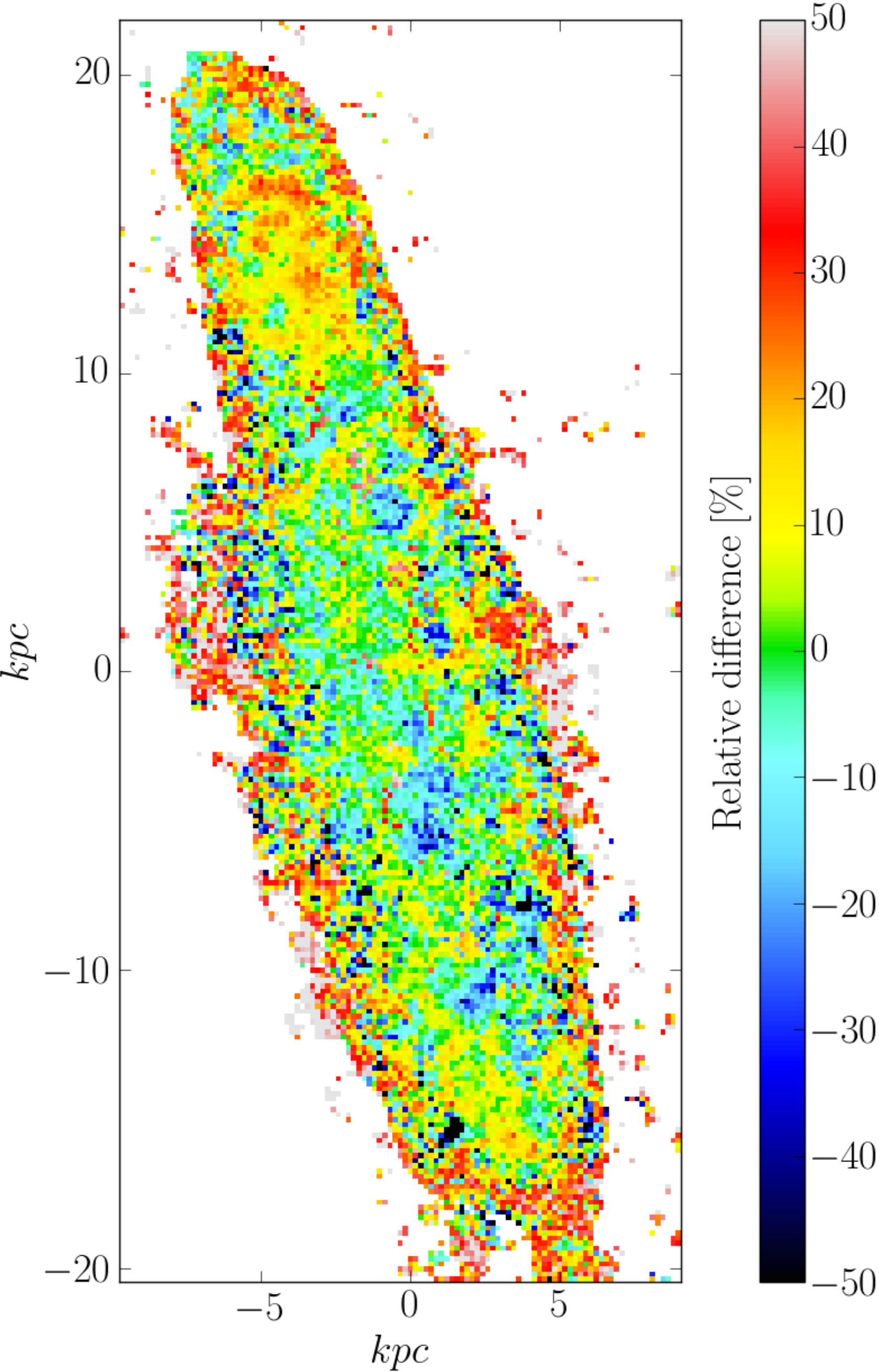}
\caption{Map of the relative difference in the H$\beta$ equivalent width values of the two models, calculated as in Eqn.~\ref{eqn:Hbdiff}, for each spaxel.}
\label{fig:Hbcomp}
\end{figure}

In the second check, we have calculated the equivalent widths of the H$\alpha$ and H$\beta$ lines from the best fits, and compared the values from the two models. On average, we found a difference of about 15\% in the value of the equivalent width of the two lines when comparing measurements in each spaxel. In Fig.~\ref{fig:Hbcomp} we show the map of the spaxel--by--spaxel differences, expressed in percentage of the H$\beta$ equivalent width value (which is the feature displaying the largest difference), calculated as:
\begin{equation}\label{eqn:Hbdiff}
\Delta_\beta = \frac{EW_s-EW_g}{EW_s}
\end{equation} 
where $EW_s$ and $EW_g$ are the equivalent width values, expressed in \AA, of the \sinopsis \ and {\sc gandalf} models, respectively.

As shown in Fig.~\ref{fig:Hbcomp}, the differences are mostly within a $\sim5$\% across all the galaxy, with very few exceptions where the discrepancy can be as high as $\sim 50$\%, but mostly in the outskirts of the disk, where the S/N is lower. We have visually inspected the fits from both codes for a sample of these spaxels with the highest discrepancies and found that differences in the line intensities are either due to the high uncertainties in the measurement or, in many other cases, to {\sc gandalf}, which seems to display some issues fitting (or measuring) the observed line (e.g. because of a poor fit of the continuum emission near the line, which hence affects the line measurement itself). 

We conclude that the two codes perform, with respect to the determination of the spectral continuum emission and of the hydrogen absorption lines intensity, in a very similar way. This gives us strong confidence in the model fits provided by \sinopsis \ in particular with respect to the correction of the absorption component in the Balmer lines, that was our major interest.

\section{Results}\label{sec:results}
We now present the results of the kinematic and stellar population analysis. The stellar and gas velocities were derived by means of external packages. In particular, the fitting and characterisation of the emission lines has been performed by exploiting the {\sc kubeviz} \citep{fossati16} code, while the stellar velocities were measured by the pPXF software \citep{cappellari04,cappellari12a}, which works in Voronoi binned regions of given S/N \citep[10 in this case; see][]{cappellari12b}.

The gas and stellar velocity information is also used to assign a proper redshift which will be used in the spectral fitting. Only spaxels with a redshift determination will be analysed by \sinopsis.

\subsection{The stellar and gas kinematic}\label{sec:kinematic}
The stellar kinematics was derived, as customary for this kind of data, from the analysis of the characteristics of absorption lines, while the kinematical properties of the gas were inferred from a similar analysis of the H$\alpha$ emission line, using the aforementioned tools. We refer to Sect. 6.1 in \cite{poggianti17} for a detailed description of how gas and stellar kinematics are derived from these tools, and of the main parameters adopted for this task.
\begin{figure*}
\setlength{\tabcolsep}{0.6em}
\begin{tabular}{ll} 
\centering
\includegraphics[height=0.37\textwidth]{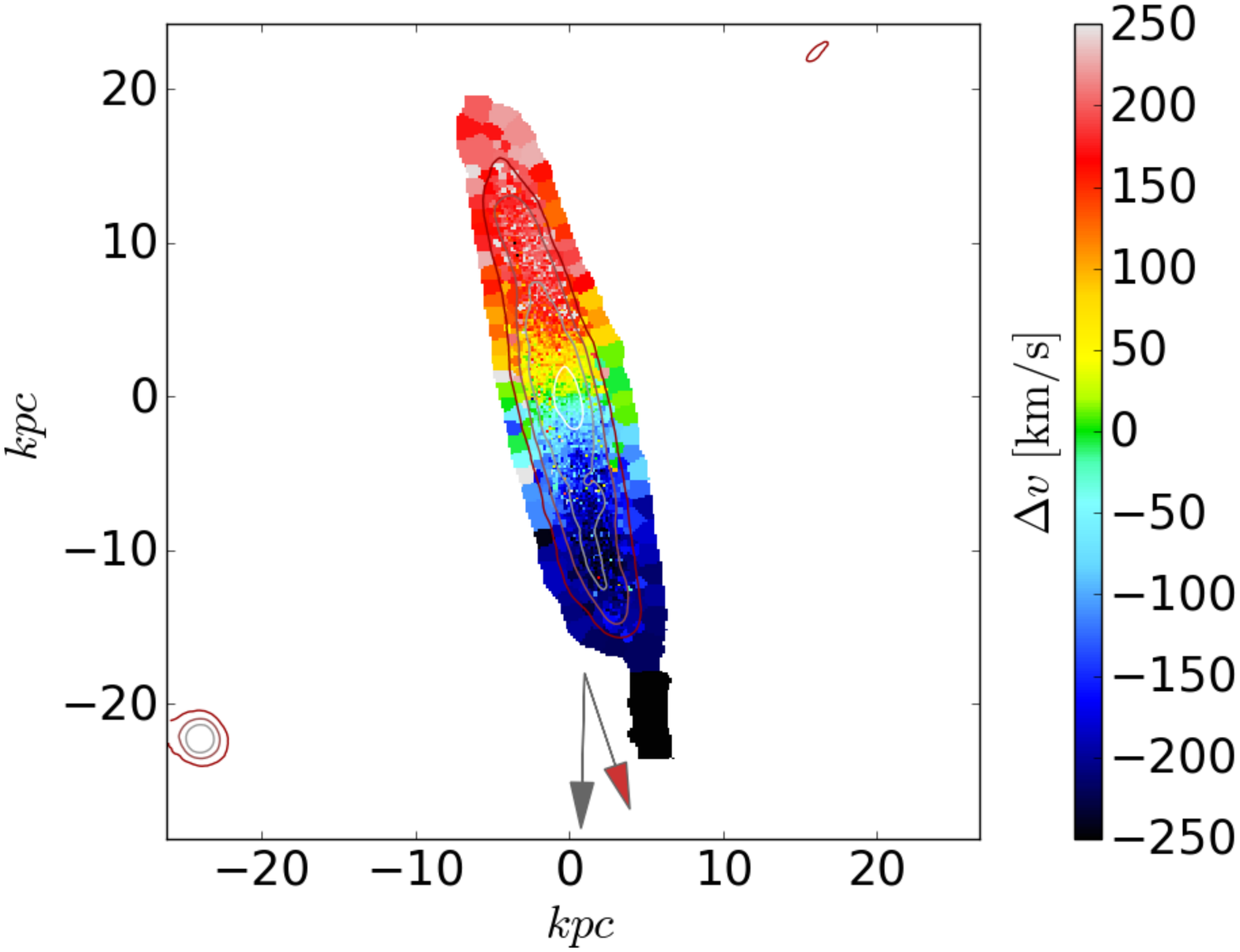} &
\includegraphics[height=0.37\textwidth]{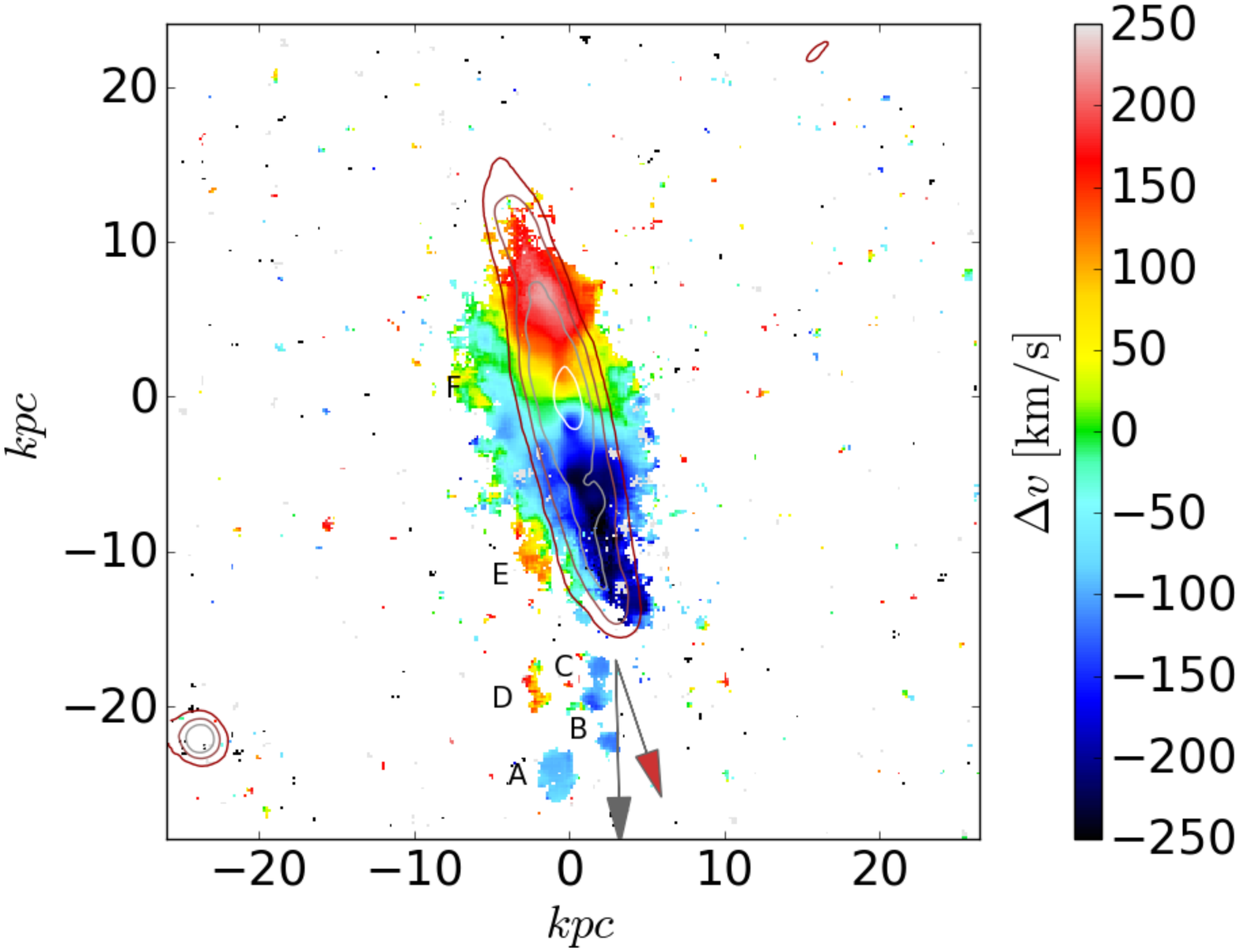} \\
\end{tabular}
\caption{Left panel: Stellar velocity map. Right panel: gas velocity map. The solid lines in both figures are H$\alpha$ continuum surface brightness contours in four logarithmically spaced levels. Regions labelled with letters from ``A'' to ``F'' are described in the text. The grey and red arrows point toward the BCG and the cluster X--ray emission, respectively. A cut of 4 in S/N was applied in the gas velocity map. North is up, east is left.}
\label{fig:kinematic}
\end{figure*}

\begin{figure}
\includegraphics[height=0.47\textwidth]{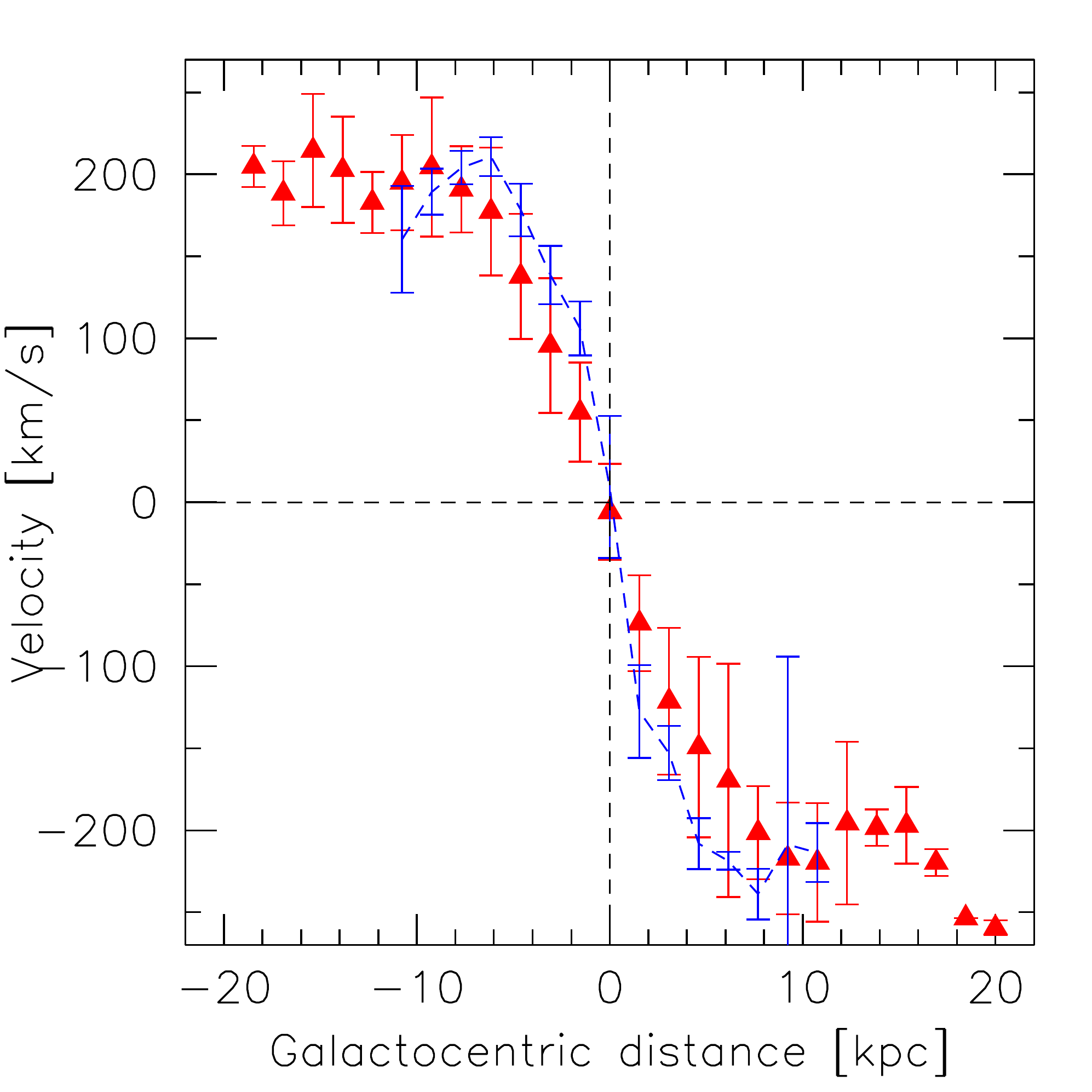}
\caption{Stellar (red triangles) and gas (blue, dashed line) radial velocity profiles, taken along the major axis of the galaxy.}
\label{fig:vprof}
\end{figure}

In Fig.~\ref{fig:kinematic} we show the velocity map of the stellar and gas components, while Fig.~\ref{fig:vprof} presents the radial velocity profiles along the major axis for stars (red triangles) and gas (blue, dashed line). Differences between the velocities of the two components are shown as green diamonds on the lower panel of the same Figure. At radii larger than $\sim 10$ kpc the trend becomes much noisier due to the fewer usable spaxels and to the more uncertain velocity determination. A cut at S/N=4 measured on H$\alpha$ has been applied in the gas velocity map. In order to obtain more reliable gas velocities, the original datacube was filtered with a $5\times5$ spaxels boxcar filter, to increase the S/N level.

While following the same pattern in the velocity profiles, the gas has velocities that are marginally higher with respect to those measured in the stellar component, even though this difference is in most of the cases within the measured uncertainties (see Fig.~\ref{fig:vprof}).

The radial distribution of the stellar velocities displays a monotonic gradient out to radii of about 10 kpc, with values as high as $\sim 200$ km/s, as expected from a nearly edge--on galaxy, and is a clear indication of a rotationally supported disk. After this radius, the velocity gradient flattens out in the north part of the disk while displaying a slight bump on the south side, reaching higher velocities further out. These velocities correspond to stars observed in a tail extending by about 5 kpc southwards, where the (stellar) radial velocities are the highest found in the disk, with (negative) values of about 270 km/s. This velocity pattern follows the trend observed in the inner disk, while the northern side shows no evidence of a similar structure, absent in both WINGS and OmegaWINGS images.

The ``rotational axis'' of the gas component (i.e. the locus of close--to--zero velocities) visible in Fig.~\ref{fig:kinematic} as a green strip, is bent in a twisted ``U'' shape, with 0--velocity gas found in the outer disk, as far as $\sim 5$ kpc away from the minor axis. Such feature is similar to that observed by \cite{merluzzi16} in the jellyfish galaxy {\it SOS90630} of the Shapley supercluster. 

Using an ad hoc N-body/hydrodynamical simulation, they found that the gas velocity field, and this very feature in particular, can be successfully reproduced when ram pressure stripping is acting on an almost edge--on geometrical configuration (see their Fig. 18 and 27), with the galaxy moving in the opposite direction with respect to the concavity. 

A careful inspection of the gas velocity map, highlights asymmetries in their values, with negative velocities extending well beyond the galaxy's center towards the north, out to a distance of about 8 kpc in the eastern side of the disk (see the region marked with ``F'' in Fig.~\ref{fig:kinematic}). Similarly, on the same side but towards the south, there is a clear inversion in the gas velocities, going from negative to positive values (region ``E'' on the same Figure).

Four H$\alpha$ blobs are visible towards the south, detected with S/N from $\sim10$ (the regions labelled as ``B'', ``C'', and ``D'') to more than 50 (region ``A'', the southernmost one). The most luminous one, region A, is clearly detected on the V band image of WINGS and OmegaWINGS data as well. The velocities of blobs A, B, and C, are quite compatible with those observed in the southern disk, while those in region D are similar to those of the northern side. A feature with similar velocities is found on the south--east side of the disk (labelled as ``E'' in Fig.~\ref{fig:kinematic}), with counter rotating velocities with respect to the gas on this side of the galaxy.

\subsection{The spatially resolved gas properties}\label{sec:gas}
\begin{figure}
\centering
\includegraphics[height=0.37\textwidth]{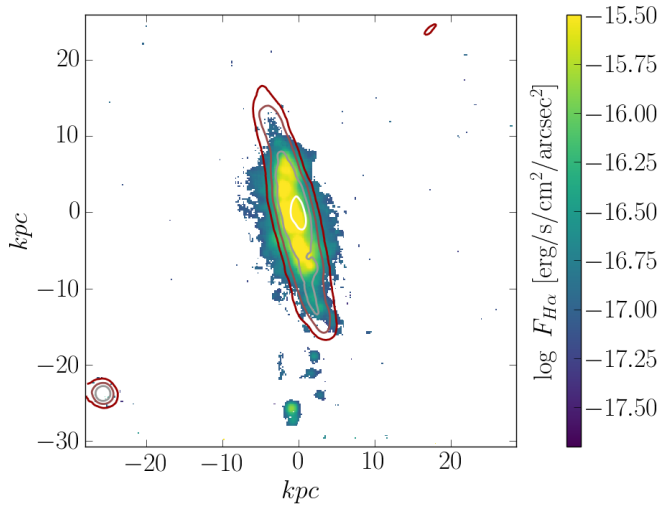}
\caption{H$\alpha$ surface brightness map. The red lines are stellar emission isocontours as derived from the H$\alpha$ continuum emission.}
\label{fig:Halpha}
\end{figure}
The H$\alpha$ surface brightness map is shown in Fig.~\ref{fig:Halpha}. The map reaches a surface brightness of $2\times 10^{-18}$ erg s$^{-1}$ cm$^{-2}$ arcsec$^{-2}$ at 3$\sigma$ limit \citep[which is the characteristic value for MUSE data of this program; see][]{poggianti17} and, when compared to the stellar emission (see e.g. the stellar velocity map) it shows evidence for the truncation of the ionized gas disk. Ionized gas is found out to galactocentric distances of about 15 kpc, while the stellar disk extends to $\sim 25$ (with a surface brightness detection limit in the $V$ band of $\sim 27$ mag arcsec$^{-2}$ at the $3\sigma$ confidence level). In Sect.~\ref{sec:discussion} we will discuss the possible origin of this truncation.

We have created diagnostic diagrams \citep[see, e.g.][]{kewley06} using emission lines lying within the observed range of our data (i.e. H$\beta$ {\sc [Oiii]} 5007 \AA, {\sc [Oi]} 6300 \AA, H$\alpha$, {\sc [Nii]} 6583 \AA, and {\sc [Sii]} 6716+6731 \AA) to derive the characteristics of the ionising sources as a function of the position, and detect the possible presence of an AGN. The lines intensities were measured after subtraction of the continuum, exploiting the pure stellar emission best fit model provided by \sinopsis, so to take into account any possible contamination from stellar photospheric absorption. 
\begin{figure*}[!t]
\begin{tabular}{cc}
\includegraphics[height=0.47\textwidth]{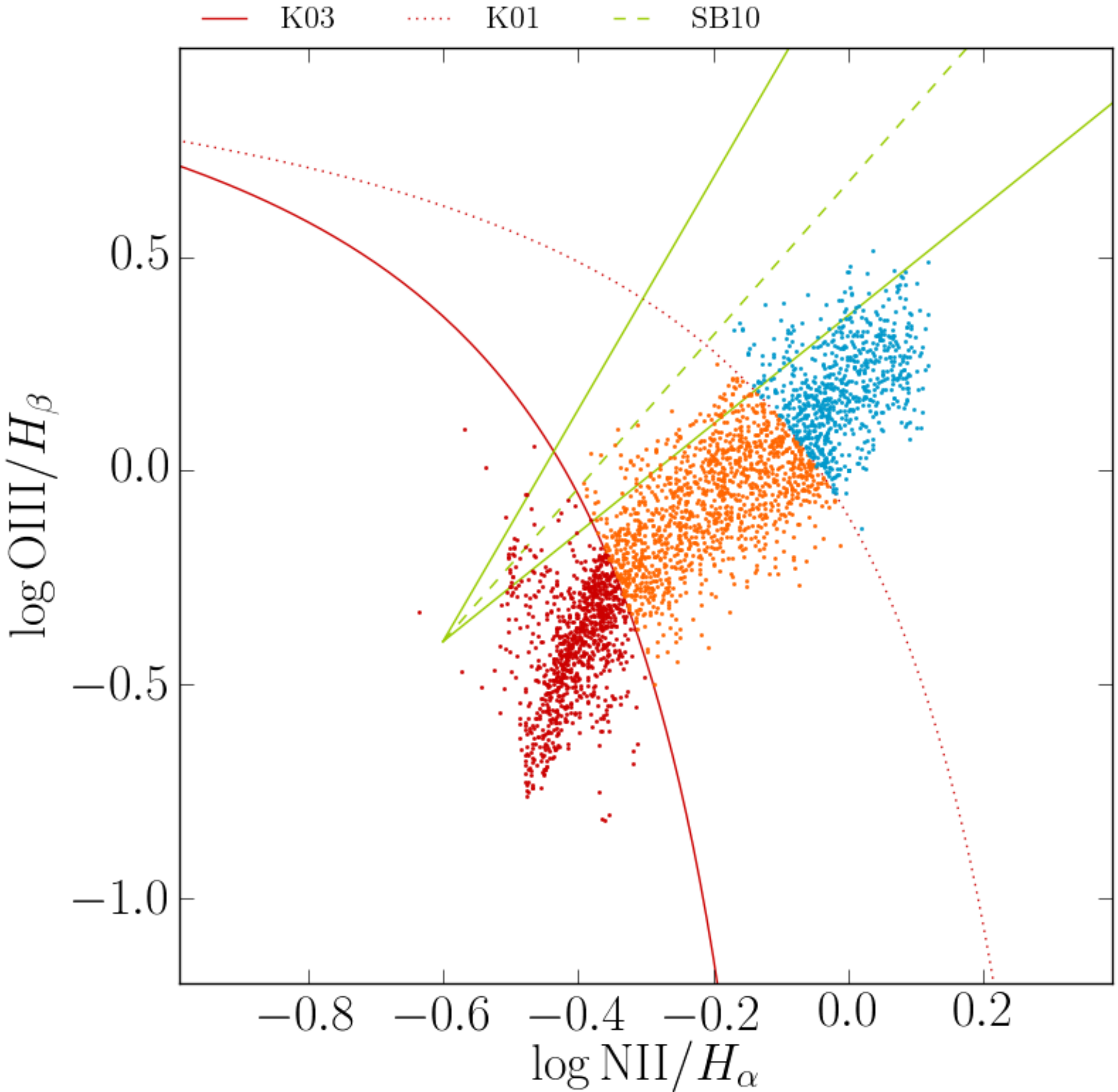} &
\includegraphics[height=0.47\textwidth]{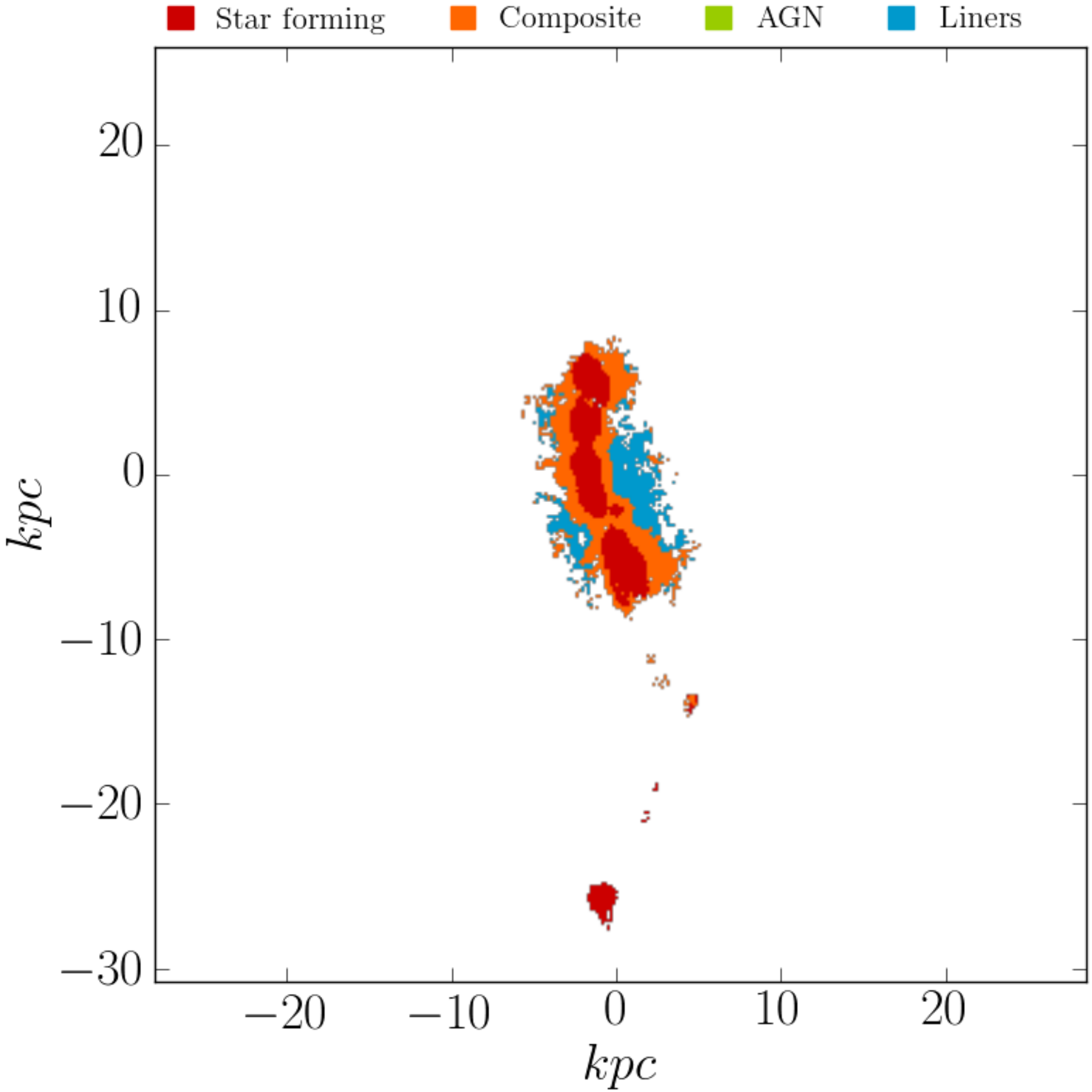} \\
\end{tabular}
\caption{Left. Diagnostic diagram of the ionising sources across the galaxy. The red dotted and continuous lines are defined as in \cite{kewley01} and \cite{kauffmann03}, respectively. The green lines are taken from \citep{sharp10} (see text for details.) Right. Spatial locations of the spaxels color--coded based on the ionising source diagnostic. H$\alpha$ continuum surface brightness contours are shown as a reference for the stellar emission.}
\label{fig:bpt}
\end{figure*}

The three diagrams we have used, namely log{\sc [Nii]}/H$\alpha$ vs log{\sc [Oiii]}/H$\beta$ (shown in Fig.~\ref{fig:bpt}), log{\sc [Oi]}/H$\alpha$ vs log{\sc [Oiii]}/H$\beta$, and log{\sc [Sii]}/H$\alpha$ vs log{\sc [Oiii]}/H$\beta$ (not presented in this paper), are concordant in excluding the presence of an AGN. We consider this result to be quite robust, given that in the center of the galaxy, where a possible AGN is likely to be located, the measured S/N is the highest.

Interestingly enough, deep Chandra archive images have detected the presence of a luminous though highly absorbed X-ray source strongly incompatible with a possible nuclear starburst, as described in Sect.~\ref{sec:xray} (and Nicastro et al. in prep.). This would imply that, if an AGN is indeed the source of this luminosity, it should be highly obscured so that it would be non--detected by optical lines diagnostics.

The results, presented in Fig.~\ref{fig:bpt}, show that the emission line luminosity is powered either by star formation or by LINER--like mechanisms such as shocks. In particular, the central parts of the disk are those dominated by star formation, while the gas at higher galactic altitudes shows characteristics of LINER emission or intermediate between the two (see the right--hand panel of Fig.~\ref{fig:bpt}).

Clear signatures of stripping along the line of sight are visible as double--peaked emission lines profile, or as departure from a gaussian profile, mostly visible in H$\alpha$. These are located in the outskirts of the disk, in regions with a LINER emission origin.

It is interesting to notice that the regions classified as ``Star Forming'' in the left panel of Fig.~\ref{fig:bpt}, are clearly displaced towards the east with respect to the center, defined by the H$\alpha$ continuum contour, and slightly bent with respect to the major axis.

\subsection{Properties of the stellar populations}\label{sec:stellarprop}
We now study both the global and spatially resolved stellar populations of this galaxy by analysing the SFR as a function of time in four age bins. These are logarithmically spaced and chosen in such a way that the differences between the spectral characteristics of the stellar populations are maximal, and are defined according to Table~\ref{tab:bins}.

The stellar populations properties were obtained by applying \sinopsis \ to the observed datacube in each spaxel with a reliable redshift determination, using three sets of SSP spectra with fixed metallicity values (namely, $Z=0.004$, $Z=0.02$, and $Z=0.04$). Whenever a stellar redshift was available, this was used for the spectral fitting, while the equivalent widths of emission lines were measured using the redshift value derived from emission lines. About 15000 observed spectra were analyzed (the runtime takes approximately 8 hours).
\begin{table}
\centering
\caption{Ages of the the stellar populations, in years, for which we calculate physical properties from the spectral fitting.}
\label{tab:bins}
\begin{tabular}{lrr} 
\hline
Bin  &  Lower age  & Upper age \\
\hline
1      &            0                  &   $2\times 10^7$        \\      
2      &   $2\times 10^7$     &   $5.72\times 10^8$   \\    
3      & $5.72\times 10^8$  &   $5.75\times 10^9$   \\
4      & $5.75\times 10^9$  &   $14\times 10^9$      \\
\hline
\end{tabular}
\end{table}

\begin{figure}[!h]
\centering
\includegraphics[height=0.45\textwidth]{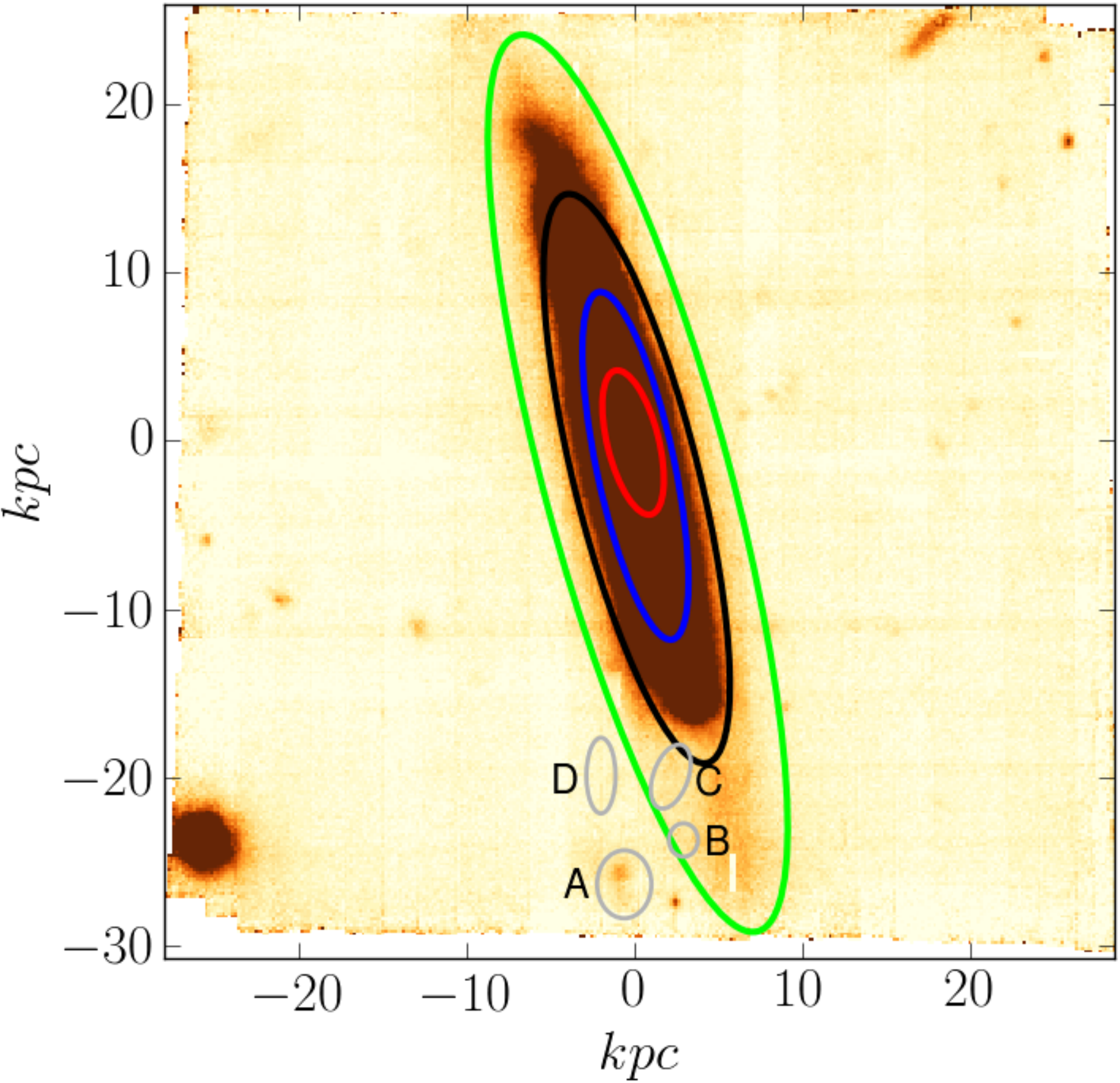}
\caption{The MUSE datacube integrated with respect to the wavelength. The ellipses are the areas where the SFR is computed, with the same color--coding as in Fig.~\ref{fig:sfh_regions}. The gray ellipses and circles on the south part are the 4 H$\alpha$ blobs as identified in Sect.~\ref{sec:kinematic}.}
\label{fig:collapsed}
\end{figure}

The total stellar mass, calculated as the sum of stellar masses in all the spaxels encompassed by region 4 (the larger ellipse in Fig.~\ref{fig:collapsed}, see also Table.~\ref{tab:ellipses}), is of $6.49^{+0.16}_{-0.19} \times 10^{10} $ M$_\odot$, and is slightly higher than $4.8\pm0.8\times 10^{10}$ M$_\odot$, which is the the mass value calculated from the WINGS integrated spectrum, after correction for aperture effects\footnote{The main source of error on this mass is probably due to the quite large aperture correction which, on top of that, is even more uncertain for edge-on galaxies.}. Using \sinopsis \ to derive the stellar mass from the integrated spectrum of region 4, yields a value of $5.89^{+0.89}_{-1.14}\times 10^{10}$ M$_\odot$, fully compatible with the value calculated from the spatially resolved data.

Similarly to the stellar mass, \sinopsis \ also provides an estimate of the recent (i.e. $\lesssim 2\times 10^7$ yr) SFR. This value is obtained by summing the SFR values in each spaxel, and it already contains a correction for dust extinction which is performed within the spectral fitting procedure. The integrated value of the recent SFR calculated in this way, is $5.88^{+1.57}_{-0.93}$ M$_\odot$/yr, about 90\% of which is concentrated within the innermost parts, where dust extinction also reaches the highest values as shown in Fig.~\ref{fig:Av} (this corresponds to the spaxels enclosed within ellipse n.2 in Fig.~\ref{fig:collapsed}). 

To find possible trends in the stellar properties as a function of the position, we have considered 4 annuli, defined as the regions in between elliptical apertures, which are chosen to roughly match the surface brightness intensity of the stellar emission at different levels. These are depicted in Fig.~\ref{fig:collapsed}. Table~\ref{tab:ellipses} reports the physical sizes of the ellipses. Furthermore, we separately analyze the star formation histories of both the stellar tail and the four H$\alpha$ emission blobs identified in Fig.~\ref{fig:kinematic}.
\begin{table}[!h]
\centering
\begin{tabular}{lcc}
\hline
Region & a  	& b  \\
\hline
1	&    4.1	&  1.4	\\
2	&    9.8	&  2.2	\\
3	&  16.2	&  3.5	\\
4	&  25.6	&   5.1	\\
\hline
\end{tabular}
\caption{Size, in kpc, of the major (a) and minor (b) semiaxes of the ellipses defining the regions in Fig.~\ref{fig:collapsed}}
\label{tab:ellipses}
\end{table}

Calculating the SFR in the previously defined annuli, is an effective way to look for broad spatial trends in the average ages of the stellar populations as a function of the galactocentric distance. After the first star formation episode, when about 65\% of the stellar mass was created, the galaxy underwent a decrease in the star forming activity, followed by a subsequent star formation episode with an intensity, relative to the previous age bin, higher in the outskirts with respect to the center.

\begin{figure}
\centering
\includegraphics[height=0.65\textwidth]{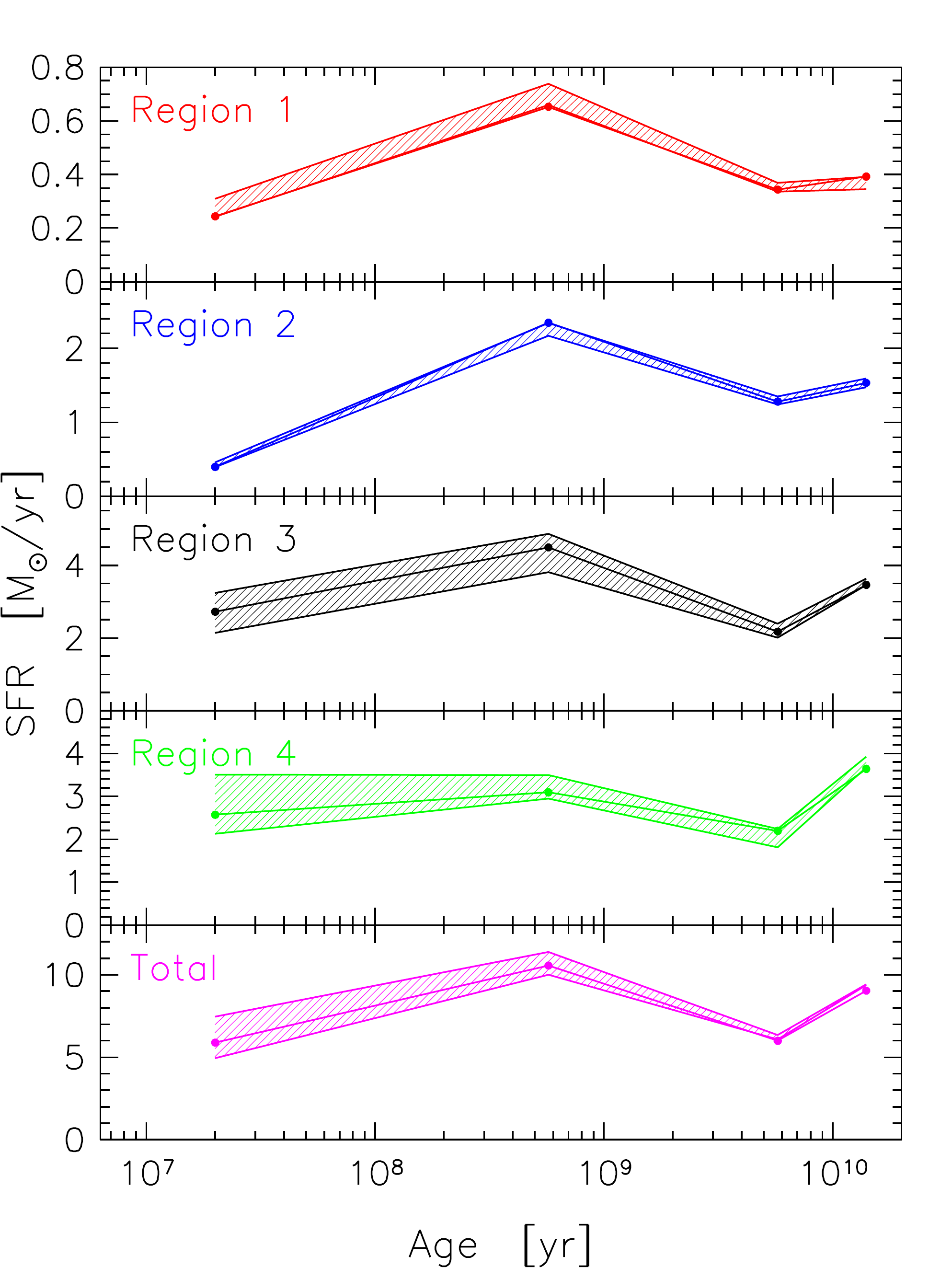}
\caption{The first 4 panels from the top show the star formation history of the galaxy calculated within the annuli defined in Fig.~\ref{fig:collapsed} (same color coding)  and Table~\ref{tab:ellipses}. These are labelled from 1 to 4 going from the innermost to the outermost region. The lowest panel displays the same quantity but for the whole galaxy.}
\label{fig:sfh_regions}
\end{figure}

This is clearly represented in Fig.~\ref{fig:sfh_regions}, where we show the SFR a function of age and position. This is indicative of an inside--out formation scenario in the early epochs of the galaxy: the SFR decreased after the initial burst more abruptly in the innermost regions, while being sustained at a higher rate in the disk outskirts. An intense star formation activity involving the whole galaxy occurred between 20 Myr and 0.5 Gyr ago, with a much higher intensity in the outer part than in the galaxy center. During this event, the SFR increased by only $\sim15$\% in the innermost region (region 1), while the outer parts (regions 3 and 4) experienced a boost by almost 50\%. 

This event converted, according to our modelling, about $10^{10}$ M$_\odot$ of gas into stars in the outer disk (i.e. the annulus between ellipses 2 and 4), an amount that represents about 15\% of the currently observed total stellar mass in the whole galaxy.

\begin{figure}
\includegraphics[height=0.41\textwidth]{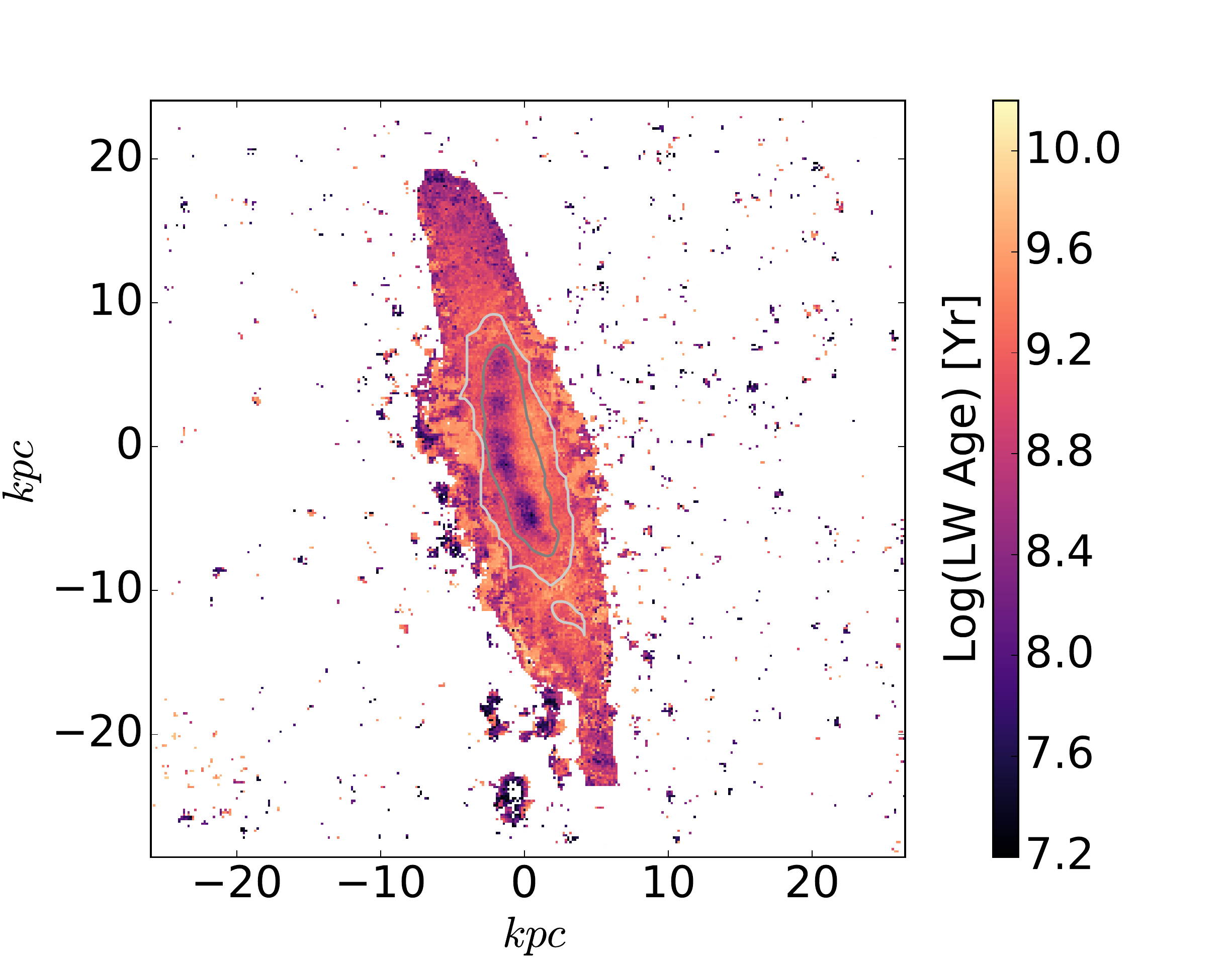}
\caption{Map of the luminosity--weighted stellar age as calculated from spectral modelling.}
\label{fig:lwa}
\end{figure}
\begin{figure*}
\begin{tabular}{ll}
\centering
\includegraphics[height=0.39\textwidth]{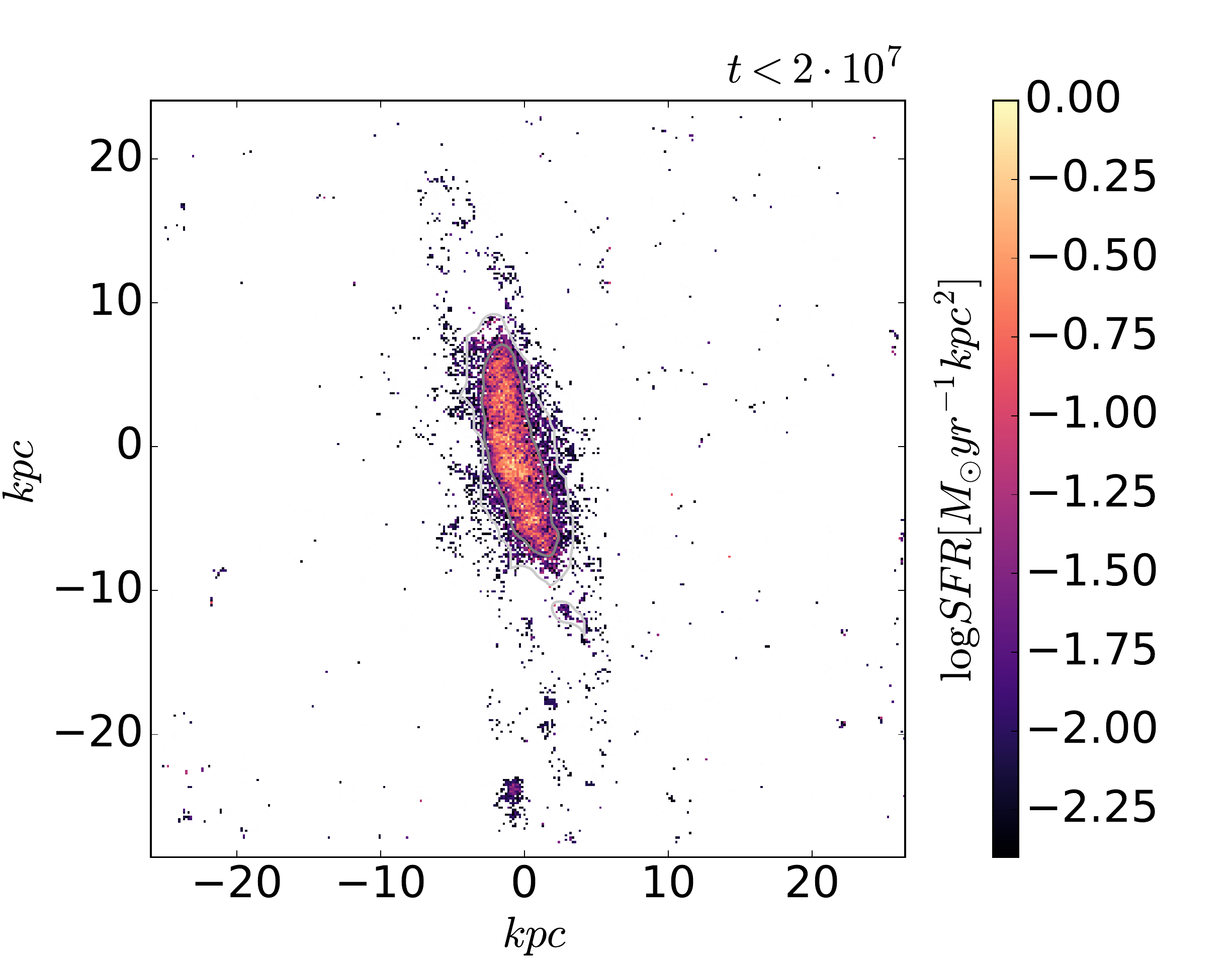} &
\includegraphics[height=0.39\textwidth]{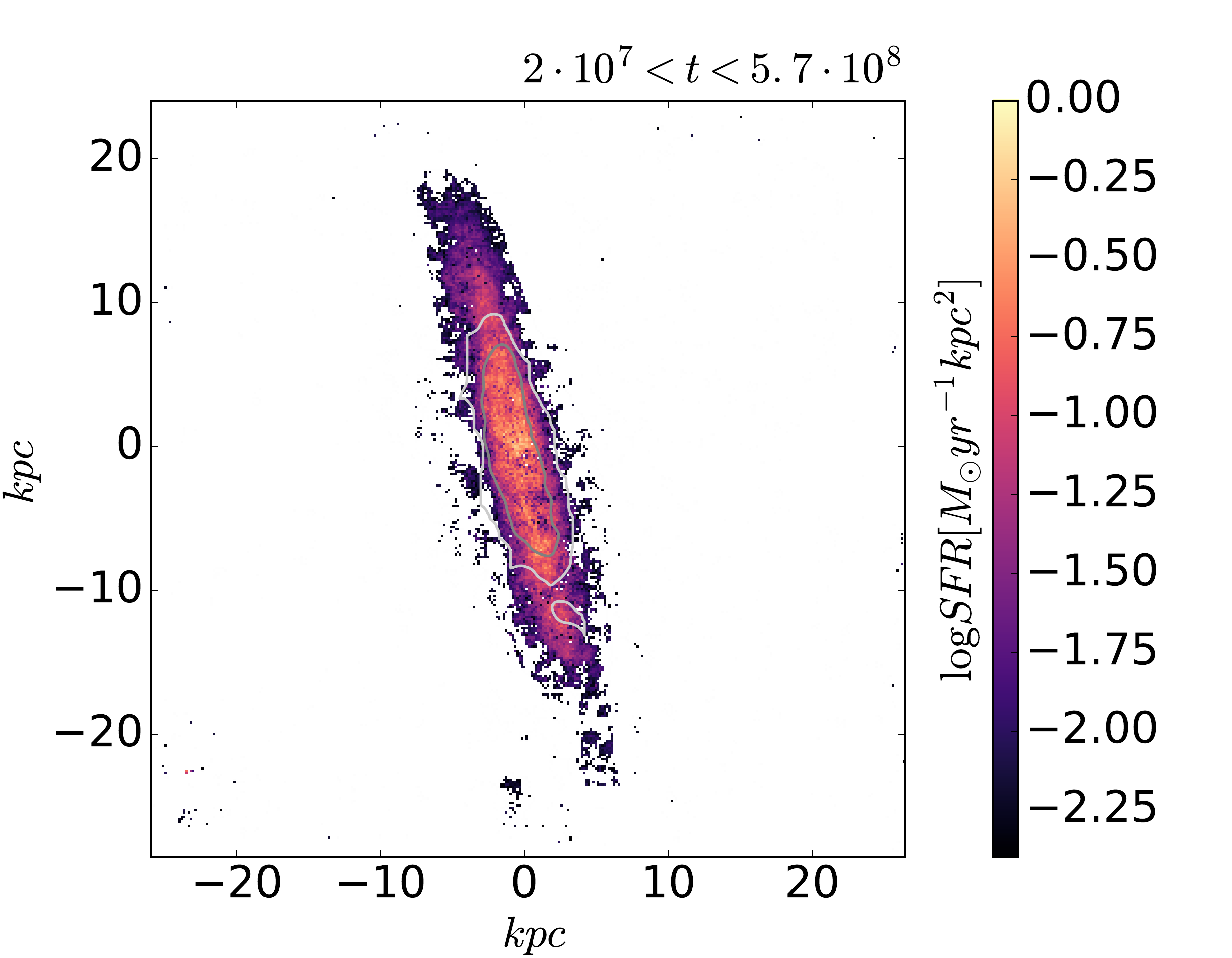} \\
\includegraphics[height=0.39\textwidth]{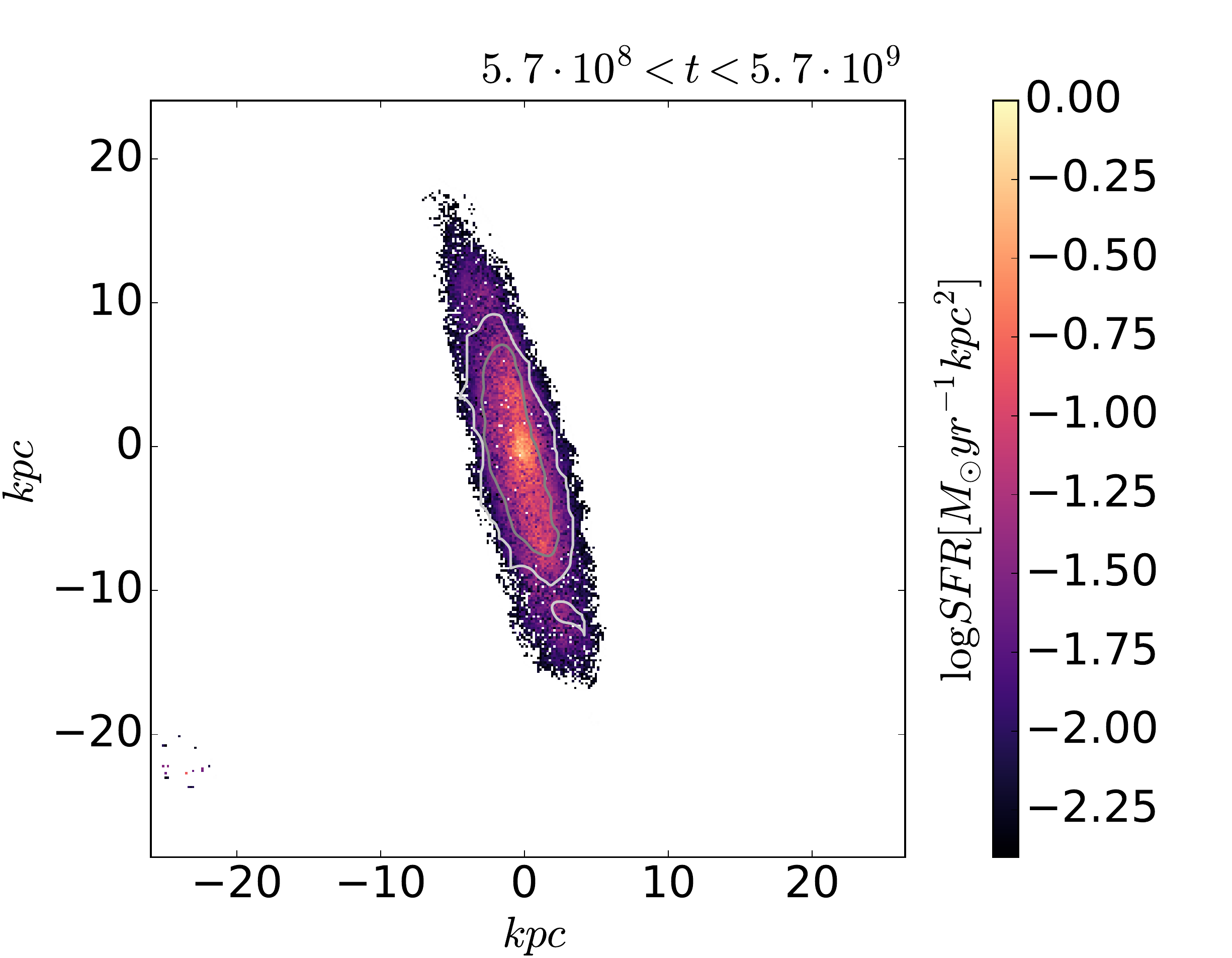} &
\includegraphics[height=0.39\textwidth]{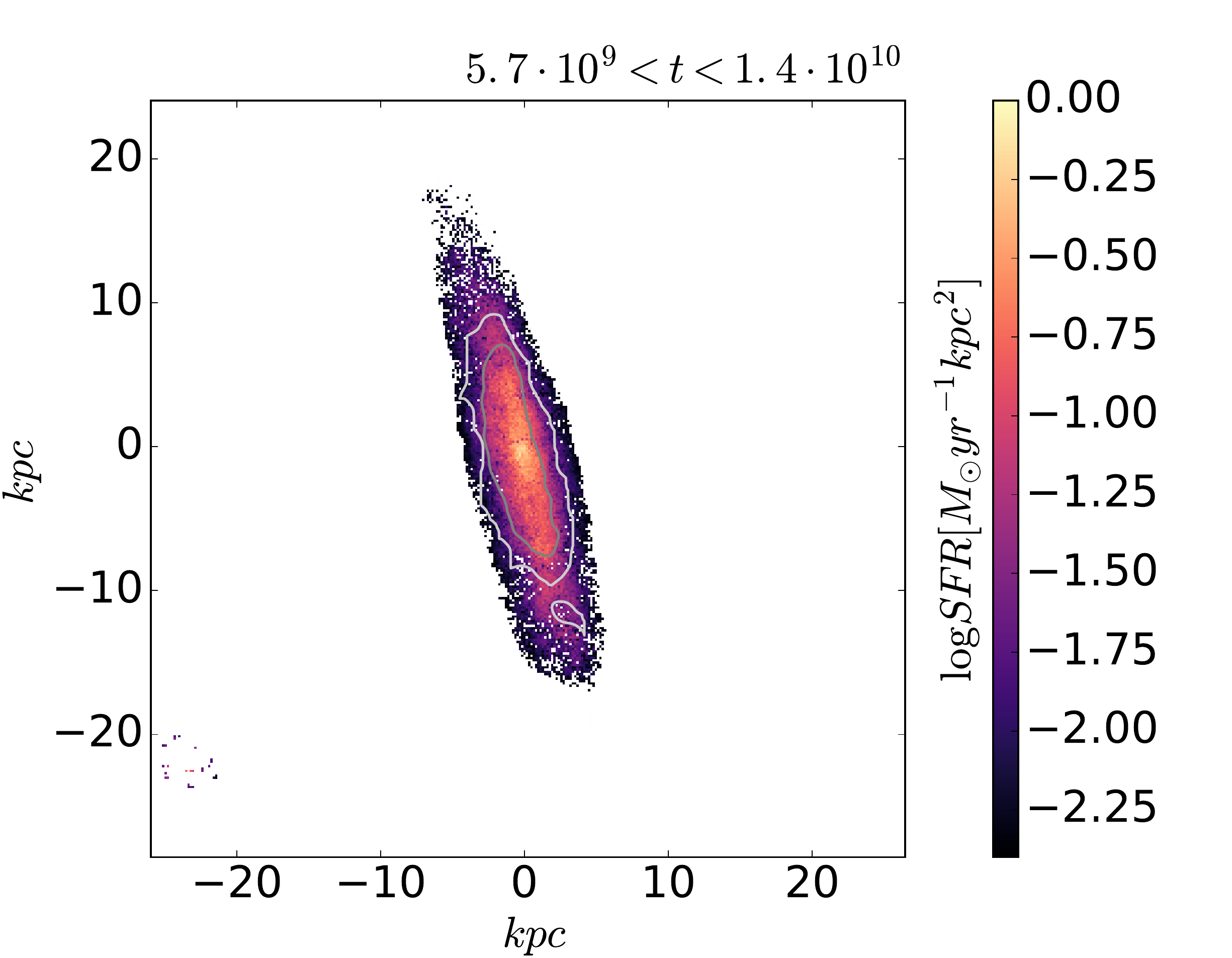} \\
\end{tabular}
\caption{Star formation rate surface density as a function of position for different epochs corresponding to the 4 main SSP age bins. The contours are defined in the same manner as for Fig.~\ref{fig:kinematic}. The ring--like structure visible in the youngest stellar population is an artefact from the code, which mistakes noisy features for an H$\alpha$ line in emission.}
\label{fig:sfh}
\end{figure*}

Fig.~\ref{fig:sfh} presents the spatially resolved star formation rate surface density in 4 age bins. These are calculated by re-binning the SFR values of the 12 SSPs used for the fit, according to the definition and ages given in \cite{fritz07}.

There are no signs of ongoing star formation outside the disk, except for the southern blobs where we clearly detect ionized gas. Indeed, the top--left panel in Fig.~\ref{fig:sfh}, shows that the most intense star forming spaxels are found within the central parts of the disk with values up to $\sim 5\times 10^{-2}$ M$_\odot$ yr$^{-1}$ kpc $^{-1}$, while outside this region, very well defined by the H$\alpha$ continuum contours, only very faint and sparse signatures of current star formation are found. 

The outermost parts of the disk are dominated by intermediate--age (i.e. between $\sim2\times 10^7$ and $\sim6\times 10^8$ years) stellar populations; these very same stars are also the main population found in the tail departing from the southern disk that was identified in Fig.~\ref{fig:kinematic}, where no emission lines were detected. The oldest stars are dominating the bulge of the galaxy, and they are the most concentrated population as depicted in the lower--right panel of Fig.~\ref{fig:sfh}.

The luminosity weighted age map, shown in Fig.~\ref{fig:lwa}, highlights the changes in the average age of the stellar populations at each location in the galaxy. This displays a minimum in the central parts of the galaxy, as expected given that it is at this location where the bulk of the star formation is happening. Very young ages are found in the blobs located in the southern outskirts as well, which are all found to be star--forming. This is consistent with the faint stellar continuum and H$\alpha$ being observed in emission, and it is further backed up by the low value of the luminosity weighted ages.

In the most luminous, blob ``A'', H$\alpha$ equivalent width reaches a value of $-64$ \AA. The star formation rates derived from \sinopsis \ from the integrated spectra of the blobs, range from $3\times 10^{-3}$ (blob B) to $1.2\times 10^{-2}$ M$_\odot$/yr (blob A), while the stellar masses have values in the range between $5.1\times 10^6$ (blob D) and $1.7\times 10^8$ M$_\odot$ (blob A). Relatively young ($\lesssim 500$ Myr) stars are present throughout all the disk. 

We point out that the spatial trends we observe in the stellar population properties are very likely weakened by projection effects, given the high inclination angle of the galaxy, and might be actually even stronger. 

\subsection{The Chandra View of the Nuclear Region of JO36}\label{sec:xray}
JO36 was serendipitously observed by {\em Chandra} on October 18th, 2002, as part of the targeted observation of the cluster Abell~160, and for a total of 58.5 ks. The galaxy is located 5.8 arcmin off-axis, with respect to the {\em Chandra} ACIS-I aimpoint, where the 2 keV off-axis/on-axis effective area ratio (i.e. vignetting) is $\sim 0.9$, and the Point Spread Function (PSF) Encircled Energy Radius is $\sim 1.5-2$ arcsec (cf. with $\sim 0.5$ arcsec on-axis; ``The {\em Chandra} Proposal Observatory Guide'', v. 19.0, \url{http://cxc.harvard.edu/proposer/POG/html/chap6.html}). 

\begin{figure}[!t]
\centering
\includegraphics[width=\columnwidth]{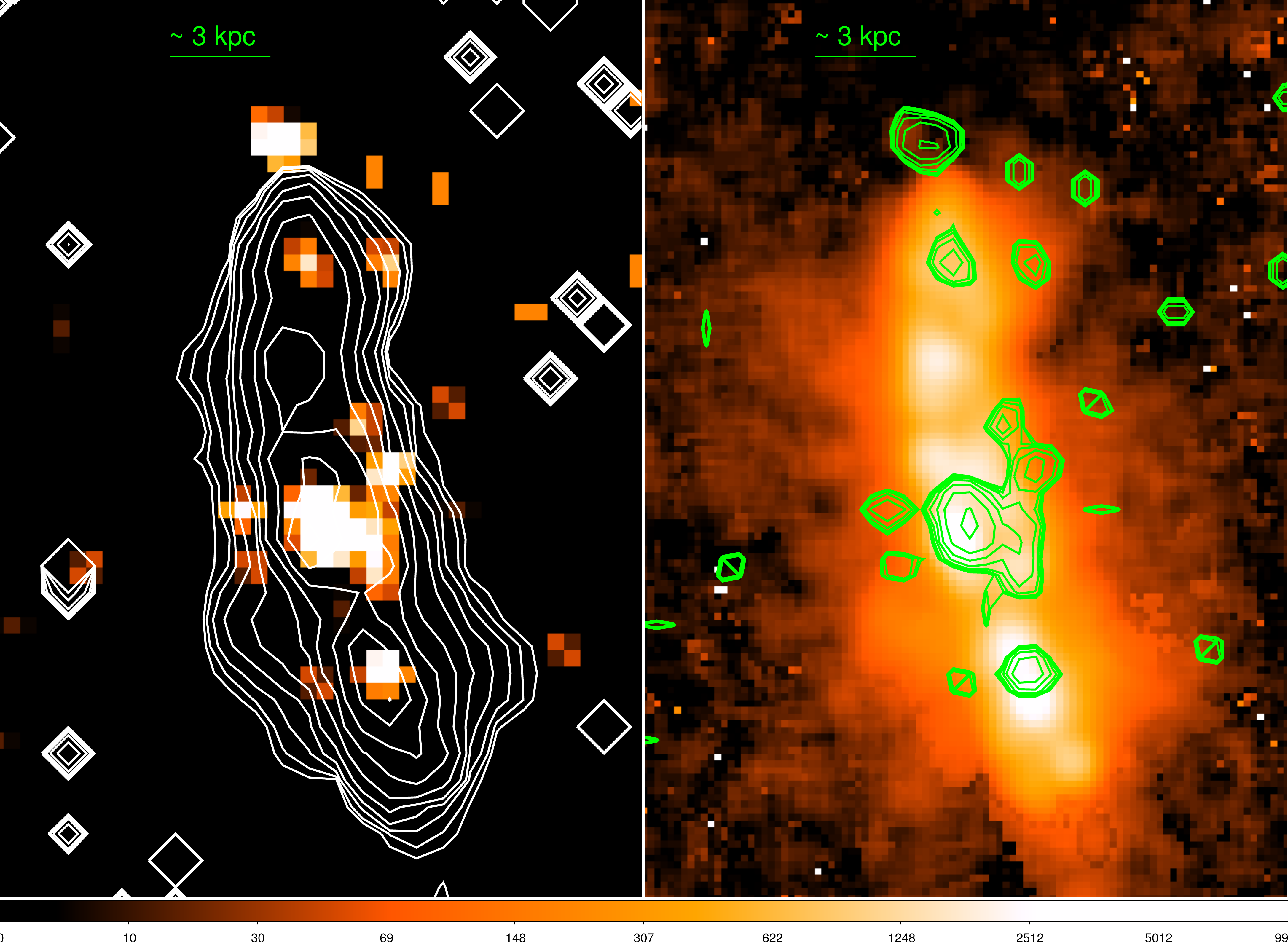}
\caption{{\em Chandra} 0.3-10 keV (left panel) and MUSE H$\alpha$ (right panel) images of the GASP cluster galaxy JO36: X--ray (green) and H$\alpha$ (white) contours are superimposed on the H$\alpha$ and X--ray images, respectively.}
\label{fig:xray1}
\end{figure}

A bright X--ray nucleus is clearly detected (Fig.~\ref{fig:xray1}, left panel), at a position coincident with that of the bright H$\alpha$ nucleus (Fig.~\ref{fig:xray1}, right panel), together with several fainter point-like X--ray sources (most likely Ultra-Luminous X--ray - ULX - sources; Nicastro et al., in preparation), aligned with the galaxy's edge-on disk seen in H$\alpha$ (white contours superimposed to the X--ray image in the left panel of Fig.~\ref{fig:xray1}). Interestingly, the brightest of these off-nuclear X--ray sources is located just at the north edge of the truncated gas disk, where little or no H$\alpha$ emission is seen. 

To estimate the X--ray luminosity of the nucleus, we extracted source and background X--ray counts respectively from a 3 arcsec radius circular region centred on the source centroid (RA=18.24788, DEC=15.59122) and from 4 additional 3 arcsec radius source-free circular regions located $\sim 15$ arcsec north-east and south-east of the nucleus. The nuclear region contains 32 full band {\em Chandra} counts, while the 4 background regions contain a total of 7 counts. Rescaling by the 4--times source to background smaller extraction area, this gives a net number of 0.3--10 keV source counts of $30.8 \pm 5.6$, or a count rate of $(5.3 \pm 1.0)\times 10^{-4}$ cts s$^{-1}$. 

\begin{figure}
\centering
\includegraphics[width=\columnwidth]{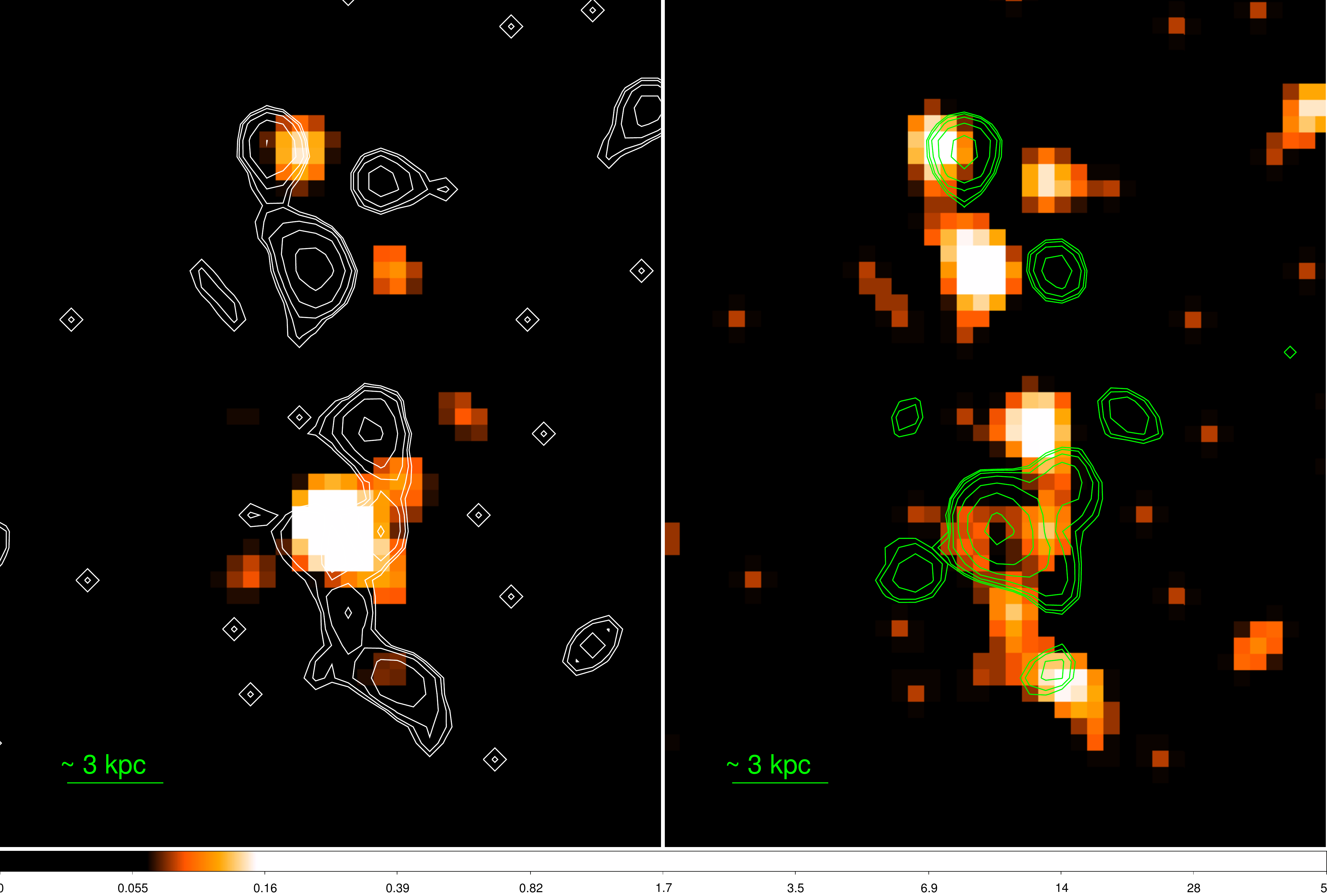}
\caption{{\em Chandra} 2-10 keV (left) and 0.3-2 keV (right) images of the GASP cluster galaxy JO36.}
\label{fig:xray2}
\end{figure}

The nuclear X--ray counts are all detected above 2 keV (compare left and right panels of Fig.~\ref{fig:xray2}), which suggests that the X-ray emission is highly absorbed. Indeed, binning the $\sim$ 31 source net counts into bins with $\ge 10$ counts, leaves a three-bin spectrum (E$_{bin} = 1.8, 4.2$ and 6.7 keV) peaked at 4.2 keV. Modeling the spectrum with a simple power-law (F$ = A (E/E_0)^{\Gamma}$) yields an extremely flat photon spectral index $\Gamma = -0.9$, which also underestimates the spectrum peak count rate. Including a column N$_H$ of intrinsic nuclear cold gas surrounding the X--ray source and attenuating the soft X--rays along our line of sight, and freezing the photon spectral index to the commonly observed AGN value of $\Gamma = 2$ \citep[e.g.][]{piconcelli05}, yields instead flat residuals and a best-fitting N$_H = 1.1^{+0.7}_{-0.4} \times 10^{23}$ cm$^{-2}$, as typically observed in highly obscured type 2 Seyfert galaxies \citep[e.g.][]{risaliti99}. 

From the best--fitting spectral model we derive an observed (i.e. absorbed) 2--10 keV flux F$_{2-10} = (3.5 \pm 1.5) \times 10^{-14}$ erg s$^{-1}$ cm$^{-2}$, which translates (at the distance of JO36) into an observed luminosity L$_{2-10} = (1.4 \pm 0.6) \times 10^{41}$ erg s$^{-1}$ and an intrinsic (i.e. unabsorbed) luminosity of $L^{Unabs}_{2-10} = (2.8 \pm 1.1) \times 10^{41}$ erg s$^{-1}$. By factoring a bolometric correction factor of $\simeq 10$ \citep[appropriate for L$_{2-10} \simeq 3 \times 10^{41}$ erg s$^{-1}$, i.e.][]{marconi04}, we get: L$_{Bol} \simeq 4 \times 10^{42}$ erg s$^{-1}$, consistent with the low luminosity end of Seyfert galaxies and thus pointing towards the presence of a buried AGN in the nucleus of JO36, which was non detected by optical diagnostic diagrams (Sect.\ref{sec:gas}). 

An additional independent (but indirect) check of the presence of an AGN in the nucleus of JO36 can be done by comparing the star formation rate density (SFRD) derived in the nuclear region through the H$\alpha$ diagnostics (SFRD$_{H\alpha} \simeq 0.14$ M$_{\odot}$ yr$^{-1}$ kpc$^{-2}$), with the estimate derived by assuming that all the 31 nuclear {\em Chandra} counts be uniformly distributed over a compact, $\ls 2.7$ kpc (i.e. 3 arcsec) radius, nuclear star-bursting region and be due to a large (i.e. $\gs 100$) number of unresolved luminous X-ray binaries. This gives an observed 2--10 keV luminosity density of $\mathcal{L}_{2-10} \gs  7.6 \times 10^{39}$ ergs s$^{-1}$ kpc$^{-2}$, which translates into SFRD$\gs 1.5$ M$_{\odot}$ yr$^{-1}$ kpc$^{-2}$ \citep[e.g.][]{ranalli03}. This is more than 15 times larger than that observed in H$\alpha$, again suggesting the presence of an AGN in the nucleus of JO36. 

\section{The Interstellar Medium in JO36}\label{sec:ism}
\begin{figure}[!t]
\includegraphics[height=0.50\textwidth]{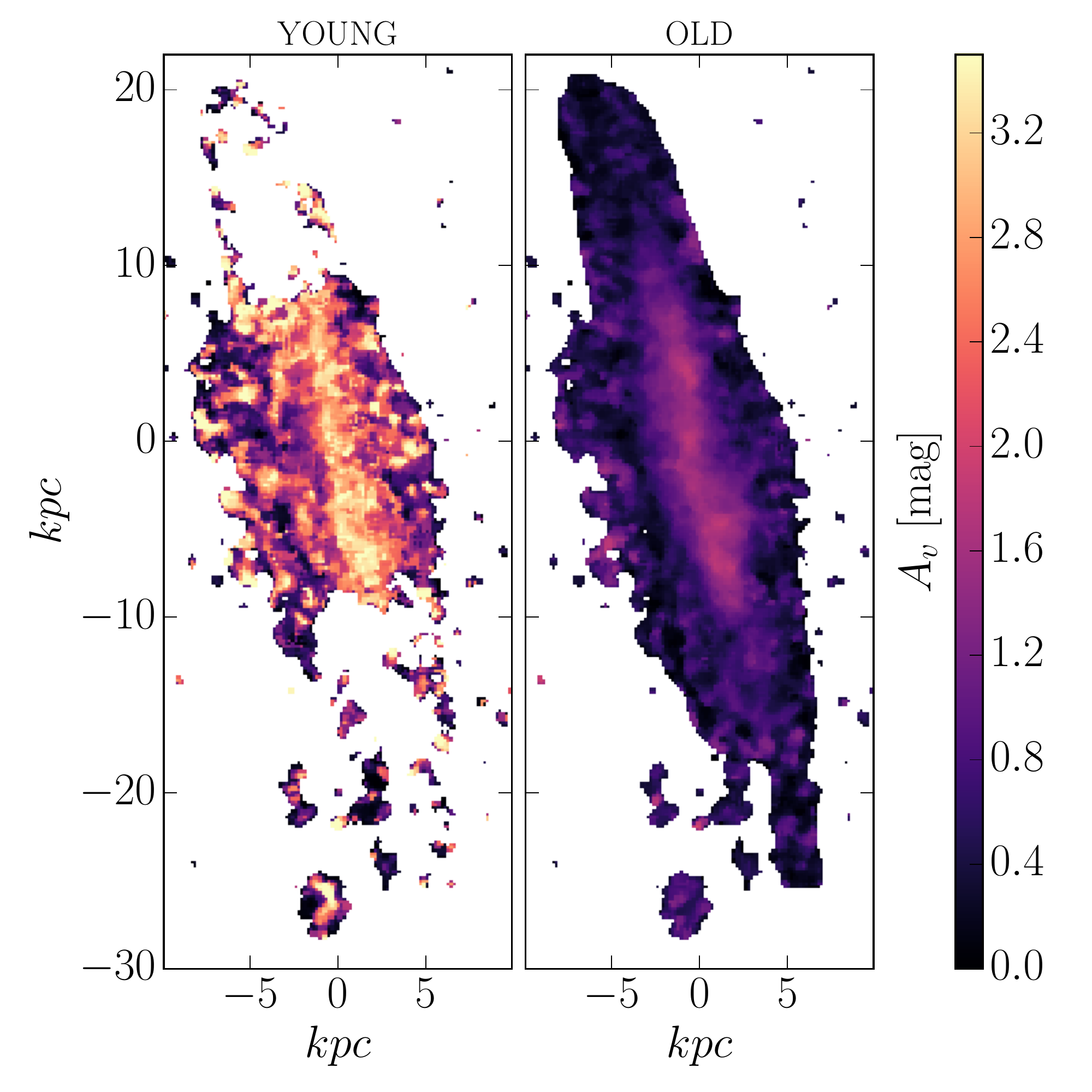}
\caption{On the left--hand panel, the extinction map for the youngest stellar populations (i.e. with age $\lesssim 2\times 10^7$ years), as derived by spectral fitting, is presented. On the right--hand panel, the same quantity is shown, but for older stars. Both maps were smoothed to improve readability.}
\label{fig:Av}
\end{figure}

\sinopsis \ provides values for the emission lines dust attenuation, whose map we show in the left--hand panel of Fig.~\ref{fig:Av}. A comparison with the same quantity calculated from the observed Balmer decrement (H$\alpha$/H$\beta$) shows excellent agreement. Values of $A_V$ as high as 4 are found towards the central parts of the galaxy, this being partly due to the high inclination of the galaxy disk with respect to the line of sight. 

In the right--hand panel of Fig.~\ref{fig:Av} we show the dust extinction calculated by \sinopsis \ for the stars older than $2\times 10^7$ yr. The highest values are reached in the same position where the extinction from emission lines is also maximum. The ratio between the two extinction values is in general $<2$, as found e.g. in \cite{calzetti00} \citep[but see also][for slightly lower values]{wuyts13}, but it reaches higher values in a small fraction of pixels, probably due to the galaxy geometry.

These maps only give a proxy for the presence of dust, as they do not take into account the 3D structure of the galaxy, projection effects, and the fact that the most dusty regions can be completely invisible at optical wavelengths. 

Deriving the amount of dust from attenuation maps in the optical is doable, but prone to the aforementioned uncertainties and is best done by means of radiative transfer models \citep[see e.g.][and references therein]{popescu02,baes10,degeyter14,saftly15}, which are well beyond the goals of this work.

A much more reliable way, as opposed to the extinction map, is to look at the dust thermal emission, showing up at far infrared and sub--millimetre wavelengths.

JO36 is located within a field recently observed with the infrared space observatory {\it Herschel} \citep{pilbratt10} as part of the program \texttt{KPOT\_mjuvela\_1} \citep[P.I. Mika Juvela,][]{juvela07}. These observations, taken with both the PACS \citep{poglitsch10} and SPIRE \citep{griffin10} instruments, reveal an intense infrared emission detected at all wavelengths (100, 160, 250, 350, and 500 $\mu$m). 

We have reduced both PACS and SPIRE data in two steps, with the first one making use of the latest version of hipe (v14.2.0) to get the data to Level1, while the map making, de--glitching and baseline removal were performed with the latest version of the IDL package SCANAMORPHOS \citep[v25,][]{roussel13}. We have measured fluxes in apertures encompassing the whole galaxy in all maps, performing background subtraction as customary for such kind of data \citep[see, e.g.][]{ciesla12,verstappen13,cortese14}.

The much lower spatial resolution of {\it Herschel} data (the highest resolution is reached for PACS at 100 $\mu$m, and is about 6\arcsec), when compared to optical images, makes it very hard to establish a spatial connection between the geometrical distribution of the dust and that of the ionized gas, as derived from MUSE data. Nevertheless, we can calculate a global estimate of the total dust mass and use this value to infer the gas mass. 

Dust mass can be derived by means of SED fitting using a modified black body model emission. In Table~\ref{tab:irfluxes} we report the measured infrared fluxes used for the modeling.
\begin{table}[h]
\centering
\caption{Fluxes densities and corresponding uncertainties, in Jy, measured on the five {\it Herschel} bands from archival images.}
\label{tab:irfluxes}
\begin{tabular}{lcc} 
\hline
$\lambda$ [$\mu$m] &  Flux  & Error \\
\hline
100      &     0.77    &    0.05     \\      
160      &     1.01    &    0.08     \\    
250      &      0.47   &    0.04     \\
350      &      0.20   &    0.02       \\
500      &      0.07   &    0.01       \\
\hline
\end{tabular}
\end{table}

\begin{figure}[!t]
\centering
\includegraphics[width=\columnwidth]{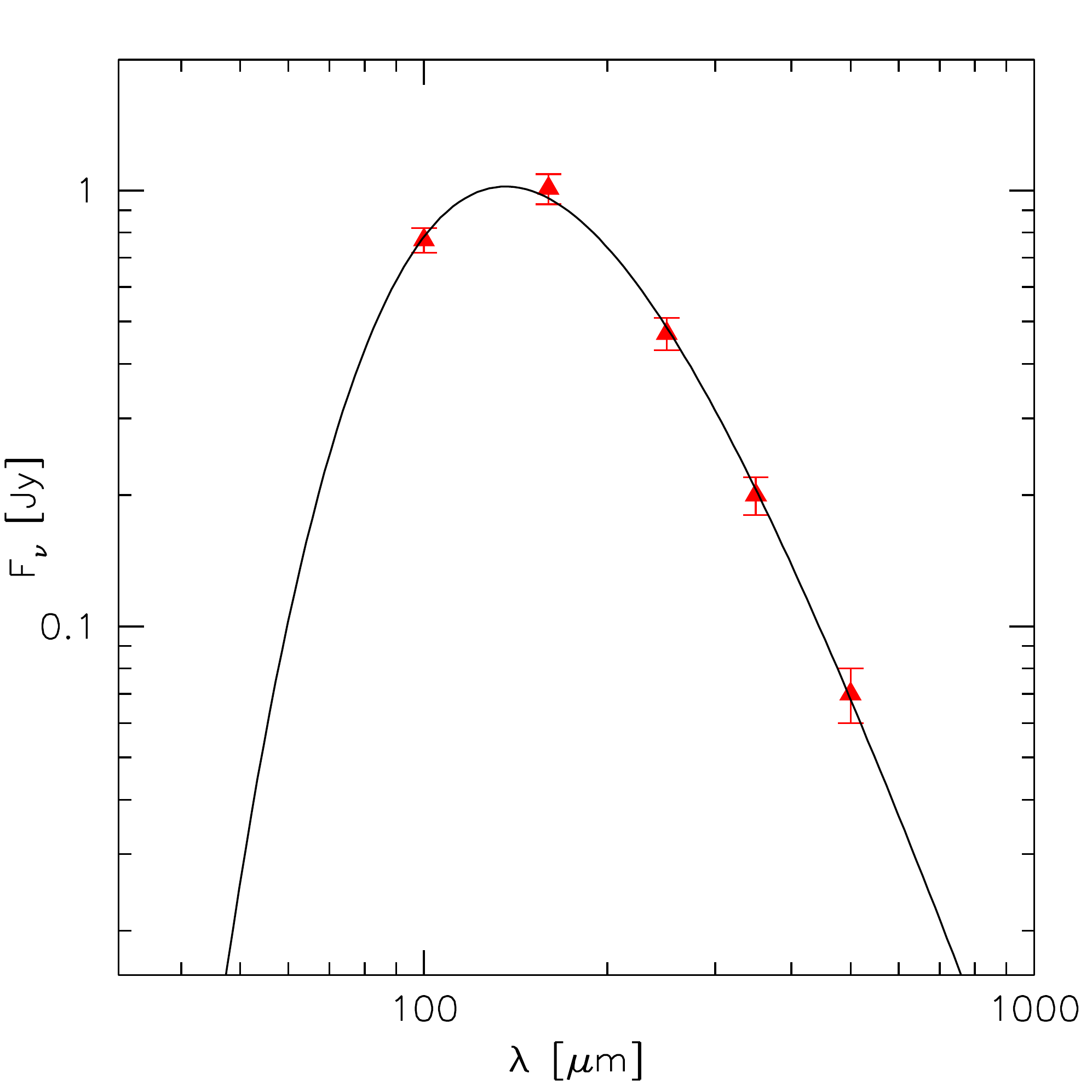}
\caption{Modified black body fit (black line) to the observed {\it Herschel} datapoints (red triangles). A dust temperature of 20.63 K, a $\beta =2.15$, and a dust mass of $9.8\times 10^7$ M$_\odot$ are derived from this model.}
\label{fig:bbfit}
\end{figure}

Fig.\ref{fig:bbfit} shows the infrared (IR) datapoints and the fit by means of a standard modified black body model, whose parameters are the mass of dust (i.e. the normalisation), the dust temperature, and the dust emissivity. The latter is parametrised through the emissivity index, $\beta$, as defined in the following:
\begin{equation}
F_\nu=M_D k_{\nu_0} \left(\frac{\nu}{\nu_0}\right)^\beta \frac{B_\nu(T)}{D^2} \mbox{,}
\end{equation}
where $M_D$ is the dust mass, $k_{\nu_0}$ is the dust emissivity coefficient at a reference frequency $\nu_0$, $D$ is the distance to the galaxy, and $B_\nu$ is the Planck function \citep[see e.g.][for an application of this method to local galaxies]{smith10}. While simplistic, this fitting approach has been widely used in the literature, and has proven to give a fair physical approximation to the dust emission characteristics \citep{bianchi13}. 

As for the dust emissivity coefficient, we adopted the standard one from \cite{draine03}, which has a value of 0.192 m$^2$ kg$^{-1}$ at 350 $\mu$m. The dust mass derived in this way ranges from $\sim 6\times 10^7$ to $\sim10^8$ M$_\odot$. More specifically, if we leave the emissivity index $\beta$ as a free parameter, we find a best fit for a dust temperature of $21.42 \pm 1.80$ K, an emissivity index $\beta=2.17 \pm 0.32$, and a dust mass of $9.8^{+2.2}_{-1.8} \times 10^7$ M$_\odot$. Integrating the black body model SED over the 10 to 1000 $\mu$m range, we get an IR luminosity of $2.59\times 10^{10}$ L$_\odot$. We convert this into a SFR using the \cite{kennicutt98} relation, rescaled to the \cite{chabrier03} IMF by using the conversion factor as in \cite{hayward14}:
\begin{equation}\label{eqn:ir_sfr}
SFR_{IR}=3.0\times10^{-37} L_{IR}\;\;\mbox{M}_\odot\; \mbox{yr}^{-1}
\end{equation}
where $L_{IR}$ is expressed in W  (note that the aforementioned conversion factor is actually calculated for a \citealt{kroupa01} IMF which has anyway a minimal difference with respect to IMF we adopt here, see e.g. \citealt{madau14}). The SFR calculated in this way is 2.98 M$_\odot$ yr$^{-1}$.

Using a lower value for $\beta$ (1.5), the dust mass we obtain is slightly lower, $5.8\times 10^7$ M$_\odot$ (and has a higher temperature compared to the previously found value, namely $T=26.72$ K), but the 160 $\mu$m point is underestimated by more than 20\%, well beyond the flux uncertainty in this band.

Following \cite{eales12}, and using their equation 2, we can derive the total gas mass (i.e. the mass of the gas in all phases) from sub-mm fluxes. Using the 500 $\mu$m flux, and assuming the Galactic gas--to--dust ratio, we get a value of $3.2\times 10^9$ M$_\odot$. Similarly, following a much direct and straightforward approach, we can simply convert the dust mass into a gas mass of about $10^{10}$ M$_\odot$ assuming the same Galactic gas--to--dust ratio of 100. Both values of the gas mass are quite consistent to those expected, within the observed dispersion, in normal, non star--bursting galaxies of similar stellar mass as JO36 \citep[see e.g.][]{magdis12,morokuma15}, and might give an indication that the majority of the gas is still retained by the galaxy.
\begin{figure}[!t]
\centering
\includegraphics[height=0.46\textwidth]{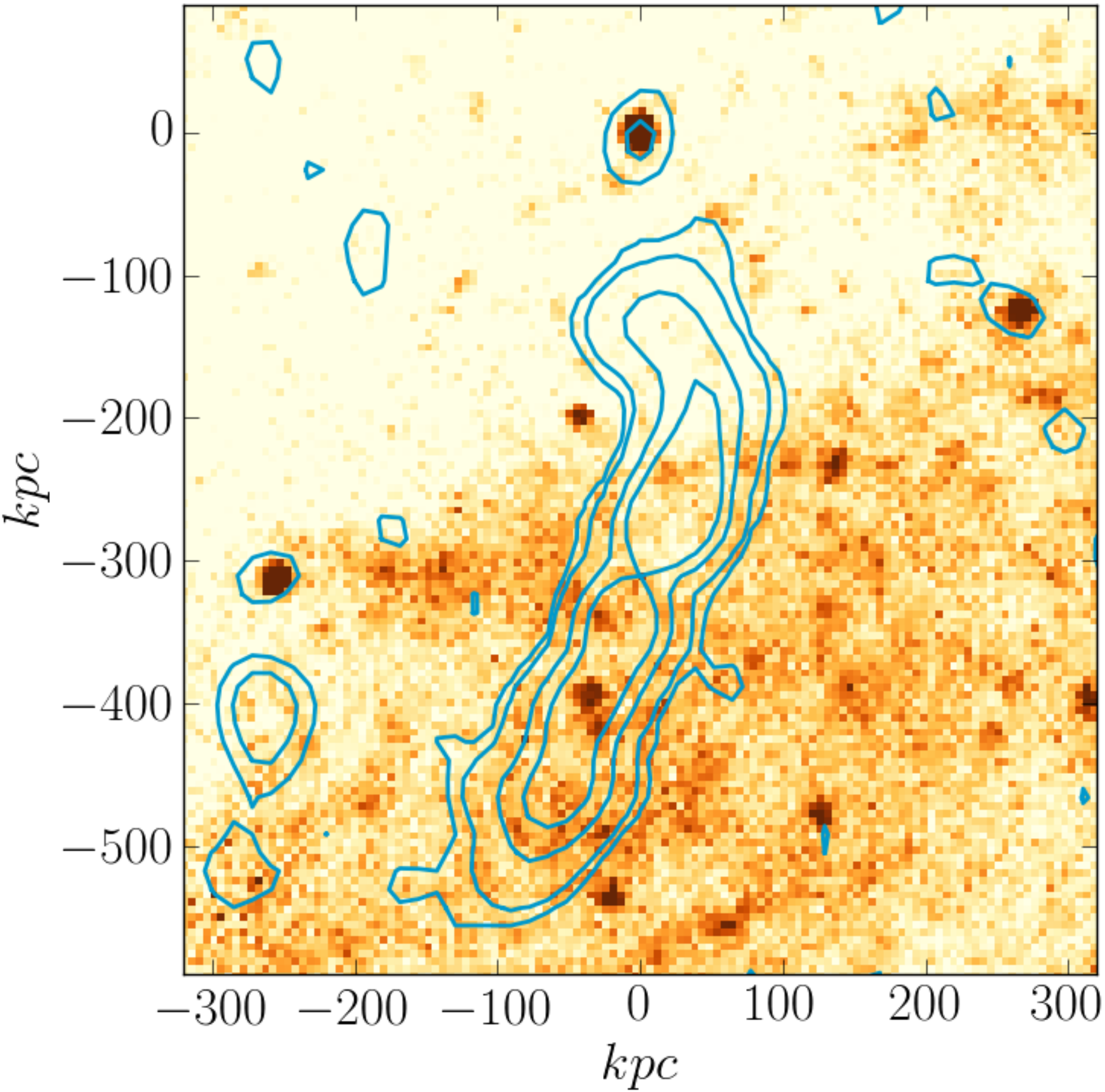}
\caption{SPIRE 250 $\mu$m emission and radio contours \citep[continuum emission at 1.4 GHz from][]{condon98} of the region of the sky surrounding JO36.}
\label{fig:spire250}
\end{figure}

This, of course, heavily relies on the assumption of a given gas--to--dust ratio that, for a galaxy in a cluster environment, might not be strictly true. \cite{cortese10}, studying the spatially--resolved dust emission versus the gas content on a sample of galaxies in the Virgo cluster, found evidences of dust truncated disks in highly {\sc Hi}--deficient galaxies ($def_{H\scriptscriptstyle{I}}>0.87$). The fact that we observe such a high value of the dust mass, can be hence taken as an indication that the amount of atomic gas that has been stripped must yield a deficiency value smaller than 0.87. Using the definition of {\sc Hi} deficiency given by \cite{chung09}, and assuming the aforementioned value for $def_{H\scriptscriptstyle{I}}$, we can calculate a lower limit for the {\sc Hi} mass in JO36. The value we derive in this way is $\sim1.4\times 10^9$ M$_\odot$. Again, this value very well compares to the {\sc Hi} mass expected for galaxies with similar stellar masses \citep[see e.g.][]{popping14,jaskot15}.

Using PACS data at 100 $\mu$m, which are those with the highest spatial resolution at these wavelengths, we have looked for evidences of a possible truncation in the dust disk. Convolving the H$\alpha$ image to the same 6\arcsec \ resolution, and regridding the map obtained in this way to the same pixel size, we found a somewhat good match between its extension and the one observed from 100 $\mu$m emission. Despite this, we cannot claim that there is a truncated dust disk, as {\it Herschel} images for this dataset are made with only one cross--scan, and the data are quite shallow. For this reason, it is more difficult to detect IR emission in the galaxy's outskirts, where dust is not only less abundant (in part due to weaker projection effects as well), but also colder.

If, instead, we look at the total extinction map (i.e. the extinction value calculated over all the stellar ages, whose detection is not relying on the presence of emission lines) as derived by \sinopsis, we note that dust seems to be present throughout all the disk, affecting the starlight to different degrees depending on the position.

To determine the mass of the ionized gas, we have used the relation between H$\alpha$ luminosity and the mass of ionized hydrogen, as described in \cite{poggianti17}. This also depends on the electron density, which we have calculated from the ratio of the sulfur forbidden doublet at 6714 and 6731 \AA. To calculate it we have adopted the prescription given in \cite{proxauf14}:
\begin{eqnarray}\label{eqn:n_e}
n_e = 0.0543\cdot\tan(-3.0553\times R+2.8506)+6.98- \nonumber  \\
 10.6905\times R+9.9186\times R^2- 3.5442\times R^3
\end{eqnarray}
where $R=F_{6714}/F_{6731}$ is the ratio between the fluxes of the two lines \citep{poggianti17}. Eqn.~\ref{eqn:n_e} is valid in the range $0.436\leq R \leq 1.435$. We have used line fluxes measured by {\sc kubeviz} and, when R assumed a value outside the two limits, we have adopted a value equal to the closer limit. In case neither of the two lines were measurable, we took $R=0.966$, which is the average between the upper and lower limit. As for the H$\alpha$ flux, we have used the value measured by {\sc kubeviz} on the absorption--corrected spectra. The effect of dust attenuation was also corrected for, by using the value $A_V$ that \sinopsis \ provides for the young (i.e. lines--emitting) stellar populations. This has the advantage that an extinction value is given also when H$\beta$ is not available because being too faint. No extinction correction was applied in case $A_V$ was not calculated for a given spaxel. The total ionized gas mass computed in this way amounts to $6.9\times 10^8$ M$_\odot$. This mass is $2\sigma$ lower than the average value expected for galaxies of similar stellar mass \citep{popping14}.

JO36 is also detected by the NVSS radio survey \citep{condon98}, and has a (broad band) flux density of $4.3 \pm 0.5$ mJy at 1.4 GHz. The emission in this band is dominated by the radio continuum and it is therefore another tracer of the ionized gas. In Fig.~\ref{fig:spire250} we present infrared data together with the radio (1.4 GHz) contours superimposed. The long radio tail is likely related to the nearby BCG (VV 382 or GIN 049), and there is a clear detection at the position of JO36.

These data can be used to derive another, independent estimate of the ionized gas mass. Using the prescription given in \cite{galvan08}, which assumes that the gas is homogeneously distributed within a sphere, we find a value which is two orders of magnitude higher with respect to the previously calculated one, meaning that assumptions made regarding the geometrical distribution of gas are probably too strong, and this method cannot be applied to a jellyfish like JO36 to derive the ionized gas mass. Emission from supernovae at these frequencies might also bias the result.

A further estimate of the SFR can be given using this data, with the advantage that this tracer is insensitive to dust extinction. With a luminosity $L_{1.4}=1.67\times 10^{22}$ W$\cdot$Hz$^{-1}$, and using the prescription from \cite{hopkins03} (see their Equations 1 and 2), we find a SFR of 9.2 M$_\odot$ yr$^{-1}$.

To summarise, we have derived total gas masses from IR and sub--mm data in the range between $3.2\times 10^9$ and $10^{10}$ M$_\odot$. Both estimates rely on the assumption that a Galactic gas--to--dust ratio can be used for this galaxy. A lower limit of $1.4\times 10^9$ M$_\odot$ to the {\sc Hi} mass was extrapolated from the substantial presence of dust that we used as an indicator of the maximum degree of {\sc Hi} deficiency. All of these values agree with the gas mass expectations in galaxies of similar stellar mass. The exception to this is the ionized gas mass, which is lower by more than $2\sigma$ when compared to the average relation for similar galaxies.

The SFR calculated from the spectral fitting is $\sim 5.9$ M$_\odot$ $yr^{-1}$, and naturally takes into account and hence corrects for the effect of dust attenuation. This SFR value depends on the intensity of the H$\alpha$ line and, even when corrected for attenuation, might miss a completely embedded star formation component \citep[e.g.][]{leroy08}. \cite{saftly15}, for example, have demonstrated that small scale inhomogeneities and structures in the ISM distribution (which could host severely obscured star formation), can have a negligible effect on the optical extinction, but their presence is revealed from their mid-- and far--infrared emission. 

On the other hand, converting the IR emission into a SFR yielded $\sim3$ M$_\odot$ yr$^{-1}$. To derive a value of the integrated SFR that includes both components (i.e., the extinction--corrected plus the completely obscured), is not straightforward: the timescales of star formation that they sample are quite different, with H$\alpha$ being a tracer sensitive to the ``instantaneous'' star formation (i.e. stars younger than $\sim 10^7$ yr), and the IR tracing star forming activity within $10^8$ yr. 

To be able to properly take this two components into account, we have calculated the UV flux expected from the \sinopsis \ model (no GALEX data are available for this galaxy), and exploit it to derive the unobscured SFR component. Using the prescription given in \cite{kennicutt12}, we calculate a non--obscured SFR value of 3.4 M$_\odot$ yr$^{-1}$ which, as UV bands typically sample timescales very close to those of the IR, we can add to the value calculated from the dust emission. Doing so, we get a SFR of 6.4  M$_\odot$ yr$^{-1}$, over a 100 Myr timescale.

An extinction independent value for the SFR is given by the radio continuum, from which we have calculated a value of 9.2 M$_\odot$ yr$^{-1}$, significantly higher with respect to the aforementioned estimates. The discrepancy with respect to the previously calculated values, might come from the presence of an AGN (see Sect.~\ref{sec:xray}), even though of relatively low luminosity, which could indeed boost the radio emission.

\section{Discussion}\label{sec:discussion}
\begin{figure*}
\centering
\hspace{-1cm}
\includegraphics[height=0.50\textwidth]{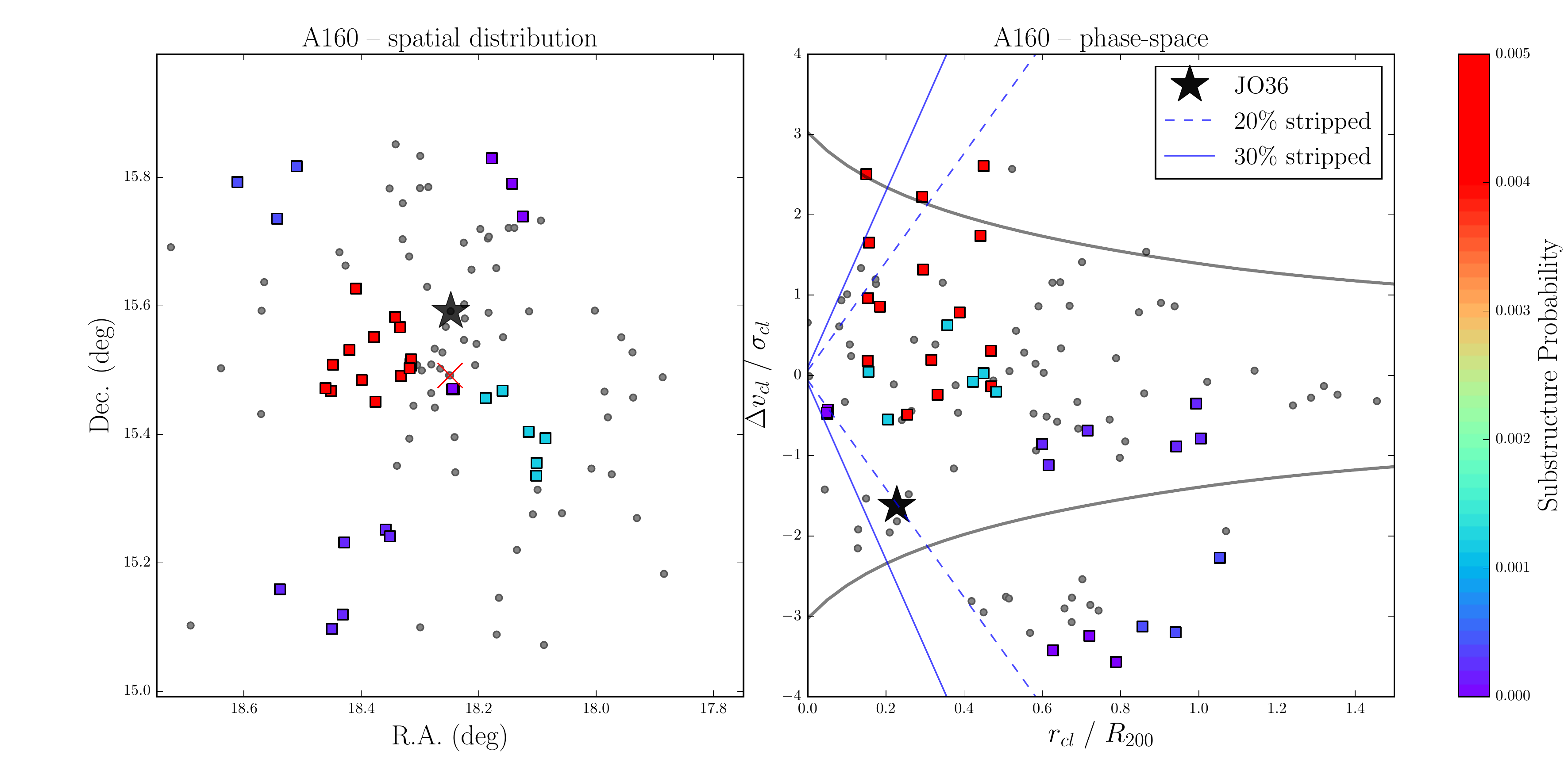} 
\caption{Left: The position in the sky of JO36 spectroscopic members from OmegaWINGS (small gray points), JO36 (star), and the BCG (red cross). Squares correspond to identified substructures, which have been color-coded according to their probability to be random fluctuations \citep[i.e. values close to zero indicate highly significant substructure detections][]{biviano17}. Right: Phase-space diagram with symbols as in the left panel. Curves show the escape velocity in a \cite{navarro97} halo. The dashed and solid blue lines correspond to 20 and 30\% of total gas mas stripped in JO36 by the ICM in a Virgo-like cluster (see text for details).}
\label{fig:substructures}
\end{figure*}
The most important results we have obtained so far can be summarized as follows.
\begin{itemize}
\item the stellar disk extends out to a radius of about 25 kpc, while the ionized gas only reaches galactrocentric distances of about 15 kpc. We interpret this as a clear evidence of a truncated ionized gas disk;
\item a stellar tail, extending by $\sim 5$ kpc with respect to the main body of the disk, is observed towards the south;
\item four H$\alpha$ blobs are present southwards of the galaxy, close to the aforementioned tail;
\item the (ionized) gas velocity field is noticeably distorted, especially when compared to the stellar one;
\item the dust mass is compatible with that expected in ``normal'' field galaxies having similar stellar masses. This strongly suggests that dust has not been stripped. If dust is used as a tracer for the presence of gas, we infer a total gas mass in the range expected for the physical characteristics of this galaxy. The fact that no evidence is found of a significant dust stripping constraints the {\sc Hi} deficiency level of the galaxy, and was used to estimate a lower limit of the {\sc Hi} mass;
\item the mass of ionized gas is at the lower limit with respect to the value expected for galaxies of similar stellar mass;
\item star formation is currently happening only in the central region of the galaxy, within a 10 kpc radius, while the external ($r> 10$ kpc) parts of the disk are dominated by stars with ages $< 500$ Myr;
\item the star formation history of the galaxy shows evidences of an inside--out formation process. An enhancement in the SFR happened between 20 and 500 Myr ago, more deeply affecting the outer disk with respect to the central regions.
\end{itemize}

In the next Section (\ref{sec:rps}), we present various pieces of evidences of active ram pressure in this galaxy, while in the followings we try to build a self--consistent picture that can interpret the aforementioned observed features simultaneously. 

\subsection{The strength of ram-pressure in JO36}\label{sec:rps}
Given JO36's vicinity to the core of A160, and its high velocity within the cluster \citep[see the phase--space diagram, as in][shown in Fig.~\ref{fig:substructures}]{jaffe15}, it is likely that ram-pressure stripping (RPS) is or has been at play. The ram-pressure by the ICM can be estimated as $P_{ram} = \rho_{ICM} \times v_{cl}^{2}$ \citep{gunn72}, where $\rho_{ICM}(r_{cl})$ is the radial density profile of the ICM, $r_{cl}$ the clustercentric distance, and $v_{cl}$ the velocity of the galaxy with respect to the cluster.  Since A160  is a low mass cluster (velocity dispersion $=561$ km s$^{-1}$), we assume a smooth static ICM similar to that of the  Virgo cluster. Utilizing the density model used by \cite{vollmer01}, we can get an estimate of the ram--pressure at the projected $r_{cl}$ and line-of-sight velocity of JO36: 
\begin{equation} 
P_{ram} = 9.5 \times10^{-14} N m^{-2}
\end{equation}

To assess whether this is enough to strip gas from JO36, we compute the anchoring force of the galaxy assuming an exponential disk density profile for the stars and the gas components ($\Sigma_{s}$ and $\Sigma_{g}$ respectively) defined as:
\begin{equation}
\Sigma=  \left(\frac{M_{d}}{2 \pi r_d^2}\right)  e^{-r/r_d},
\label{sigma0}
\end{equation}
where $M_{d}$ is the disk mass, $r_d$ the disk scale-length and $r$ the radial distance from the center of the galaxy. For the stellar component of JO36 we adopted a disk mass $M_{d,stars}=5.2\times10^{10}M_{\odot}$ (accounting for a bulge to total ratio of 0.2), and a disk scale-length $r_{d,stars} = 4.63$~kpc, obtained by fitting the light profile of the galaxy.  For the gas component we assumed a total mass $M_{d,gas} = 0.1 \times M_{d,stars}$, and scale-length $r_{d,gas} = 1.7 \times r_{d,stars}$ \citep{boselli06}.

The anchoring force in the disk can then be computed as: $\Pi_{gal} = 2 \pi G \Sigma_{g} \Sigma_{s}$, at different radial distances from the centre of the galaxy ($r$). We find that the condition for stripping is met at $r \sim 13.4$~kpc, where $\Pi_{gal}$ drops below $P_{ram}$. This truncation radius corresponds to $\sim 21\%$ of the total gas mass stripped, (see reference dashed line in the right panel of Fig.~\ref{fig:substructures}). 

The estimated fraction of stripped gas is consistent with the lower limit of {\sc Hi} mass derived in Section 5 (from the dust content), that when compared to the gas mass in our disk model, yields an upper limit for the fraction of stripped gas of $\sim 27\%$. It is also interesting to compare the expected stripping from our modeling with the observed truncation radius. Taking the extent of H$\alpha$ emission as a good estimate, we get an observed truncation radius of $r_t=11$~kpc, which corresponds to more stripping than predicted ($\sim 27\%$ of the total gas mass; solid blue line in Fig.~\ref{fig:substructures}). We note however that the predicted stripping suffers from uncertainties in the galaxy and cluster model, projection effects, and that it does not take into account possible inhomogeneities of the ICM. 

To test for the presence of substructures within the cluster, we selected galaxies with significant deviations from the cluster velocity dispersion (coloured symbols) and found that JO36 does not belong to any clear group. A dynamical analysis of A160 \citep{biviano17} reveals several substructures,  shown with coloured squares in Fig.~\ref{fig:substructures}. However, there is no evidence for JO36 to reside in any of these substructures. On the contrary, its phase-space position suggests that this galaxy has fallen recently into the cluster as an isolated galaxy.

Overall, our analysis shows that JO36 must have lost between $\sim20$ and $30\%$ of its total gas mass via ram-pressure stripping by A160's ICM.

We now propose two mutually exclusive scenarios, each of which is successful in explaining some of the observed features listed above while, at the same time, failing to account for others. The difference in the two scenarios simply lies in the direction of the tangential velocity of the galaxy.

\subsection{Tangential velocity towards the north (1)}\label{sec:scenario1}
JO36 was selected as a possible jellyfish candidate because of the presence of a tail, pointing towards the location of the BCG, visible in WINGS V and B band images. This, together with the detection of few relatively bright H$\alpha$ spots located close to this tail, are features that we recover in MUSE data as well (see the left--hand panel of Fig.~\ref{fig:kinematic} for the tail, and the right-hand panel in the same figure for the blobs). 

These features can be explained in a scenario where the galaxy has a velocity component in the opposite direction to the location of both the tail and the blobs, moving away from the cluster center (the direction towards the X-ray center and the location of the BCG are indicated by the two arrows in Fig.~\ref{fig:kinematic}). In this picture, the denser gas in the central regions of the cluster exerted a RPS force capable of ripping part of the gas away from the galaxy, which would be now found in the form of the observed H$\alpha$ emitting blobs. Something very similar, although to a much more spectacular degree, is observed in other jellyfish galaxies \citep[e.g.][]{merluzzi13,fumagalli14,merluzzi16,poggianti17,bellhouse17}, where bright tails and star forming blobs are found in locations opposite to the direction of the galaxy motion. Just like the aforementioned cases, the blobs we observe here retain the disk velocity. 

\subsection{Tangential velocity towards the south (2)}\label{sec:scenario2}
While scenario (1) is the most likely explanation for the star--forming blobs, there is a number of other observed features that it cannot account for.

Analyzing Fig.~\ref{fig:kinematic}, we have already pointed out how the locus where the gas has a zero radial velocity component is twisted in an irregular ``U'' shape, with a concavity directed towards the north, and it reaches galactocentric radii of about 8 kpc towards the same direction. Similarly, gas with positive radial velocities is found on the same side (the region labelled as ``F'' in Fig.~\ref{fig:kinematic}). This twisting of the gas rotational axis is a feature that is predicted as a consequence of RPS by the aforementioned simulations of \cite{merluzzi16}, where the direction of the bending is directly related to the velocity of the galaxy on the plane of the sky. This, in the case of JO36, would be pointing to the cluster center.

This scenario would also explain why the star forming region, clearly visible in the right--hand panel of Fig.~\ref{fig:bpt}, is slightly bent in a ``C'' shape pointing towards the south--east and offset, with respect to the stellar continuum, in that direction. If shocks with the intracluster gas are responsible for the enhanced star formation, it is hence logical to expect that this would happen first in the direction of the interaction between the two gas components which, in this scenario, would be on this side of the disk.

\subsection{More than just one mechanism at play?}
JO36 shows clear signature of past and ongoing RPS. This is further confirmed by the phase--space diagram (Fig.~\ref{fig:substructures}) that shows that the galaxy is well within the region where ram pressure is strong enough to eventually strip all the gas.

Hence, while it is quite clear that we are observing RPS signatures, scenario (2) cannot naturally explain the presence of the four gas blobs for which we would need to appeal to other phenomena. On the other hand, scenario (1) seems to be partially in contradiction with the gas velocity map. Such a distortion naturally arises due to RPS if the galaxy would be moving in the opposite direction, i.e. towards the cluster center.

Furthermore, the stellar tail, visible in Fig.~\ref{fig:kinematic}, does not fit either of the two proposed scenarios, and needs other physical mechanisms to be invoked. While being a morphological feature clearly departing from the disk, it does not seem to have any counterpart with similar characteristics in the north side of the galaxy. Moreover, the (stellar) velocities follow the trend observed in the disk itself, as expected if this was a natural prosecution of the disk, and have the highest values found in the galaxy.

If this tail was the result of gas stripping from the outer disk by ram pressure, one would expect it retains a similar velocity with respect to the region within the disk where it came from. The measured velocities are higher by up to 50 km/s with respect to the rest of the disk (see also the rotation curve in Fig.~\ref{fig:vprof}). 

In addition, the average age of the stellar populations of this tail is compatible with a formation epoch up to 500 Myr ago, hence significantly older with respect to the ages derived for the blobs that are rich in ionized gas and still actively star forming. If this was to happen in a stripped gas component, we should not be observing it still being attached to the galaxy, as the gas would have had the time to move away and detach from the galaxy's disk. On the other side, stars are not affected by RPS.

For these reasons, we can conclude that RPS cannot be the mechanism by which this stellar tail originated, and we would need to invoke a different mechanism to explain its formation.

According to numerical simulations performed by \cite{kronberger08a}, aimed at studying the effect of ram pressure on the star formation of spiral galaxies, an observed  enhanced star formation rate in stellar populations with ages in the 20--500 Myr range, is a direct effect of the interaction between the gas in the galaxy and that in the intracluster medium. This would somehow date the beginning of the interaction between the galaxy and the hot gas in the cluster. 

Following the same authors, when the interaction is ``edge--on'', such as in our case, the gas loss is much lower compared to a face--on interaction, the main signature of ram pressure being a distortion and compression in the gas disk, which is indeed what we observe. Enhanced SFRs by up to a factor of 3 are observed in these simulations, compatible with the values we have derived by spectral fitting \citep[see also][]{koopmann04}.

Numerical simulations from the same group \citep{kronberger08b}, focussing on the effects on the rotation curves and velocity fields of the gas, show a stronger distortion of the gas distribution in edge--on interactions, as compared to the face--on case. They also observe a displacement on the rotation axis of stars and gas, something that we instead do not find.

Integrating the observed datacube with respect to the wavelength coordinate, we get a high S/N picture which better allows to view the morphology of the lowest surface brightness components of the galaxy (Fig~\ref{fig:collapsed}). By doing this, we can confirm the absence of a tidal feature in the northern disk, while 2 of the H$\alpha$ blobs appear to be almost embedded within the disk, making it unclear whether they effectively are jellyfish morphological features or nothing more but regions of residual star formation from a quenched disk. Nevertheless, the brightest and largest blobs (A and D) are, even in projection, too far away to fill in this picture.

Different methods to infer the gas mass yielded values in fairly good agreement with respect to each other, and these point to a regular gas--to--stellar mass content.

In any case, given the relatively low mass of the A160 cluster (L$_X=10^{43.6}$ erg/s, \citealt{ebeling96}, and $\sigma_{gal}=561$ km/s, \citealt{moretti17a}), and given the dependency of the RPS effect on the cluster gas density, we do not expect, at least as long as short time scales are concerned, massive gas losses, most of all given the geometry of the galaxy motion \citep[numerical simulations by][have shown that edge--on systems are much less prone to gas loss]{kronberger08a} and the mass of the galaxy.

One possible explanation for the extended stellar disk (i.e. the tail), is that it could be the result of a localized interaction. \cite{kronberger06} performed numerical simulations to study how galaxy encounters influence the kinematics of stellar disks. For given sets of simulations parameters, they find that a fly--by can affect the stellar rotation in the disk outskirts in different ways depending on the configuration of the encounter and on the line-of-sight of the observation. Some of the rotation curves they extract from their simulations resemble the asymmetry we observe in JO36 stellar kinematic, and in particular in the tail. Similarly, \cite{pedrosa08} claim that bifurcations, i.e. asymmetries in the outer parts of a rotation curve as we observe in JO36, are a clear indicator of a recent galaxy encounter. 

It would be tempting to identify in one of the blobs (e.g. blob A, the most massive one) the possible candidate for this kind of interaction. In this case, we would be witnessing the later phases of an encounter between JO36 and a dwarf galaxy. Nevertheless, the metallicity\footnote{Metallicities are calculated throughout the whole galaxy by means of the {\it pyqz} code by \cite{dopita13}. Further details on this issue can be found in \cite{poggianti17}, but see also \cite{kewley08} for absolute uncertainties assessment.}values we derive for blob A are way too high to be compatible with those of a dwarf system, being instead fully consistent, within the typical uncertainties, with the metallicity of the outer gas remained for now in the disk.

\section{Summary and conclusions}\label{sec:summary}
In this work, we have undertaken an analysis of the properties of the stellar populations and of the interstellar medium in JO36, a galaxy in the Abel 160 cluster, with slightly distorted optical morphology, possibly a signature of gas stripping. We have used these observations to validate our spectral fitting code, \sinopsis, for applications to IFU data analysis by comparing its results with those obtained from {\sc gandalf}, a well known and widely used code generally exploited to derive the properties of the emission lines and of the underlying stellar populations. This comparison indicates that our approach gives robust results fully compatible with those obtained with {\sc gandalf} on the same dataset. 

From the results of the kinematic analysis and of the stellar populations properties in this galaxy, we draw the following conclusions:
\begin{enumerate}
\item JO36 shows no spectacular morphological signatures of gas stripping such as those commonly encountered in the so--called jellyfish galaxies, but the ionized gas disk is clearly truncated with respect to the stellar one;
\item if any gas stripping has occurred in the past, it most likely involved a minor fraction of the total gas in the galaxy. A substantial gas depletion due to an intense star forming episode happened about 500 Myr ago, could have been concurred to the creation of the truncated ionized gas disk;
\item from a kinematical point of view, the rotation curve of the gas displays asymmetries in the outer parts of the disk, with a rotation axis strongly distorted and suggestive of a velocity component towards the center of the cluster. This is in agreement with numerical simulations of RPS acting with a relative velocity parallel to the galaxy plane (edge--on);
\item the presence of H$\alpha$ blobs close to the southern edge of the galaxy, might suggest a tangential velocity component in the north direction, something that seems to be incompatible with the morphological characteristics of the gas rotational axis;
\item the presence of a stellar tail in the southern disk, with no clear counterpart in the opposite direction, cannot be attributed to ram pressure effects. Its velocities follow the stellar rotation curve from the inner parts, and are higher than those measured across the whole galaxy disk. Composed by stellar populations of ages between $2\times 10^7$ and $5\times 10^8$ yr, and showing no evidences for the presence of gas, it can be the result of a gravitational interaction with a less massive galaxy, as suggested by numerical simulations;
\item there is no evidence of AGN activity, at least as far as diagnostic lines are concerned. However, the detection of a strong emission in the X-rays, strongly suggests the possible presence of a deeply obscured AGN (Nicastro et al. in prep.).
\end{enumerate}

JO36 is a moderately massive spiral which is subject to RPS as several pieces of evidence suggest. The truncated ionized gas disk, the low ratio of {\sc Hii}/M$_\ast$ with respect to similar galaxies, the disturbed gas kinematics, the presence of ionized gas regions clearly detached from disk, its location on the phase--space diagram of the cluster, and finally an episode of enhanced star formation strongly involving the outer disk, all point to ram pressure being caught on the act. 

We also speculate that the stripped gas is probably a minor fraction of the gas in the galaxy. By indirect calculations of the amount of total gas and of {\sc Hi}, we find that the gas content is quite typical, given the stellar mass of the galaxy. Furthermore, the moderately intense star formation likely induced by shocks with the intracluster gas, has consumed a substantial amount of gas. Indeed, in the analysis of their numerical simulations, \cite{kronberger08a} propose that losses of gas by RPS together with depletion due to star formation, is the reason for the decrease, and eventual quenching, of the star formation rate.

What is less clear is instead the direction of the ram pressure or, equivalently, of the galaxy motion within the cluster. In fact, we could not reconcile in a self consistent picture the presence of ionized gas in the southern part of the galaxy, indicating a velocity component towards the north, with the distsrted shape of the gas rotational axis, suggesting instead a velocity component towards the south. Dedicated numerical simulations are probably the best tool to figure out the kinematic of the galaxy and give hints on its orbit, to better understand the relation between its star formation history and the interaction with the cluster environment.

With respect to the first point in our final remarks, it should be noted that MUSE data for this galaxy basically cover all of its disk but we cannot draw any conclusion on the possible presence of stripped tails at larger distances, which passed unobserved in optical images. Furthermore, we lack {\sc Hi} data to derive the atomic mass distribution, and give a final word about the dynamical history of the galaxy.

Both \cite{poggianti16} and \cite{mcpartland16} stress the importance of spectroscopic data to unveil the occurrence of gas stripping signatures, as opposed to pure photometric detections. In this particular case, MUSE data turned out to be critical to uncover a second dynamical mechanism affecting this galaxy, most likely a gravitational interaction with a much less massive galaxy.

\acknowledgments
We would like to help the anonymous referee whose suggestions and criticism helped us improving the quality and presentation of the results of the paper.
Based on observations collected at the European Organisation for Astronomical Research in the Southern Hemisphere under ESO programme 196.B-0578.
JF warmly thanks Anna Feltre for all the advices in running {\sc cloudy}, Theodoros Bitsakis for stimulating discussions, Raul Naranjo and Daniel D\'\i az Gonzalez who helped with some of the technicalities in \sinopsis.

JF acknowledges financial support from the UNAM-DGAPA-PAPIIT IA104015 grant, M\'exico.
GB acknowledges support for this work from UNAM through grant PAPIIT IG100115. 
This work was co-funded under the Marie Curie Actions of the European Commission (FP7-COFUND)
B.V. acknowledges the support from an  Australian Research Council Discovery Early Career Researcher Award (PD0028506).
B.C.S. acknowledges financial support through PAPIIT project IA103517 from DGAPA-UNAM.

\software{\sinopsis, gandalf \citep{sarzi06}, cloudy \citep{ferland93,ferland98,ferland13}, kubeviz \citep{fossati16}, pPXF \citep{cappellari04,cappellari12a}, pyqz \citep{dopita13}, scanamorphos \citep{roussel13}, HIPE, IDL, Python.}


\vspace{5mm}

\begin{thebibliography}{}
\bibitem[Abramson et al.(2016)]{abramson16} Abramson A., Kenney J., Crowl H., Tal T., 2016, AJ, 152, 32
\bibitem[Athanassoula(2002)]{athanassoula02} Athanassoula, E.\ 2002,\apjl, 569, L83 
\bibitem[Bacon et al.(2001)]{bacon01} Bacon R., et al., 2001, MNRAS, 326, 23 
\bibitem[Baes et al.(2010)]{baes10} Baes M., et al., 2010, A\&A, 518, L39
\bibitem[Balogh, Navarro, \& Morris(2000)]{balogh00} Balogh M.~L., Navarro J.~F., Morris S.~L., 2000, ApJ, 540, 113
\bibitem[Bellhouse et al.(2017)]{bellhouse17} Bellhouse C., et al., 2017, arXiv, arXiv:1704.05087 
\bibitem[Berta et al.(2003)]{berta03} Berta S., Fritz J., Franceschini A., Bressan A., Pernechele C., 2003, A\&A, 403, 119 
\bibitem[Bianchi(2013)]{bianchi13} Bianchi S., 2013, A\&A, 552, A89
\bibitem[Bischko, Steinhauser, \& Schindler(2015)]{bischko15} Bischko J.~C., Steinhauser D., Schindler S., 2015, A\&A, 576, A76 
\bibitem[Biviano et al.(2017)]{biviano17} Biviano A., et al., 2017, arXiv, arXiv:1708.07349
\bibitem[Boselli \& Gavazzi(2006)]{boselli06} Boselli, A., \& Gavazzi, G.\ 2006, \pasp, 118, 517 
\bibitem[Boselli et al.(2016)]{boselli16} Boselli A., et al., 2016, A\&A, 596, A11 
\bibitem[Bressan et al.(2012)]{bressan12} Bressan A., Marigo P., Girardi L., Salasnich B., Dal Cero C., Rubele S., Nanni A., 2012, MNRAS, 427, 127
\bibitem[Burrows(2000)]{burrows00} Burrows, A.\ 2000, \nat, 403, 727
\bibitem[Byler et al.(2017)]{byler17} Byler N., Dalcanton J.~J., Conroy C., Johnson B.~D., 2017, ApJ, 840, 44 
\bibitem[Byrd \& Valtonen(1990)]{byrd90} Byrd G., Valtonen M., 1990, ApJ, 350, 89 
\bibitem[Calzetti, Kinney, \& Storchi-Bergmann(1994)]{calzetti94} Calzetti D., Kinney A.~L., Storchi-Bergmann T., 1994, ApJ, 429, 582
\bibitem[Calzetti et al.(2000)]{calzetti00} Calzetti D., Armus L., Bohlin R.~C., Kinney A.~L., Koornneef J., Storchi-Bergmann T., 2000, ApJ, 533, 682
\bibitem[Cappellari \& Emsellem(2004)]{cappellari04} Cappellari M., Emsellem E., 2004, PASP, 116, 138
\bibitem[Cappellari(2012)]{cappellari12a} Cappellari M., 2012, ascl.soft, ascl:1210.002 
\bibitem[Cappellari \& Copin(2012)]{cappellari12b} Cappellari M., Copin Y., 2012, ascl.soft, ascl:1211.006 
\bibitem[Cardelli, Clayton, \& Mathis(1989)]{cardelli89} Cardelli J.~A., Clayton G.~C., Mathis J.~S., 1989, ApJ, 345, 245
\bibitem[Cava et al.(2009)]{cava09} Cava A., et al., 2009, A\&A, 495, 707 
\bibitem[Cid Fernandes et al.(2005)]{cidfernandes05} Cid Fernandes R., Mateus A., Sodr{\'e} L., Stasi{\'n}ska G., Gomes J.~M., 2005, MNRAS, 358, 363
\bibitem[Cid Fernandes(2007)]{cidfernandes07} Cid Fernandes R., 2007, IAUS, 241, 461 
\bibitem[Ciesla et al.(2012)]{ciesla12} Ciesla L., et al., 2012, A\&A, 543, A161
\bibitem[Chabrier(2003)]{chabrier03} Chabrier G., 2003, PASP, 115, 763 
\bibitem[Charlot \& Longhetti(2001)]{charlot01} Charlot S., Longhetti M., 2001, MNRAS, 323, 887
\bibitem[Cheung et al.(2016)]{cheung16} Cheung E., et al., 2016, Nature, 533, 504
\bibitem[Chevallard \& Charlot(2016)]{chevellard16} Chevallard J., Charlot S., 2016, MNRAS, 462, 1415
\bibitem[Chung et al.(2009)]{chung09} Chung A., van Gorkom J.~H., Kenney J.~D.~P., Crowl H., Vollmer B., 2009, AJ, 138, 1741 
\bibitem[Condon et al.(1998)]{condon98} Condon J.~J., Cotton W.~D., Greisen E.~W., Yin Q.~F., Perley R.~A., Taylor G.~B., Broderick J.~J., 1998, AJ, 115, 1693 
\bibitem[Cortese et al.(2007)]{cortese07}Cortese L., et al., 2007, MNRAS, 376, 157
\bibitem[Cortese et al.(2010)]{cortese10} Cortese, L., Davies, J.~I., Pohlen, M., et al.\ 2010, \aap, 518, L49
\bibitem[Cortese et al.(2014)]{cortese14} Cortese L., et al., 2014, MNRAS, 440, 942
\bibitem[Cowie \& Songaila(1977)]{cowie77} Cowie L.~L., Songaila A., 1977, Natur, 266, 501 
\bibitem[Croton et al.(2006)]{croton06} Croton, D.~J., Springel,V., White, S.~D.~M., et al.\ 2006, \mnras, 365, 11
\bibitem[Crowl \& Kenney(2008)]{crowl08} Crowl H.~H., Kenney J.~D.~P., 2008, AJ, 136, 1623
\bibitem[Debattista \& Sellwood(2000)]{debattista00} Debattista, V.~P.,\& Sellwood, J.~A.\ 2000, \apj, 543, 704
\bibitem[De Geyter et al.(2014)]{degeyter14} De Geyter G., Baes M., Camps P., Fritz J., De Looze I., Hughes T.~M., Viaene S., Gentile G., 2014, MNRAS, 441, 869
\bibitem[De Looze et al.(2014)]{delooze14} De Looze I., et al., 2014, A\&A, 571, A69
\bibitem[de Zeeuw et al.(2002)]{dezeeuw02} de Zeeuw P.~T., et al., 2002, MNRAS, 329, 513 
\bibitem[Dopita et al.(2013)]{dopita13} Dopita M.~A., Sutherland R.~S., Nicholls D.~C., Kewley L.~J., Vogt F.~P.~A., 2013, ApJS, 208, 10
\bibitem[Draine(2003)]{draine03} Draine B.~T., 2003, ARA\&A, 41, 241 
\bibitem[Dressler et al.(2009)]{dressler09} Dressler A., Rigby J., Oemler A., Jr., Fritz J., Poggianti B.~M., Rieke G., Bai L., 2009, ApJ, 693, 140-151
\bibitem[Eales et al.(2012)]{eales12} Eales S., et al., 2012, ApJ, 761, 168
\bibitem[Ebeling et al.(1996)]{ebeling96} Ebeling H., Voges W., Bohringer H., Edge A.~C., Huchra J.~P., Briel U.~G., 1996, MNRAS, 281, 799
\bibitem[Ebeling, Stephenson, \& Edge(2014)]{ebeling14} Ebeling H., Stephenson L.~N., Edge A.~C., 2014, ApJ, 781, L40
\bibitem[Fabian et al.(2003)]{fabian03} Fabian, A.~C., Sanders,J.~S., Allen, S.~W., et al.\ 2003, \mnras, 344, L43
\bibitem[Falc{\'o}n-Barroso et al.(2011)]{falcon11} Falc{\'o}n-Barroso J., S{\'a}nchez-Bl{\'a}zquez P., Vazdekis A., Ricciardelli E., Cardiel N., Cenarro A.~J., Gorgas J., Peletier R.~F., 2011, A\&A, 532, A95
\bibitem[Faltenbacher \& Diemand(2006)]{faltenbacher06} Faltenbacher A., Diemand J., 2006, MNRAS, 369, 1698 
\bibitem[Fasano et al.(2006)]{fasano06} Fasano G., et al., 2006, A\&A, 445, 805
\bibitem[Ferland(1993)]{ferland93} Ferland G.~J., 1993, hbic.book
\bibitem[Ferland et al.(1998)]{ferland98} Ferland G.~J., Korista K.~T., Verner D.~A., Ferguson J.~W., Kingdon J.~B., Verner E.~M., 1998, PASP, 110, 761
\bibitem[Ferland et al.(2013)]{ferland13} Ferland G.~J., Kisielius R., Keenan F.~P., van Hoof P.~A.~M., Jonauskas V., Lykins M.~L., Porter R.~L., Williams R.~J.~R., 2013, ApJ, 767, 123 
\bibitem[Fielding et al.(2017)]{fielding17} Fielding, D., Quataert,E., McCourt, M., \& Thompson, T.~A.\ 2017, \mnras, 466, 3810
\bibitem[Fitzpatrick(1986)]{fitzpatrick86} Fitzpatrick E.~L., 1986, AJ, 92, 1068 
\bibitem[Fossati et al.(2016)]{fossati16} Fossati M., Fumagalli M., Boselli A., Gavazzi G., Sun M., Wilman D.~J., 2016, MNRAS, 455, 2028 
\bibitem[France et al.(2010)]{france10} France, K., McCray, R.,Heng, K., et al.\ 2010, Science, 329, 1624 
\bibitem[Franzetti et al.(2008)]{franzetti08} Franzetti P., Scodeggio M., Garilli B., Fumana M., Paioro L., 2008, ASPC, 394, 642 
\bibitem[Fritz et al.(2007)]{fritz07} Fritz J., et al., 2007, A\&A, 470, 137 
\bibitem[Fritz et al.(2011)]{fritz11} Fritz J., et al., 2011, A\&A, 526, A45
\bibitem[Fritz et al.(2014)]{fritz14} Fritz J., et al., 2014, A\&A, 566, A32 
\bibitem[Fumagalli et al.(2014)]{fumagalli14} Fumagalli M., Fossati M., Hau G.~K.~T., Gavazzi G., Bower R., Sun M., Boselli A., 2014, MNRAS, 445, 4335 
\bibitem[Galv{\'a}n-Madrid et al.(2008)]{galvan08} Galv{\'a}n-Madrid R., Rodr{\'{\i}}guez L.~F., Ho P.~T.~P., Keto E., 2008, ApJ, 674, L33
\bibitem[Griffin et al.(2010)]{griffin10} Griffin M.~J., et al., 2010, A\&A, 518, L3
\bibitem[Guglielmo et al.(2015)]{guglielmo15} Guglielmo V., Poggianti B.~M., Moretti A., Fritz J., Calvi R., Vulcani B., Fasano G., Paccagnella A., 2015, MNRAS, 450, 2749
\bibitem[Gullieuszik et al.(2015)]{gullieuszik15} Gullieuszik M., et al., 2015, A\&A, 581, A41 
\bibitem[Gunn \& Gott(1972)]{gunn72} Gunn J.~E., Gott J.~R., III, 1972, ApJ, 176, 1
\bibitem[Gutkin, Charlot, \& Bruzual(2016)]{gutkin16} Gutkin J., Charlot S., Bruzual G., 2016, MNRAS, 462, 1757 
\bibitem[Hayashi et al.(2013)]{hayashi13} Hayashi M., Sobral D., Best P.~N., Smail I., Kodama T., 2013, MNRAS, 430, 1042
\bibitem[Haynes \& Giovanelli(1984)]{haynes84} Haynes M.~P., Giovanelli R., 1984, AJ, 89, 758 
\bibitem[Hayward et al.(2014)]{hayward14} Hayward C.~C., et al., 2014, MNRAS, 445, 1598
\bibitem[Hohl(1971)]{hohl71} Hohl, F.\ 1971, \apj, 168, 343 
\bibitem[Hopkins et al.(2003)]{hopkins03} Hopkins A.~M., et al., 2003, ApJ, 599, 971
\bibitem[Jaff{\'e} et al.(2015)]{jaffe15} Jaff{\'e} Y.~L., Smith R., Candlish G.~N., Poggianti B.~M., Sheen Y.-K., Verheijen M.~A.~W., 2015, MNRAS, 448, 1715 
\bibitem[Jaskot et al.(2015)]{jaskot15} Jaskot A.~E., Oey M.~S., Salzer J.~J., Van Sistine A., Bell E.~F., Haynes M.~P., 2015, ApJ, 808, 66
\bibitem[Juvela(2007)]{juvela07} Juvela M., 2007, hers.prop, 102 
\bibitem[Kauffmann et al.(2003)]{kauffmann03} Kauffmann G., et al., 2003, MNRAS, 346, 1055
\bibitem[Kennicutt(1998)]{kennicutt98} Kennicutt R.~C., Jr., 1998, ARA\&A, 36, 189
\bibitem[Kennicutt \& Evans(2012)]{kennicutt12} Kennicutt R.~C., Evans N.~J., 2012, ARA\&A, 50, 531
\bibitem[Kewley et al.(2001)]{kewley01} Kewley L.~J., Dopita M.~A., Sutherland R.~S., Heisler C.~A., Trevena J., 2001, ApJ, 556, 121
\bibitem[Kewley et al.(2006)]{kewley06} Kewley L.~J., Groves B., Kauffmann G., Heckman T., 2006, MNRAS, 372, 961
\bibitem[Kewley \& Ellison(2008)]{kewley08} Kewley L.~J., Ellison S.~L., 2008, ApJ, 681, 1183-1204
\bibitem[Kronberger et al.(2006)]{kronberger06} Kronberger T., Kapferer W., Schindler S., B{\"o}hm A., Kutdemir E., Ziegler B.~L., 2006, A\&A, 458, 69 
\bibitem[Kronberger et al.(2008a)]{kronberger08a} Kronberger T., Kapferer W., Ferrari C., Unterguggenberger S., Schindler S., 2008, A\&A, 481, 337 
\bibitem[Kronberger et al.(2008b)]{kronberger08b} Kronberger, T., Kapferer, W., Unterguggenberger, S., Schindler, S., \& Ziegler, B.~L.\ 2008, \aap, 483, 783
\bibitem[Koleva et al.(2009)]{koleva09} Koleva M., Prugniel P., Bouchard A., Wu Y., 2009, A\&A, 501, 1269
\bibitem[Koopmann \& Kenney(2004)]{koopmann04} Koopmann, R.~A., \& Kenney, J.~D.~P.\ 2004, \apj, 613, 866
\bibitem[Kroupa(2001)]{kroupa01} Kroupa P., 2001, MNRAS, 322, 231 
\bibitem[Larson, Tinsley, \& Caldwell(1980)]{larson80} Larson R.~B., Tinsley B.~M., Caldwell C.~N., 1980, ApJ, 237, 692
\bibitem[Leroy et al.(2008)]{leroy08} Leroy A.~K., Walter F., Brinks E., Bigiel F., de Blok W.~J.~G., Madore B., Thornley M.~D., 2008, AJ, 136, 2782
\bibitem[Liu et al.(2013)]{liu13} Liu G., et al., 2013, ApJ, 778, L41 
\bibitem[MacArthur, Gonz{\'a}lez, \& Courteau(2009)]{macarthur09} MacArthur L.~A., Gonz{\'a}lez J.~J., Courteau S., 2009, MNRAS, 395, 28
\bibitem[Madau \& Dickinson(2014)]{madau14} Madau P., Dickinson M., 2014, ARA\&A, 52, 415
\bibitem[Magdis et al.(2012)]{magdis12} Magdis G.~E., et al., 2012, ApJ, 760, 6
\bibitem[Marasco et al.(2015)]{marasco15} Marasco, A., Debattista,V.~P., Fraternali, F., et al.\ 2015, \mnras, 451, 4223 
\bibitem[Marconi et al.(2004)]{marconi04} Marconi A., Risaliti G., Gilli R., Hunt L.~K., Maiolino R., Salvati M., 2004, MNRAS, 351, 169 
\bibitem[Martinez-Valpuesta et al.(2006)]{martinez06}Martinez-Valpuesta, I., Shlosman, I., \& Heller, C.\ 2006, \apj, 637, 214 
\bibitem[McNamara \& Nulsen(2007)]{mcnamara07} McNamara, B.~R., \&Nulsen, P.~E.~J.\ 2007, \araa, 45, 117
\bibitem[McPartland et al.(2016)]{mcpartland16} McPartland C., Ebeling H., Roediger E., Blumenthal K., 2016, MNRAS, 455, 2994
\bibitem[Merluzzi et al.(2013)]{merluzzi13} Merluzzi P., et al., 2013, MNRAS, 429, 1747
\bibitem[Merluzzi et al.(2016)]{merluzzi16} Merluzzi P., Busarello G., Dopita M.~A., Haines C.~P., Steinhauser D., Bourdin H., Mazzotta P., 2016, MNRAS, 460, 3345
\bibitem[Mihos \& Hernquist(1994)]{mihos94} Mihos J.~C., Hernquist L., 1994, ApJ, 425, L13
\bibitem[Moore et al.(1996)]{moore96} Moore B., Katz N., Lake G., Dressler A., Oemler A., 1996, Natur, 379, 613 
\bibitem[Moretti et al.(2017a)]{moretti17a} Moretti A., et al., 2017, A\&A, 599, A81
\bibitem[Moustakas, Kennicutt, \& Tremonti(2006)]{moustakas06} Moustakas J., Kennicutt R.~C., Jr., Tremonti C.~A., 2006, ApJ, 642, 775
\bibitem[Morokuma-Matsui \& Baba(2015)]{morokuma15} Morokuma-Matsui K., Baba J., 2015, MNRAS, 454, 3792
\bibitem[Navarro, Frenk, \& White(1997)]{navarro97} Navarro J.~F., Frenk C.~S., White S.~D.~M., 1997, ApJ, 490, 493
\bibitem[Ocvirk et al.(2006)]{ocvirk06} Ocvirk P., Pichon C., Lan{\c c}on A., Thi{\'e}baut E., 2006, MNRAS, 365, 46
\bibitem[Osterbrock \& Ferland(2006)]{osterbrock06} Osterbrock D.~E., Ferland G.~J., 2006, agna.book,
\bibitem[Paccagnella et al.(2016)]{paccagnella16} Paccagnella A., et al., 2016, ApJ, 816, L25 
\bibitem[Pacifici et al.(2012)]{pacifici12} Pacifici C., Charlot S., Blaizot J., Brinchmann J., 2012, MNRAS, 421, 2002
\bibitem[Pedrosa et al.(2008)]{pedrosa08} Pedrosa S., Tissera P.~B., Fuentes-Carrera I., Mendes de Oliveira C., 2008, A\&A, 484, 299
\bibitem[Piconcelli et al.(2005)]{piconcelli05} Piconcelli E., Jimenez-Bail{\'o}n E., Guainazzi M., Schartel N., Rodr{\'{\i}}guez-Pascual P.~M., Santos-Lle{\'o} M., 2005, A\&A, 432, 15
\bibitem[Pilbratt et al.(2010)]{pilbratt10} Pilbratt G.~L., et al., 2010, A\&A, 518, L1 
\bibitem[Poggianti, Bressan, \& Franceschini(2001)]{poggianti01} Poggianti B.~M., Bressan A., Franceschini A., 2001, ApJ, 550, 195
\bibitem[Poggianti et al.(2016)]{poggianti16} Poggianti B.~M., et al., 2016, AJ, 151, 78
\bibitem[Poggianti et al.(2017)]{poggianti17} Poggianti B.~M., et al., 2017, arXiv, arXiv:1704.05086 
\bibitem[Poglitsch et al.(2010)]{poglitsch10} Poglitsch A., et al., 2010, A\&A, 518, L2
\bibitem[Popescu \& Tuffs(2002)]{popescu02} Popescu C.~C., Tuffs R.~J., 2002, MNRAS, 335, L41
\bibitem[Popping, Somerville, \& Trager(2014)]{popping14} Popping G., Somerville R.~S., Trager S.~C., 2014, MNRAS, 442, 2398 
\bibitem[Poznanski, Prochaska, \& Bloom(2012)]{poznanski12} Poznanski D., Prochaska J.~X., Bloom J.~S., 2012, MNRAS, 426, 1465 
\bibitem[Proxauf, {\"O}ttl, \& Kimeswenger(2014)]{proxauf14} Proxauf B., {\"O}ttl S., Kimeswenger S., 2014, A\&A, 561, A10 
\bibitem[Ranalli, Comastri, \& Setti(2003)]{ranalli03} Ranalli P., Comastri A., Setti G., 2003, A\&A, 399, 39
\bibitem[Risaliti, Maiolino, \& Salvati(1999)]{risaliti99} Risaliti G., Maiolino R., Salvati M., 1999, ApJ, 522, 157
\bibitem[Roussel(2013)]{roussel13} Roussel H., 2013, PASP, 125, 1126
\bibitem[Saftly et al.(2015)]{saftly15} Saftly W., Baes M., De Geyter G., Camps P., Renaud F., Guedes J., De Looze I., 2015, A\&A, 576, A31
\bibitem[S{\'a}nchez et al.(2016)]{sanchez16} S{\'a}nchez S.~F., et al., 2016, RMxAA, 52, 21 
\bibitem[S{\'a}nchez-Bl{\'a}zquez et al.(2006)]{sanchez06} S{\'a}nchez-Bl{\'a}zquez P., et al., 2006, MNRAS, 371, 703
\bibitem[Sarzi et al.(2006)]{sarzi06} Sarzi M., et al., 2006, MNRAS, 366, 1151 
\bibitem[Sharp \& Bland-Hawthorn(2010)]{sharp10} Sharp R.~G., Bland-Hawthorn J., 2010, ApJ, 711, 818 
\bibitem[Silk \& Rees(1998)]{silk98} Silk, J., \& Rees, M.~J.\ 1998, \aap, 331, L1 
\bibitem[Smith et al.(2010)]{smith10_1} Smith R.~J., et al., 2010, MNRAS, 408, 1417
\bibitem[Smith et al.(2010)]{smith10} Smith M.~W.~L., et al., 2010, A\&A, 518, L51 
\bibitem[Springel(2000)]{springel00} Springel V., 2000, MNRAS, 312, 859 
\bibitem[Steinacker, Baes, \& Gordon(2013)]{steinacker13} Steinacker J., Baes M., Gordon K.~D., 2013, ARA\&A, 51, 63
\bibitem[Steinhauser et al.(2012)]{steinhauser12} Steinhauser D., Haider M., Kapferer W., Schindler S., 2012, A\&A, 544, A54 
\bibitem[Steinhauser, Schindler, \& Springel(2016)]{steinhauser16} Steinhauser D., Schindler S., Springel V., 2016, A\&A, 591, A51
\bibitem[Takeda, Nulsen, \& Fabian(1984)]{takeda84} Takeda H., Nulsen P.~E.~J., Fabian A.~C., 1984, MNRAS, 208, 261
\bibitem[Thomas, Greggio, \& Bender(1999)]{thomas99} Thomas D., Greggio L., Bender R., 1999, MNRAS, 302, 537 
\bibitem[Tinsley \& Larson(1979)]{tinsley79} Tinsley B.~M., Larson R.~B., 1979, MNRAS, 186, 503
\bibitem[Toomre(1977)]{toomre77} Toomre A., 1977, egsp.conf, 401
\bibitem[Tojeiro et al.(2007)]{tojeiro07} Tojeiro R., Heavens A.~F., Jimenez R., Panter B., 2007, MNRAS, 381, 1252 
\bibitem[Valluri(1993)]{valluri93} Valluri M., 1993, ApJ, 408, 57 
\bibitem[Vazdekis et al.(2010)]{vazdekis10} Vazdekis A., S{\'a}nchez-Bl{\'a}zquez P., Falc{\'o}n-Barroso J., Cenarro A.~J., Beasley M.~A., Cardiel N., Gorgas J., Peletier R.~F., 2010, MNRAS, 404, 1639
\bibitem[Verstappen et al.(2013)]{verstappen13} Verstappen J., et al., 2013, A\&A, 556, A54 
\bibitem[Vollmer et al.(2001)]{vollmer01} Vollmer B., Cayatte V., Balkowski C., Duschl W.~J., 2001, ApJ, 561, 708
\bibitem[Vulcani et al.(2015)]{vulcani15} Vulcani B., Poggianti B.~M., Fritz J., Fasano G., Moretti A., Calvi R., Paccagnella A., 2015, ApJ, 798, 52 
\bibitem[Wallerstein(1962)]{wallerstein62} Wallerstein G., 1962, ApJS, 6, 407
\bibitem[Weinberg(1985)]{weinberg85} Weinberg, M.~D.\ 1985, \mnras,213, 451
\bibitem[Weiner et al.(2007)]{weiner07} Weiner B.~J., et al., 2007, ApJ, 660, L39
\bibitem[Wilkinson et al.(2015)]{wilkinson15} Wilkinson D.~M., et al., 2015, MNRAS, 449, 328
\bibitem[Wuyts et al.(2013)]{wuyts13} Wuyts S., et al., 2013, ApJ, 779, 135
\end{thebibliography}
\end{document}